\documentclass[runningheads]{llncs}
\usepackage[mobile]{eccv}
\usepackage[dvipsnames]{xcolor}

\usepackage{url}
\usepackage{booktabs}       %
\usepackage{amsfonts}       %
\usepackage{nicefrac}       %
\usepackage{microtype}      %
\usepackage{xcolor}         %
\usepackage{tikz}
\usetikzlibrary{calc, decorations.pathreplacing, positioning}

\newtheorem{assumption}{Assumption}

\usepackage{comment}
\usepackage{algorithm}
\usepackage[noend]{algpseudocode}
\algrenewcommand\algorithmicrequire{\textbf{Input:}}
\algrenewcommand\algorithmicensure{\textbf{Output:}}
\usepackage{graphicx}

\usepackage{subcaption}
\usepackage{diagbox}
\usepackage{multirow}
\usepackage{calc}
\usepackage{mathtools, nccmath}
\usepackage{pgfplots}
\pgfplotsset{compat=1.17}

\usepackage{enumitem}

\usepackage{colortbl}
\usepackage{tabularx}

\definecolor{tabfirst}{rgb}{1, 0.7, 0.7} %
\definecolor{tabsecond}{rgb}{1, 0.85, 0.7} %
\definecolor{tabthird}{rgb}{1, 1, 0.7} %

\usepackage{bbding}
\newcommand{\cmark}{\textcolor{green}{\Checkmark}}%
\newcommand{\xmark}{\textcolor{red}{\XSolidBrush}}

\usepackage{amsmath,amsfonts,bm}

\def\eqref#1{equation~\ref{#1}}

\def\1{\bm{1}}

\DeclareMathAlphabet{\mathsfit}{\encodingdefault}{\sfdefault}{m}{sl}
\SetMathAlphabet{\mathsfit}{bold}{\encodingdefault}{\sfdefault}{bx}{n}

\DeclareMathOperator*{\argmax}{arg\,max}
\DeclareMathOperator*{\argmin}{arg\,min}

\usepackage{eccvabbrv}

\usepackage{graphicx}
\usepackage{booktabs}

\usepackage[accsupp]{axessibility}  %

\usepackage{setspace}

\usepackage{hyperref}

\usepackage{orcidlink}

\pgfmathsetseed{1}
\pgfmathsetmacro{\rndA}{rnd}
\pgfmathsetmacro{\rndB}{rnd}

\begin{document}

\title{Adaptive Selection of Sampling-Reconstruction in Fourier Compressed Sensing} 

\titlerunning{Adaptive Selection of Sampling-Reconstruction in FCS}

\author{Seongmin Hong\inst{1}\orcidlink{0000-0002-4723-4547} \and
Jaehyeok Bae\inst{1}\orcidlink{0009-0005-5709-1374} \and
Jongho Lee\inst{1,2}\thanks{\ Corresponding authors.}\orcidlink{0000-0002-9485-5434} \and\\
Se Young Chun\inst{1,2,3\star}\orcidlink{0000-0001-8739-8960}}

\authorrunning{S.~Hong et al.}

\institute{$^1$Dept. of Electrical and Computer Engineering, $^2$INMC, $^3$IPAI  \\
Seoul National University, Republic of Korea \\
\email{\{smhongok, wogur110, jonghoyi, sychun\}@snu.ac.kr}\\
Project page: \url{https://smhongok.github.io/ada-sel.html}}

\maketitle

\begin{abstract}
Compressed sensing (CS) has emerged to overcome the inefficiency of Nyquist sampling. However, traditional optimization-based reconstruction is slow and may not yield a high-quality image in practice. Deep learning-based reconstruction has been a promising alternative to optimization-based reconstruction, outperforming it in accuracy and computation speed. Finding an efficient sampling method with deep learning-based reconstruction, especially for Fourier CS remains a challenge. Existing joint optimization of sampling-reconstruction works ($\mathcal{H}_1$) optimize the sampling mask but yield suboptimal results because it is not adaptive to each data point. Adaptive sampling ($\mathcal{H}_2$) has also disadvantages of difficult optimization and Pareto sub-optimality. Here, we propose a novel adaptive selection of sampling-reconstruction ($\mathcal{H}_{1.5}$) framework that selects the best sampling mask and reconstruction network for each input data. We provide theorems that our method has a lower infimum of the true risk compared to  $\mathcal{H}_1$ and effectively solves the Pareto sub-optimality problem in sampling-reconstruction by using separate reconstruction networks for different sampling masks. To select the best sampling mask, we propose to quantify the high-frequency Bayesian uncertainty of the input, using a super-resolution space generation model. Our method outperforms joint optimization of sampling-reconstruction ($\mathcal{H}_1$) and adaptive sampling ($\mathcal{H}_2$) by achieving significant improvements on several Fourier CS problems.
\end{abstract}

\section{Introduction}
\begin{figure}[!t]
  \centering
  \begin{subfigure}[b]{0.7\linewidth}
    \centering
    \begin{tikzpicture}
      \node at (0,0) {\includegraphics[width=0.97\linewidth]{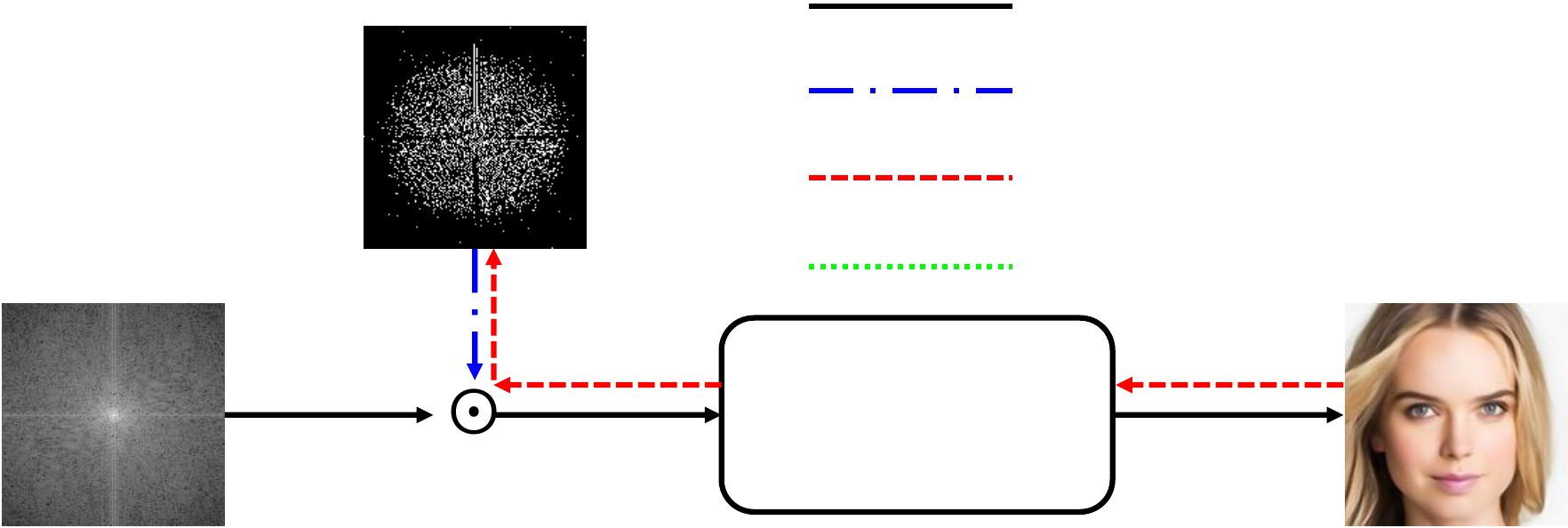}};
      \node[text width=\textwidth, font=\small, align=left] at (5.55,1.38) {Data};
      \node[text width=\textwidth, font=\small, align=left] at (5.55, 0.92) {Mask};
      \node[text width=\textwidth, font=\small, align=left] at (5.55,0.43) {Backpropagation};
      \node[text width=\textwidth, font=\small, align=left] at (5.55, 0) {Selection};
      \node[text width=\textwidth, font=\small, align=left] at (7.15,0) {Output $I'$};
      \node[text width=\textwidth, font=\small, align=left] at (0.22,0) {Input $k$};
      \node[text width=\textwidth, font=\small, align=left] at (0.75,1) {Mask $M$};
      \node[text width=\textwidth, font=\small, align=center] at (0.71,-0.78) {Recon. \\ Network $\theta$};
    \end{tikzpicture}
    \caption{Joint optimization of sampling-reconstruction ($\mathcal{H}_1$~\cite{bahadir2020deep, zhang2020extending, lazarus2019sparkling, chaithya2021learning, chaithya2022optimizing})}
    \label{fig:absa}
  \end{subfigure}

  \par\bigskip

  \begin{subfigure}[b]{0.7\linewidth}
    \centering
    \begin{tikzpicture}
      \node at (0,0) {\includegraphics[width=0.97\linewidth]{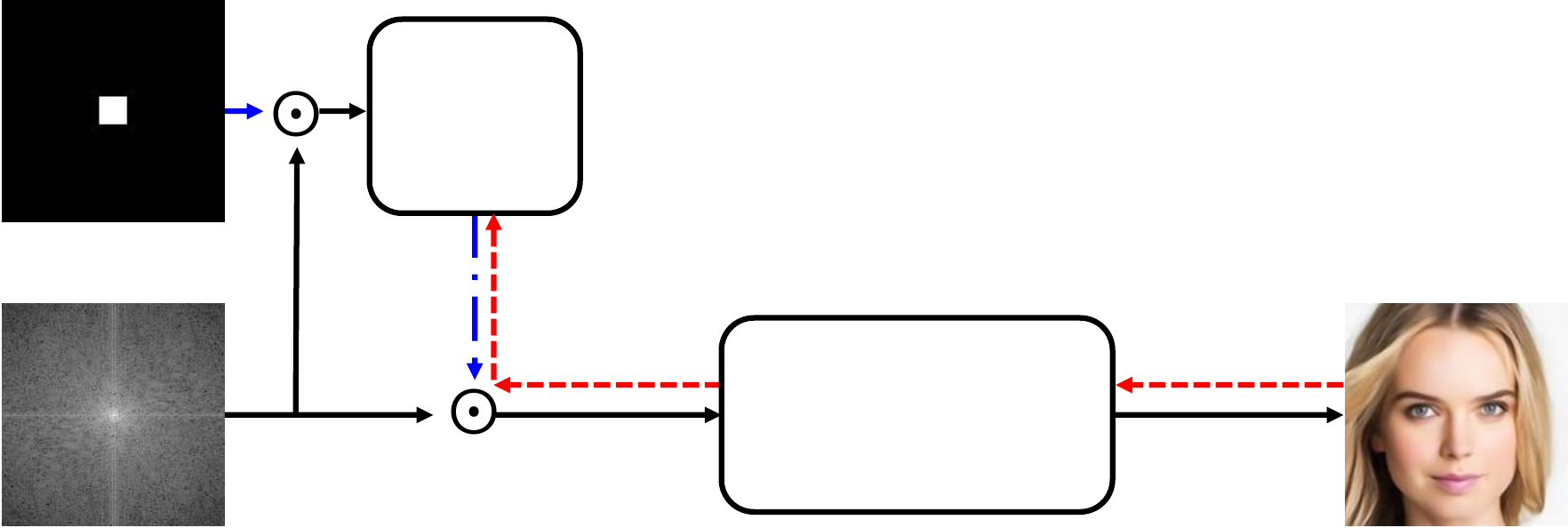}};
      \node[text width=\textwidth, font=\small, align=left] at (7.15,0) {Output $I'$};
      \node[text width=\textwidth, font=\small, align=left] at (0.21,-0.03) {Input $k$};
      \node[text width=\textwidth, font=\small, align=left] at (0.13,1.6) {Mask $M_0$};
      \node[text width=\textwidth, font=\small, align=left] at (2.5,0.77) {$\pi_\phi$ \quad $:M_0\mathcal{K} \rightarrow \mathcal{M}$};
      \node[text width=\textwidth, font=\small, align=center] at (0.71,-0.78) {Recon. \\ Network $\theta$};
    \end{tikzpicture}
    \caption{Adaptive sampling ($\mathcal{H}_2$~\cite{ bakker2020experimental, yin2021end, yang2022l2sr, bakker2022learning})}
    \label{fig:absb}
  \end{subfigure}

  \par\bigskip
  
  \begin{subfigure}[b]{0.7\linewidth}
    \centering
    \begin{tikzpicture}
      \node at (0,0) {\includegraphics[width=0.97\linewidth]{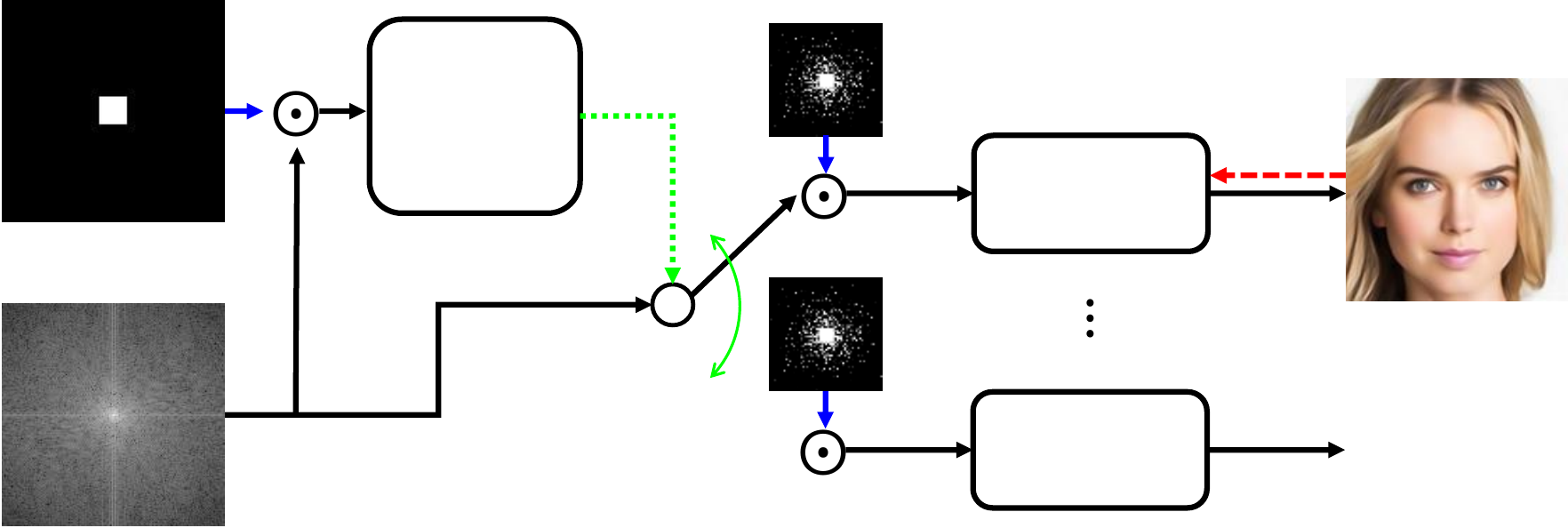}};
      \node[text width=\textwidth, font=\small, align=left] at (7.15,1.2) {Output $I'$};
      \node[text width=\textwidth, font=\small, align=left] at (0.21,-0.03) {Input $k$};
      \node[text width=\textwidth, font=\small, align=left] at (0.13,1.6) {Mask $M_0$};
      \node[text width=\textwidth, font=\small, align=left] at (2.5,0.75) {$e_\psi$};
      \node[text width=\textwidth, font=\small, align=left] at (4.85,0.95) {$M_1$};
      \node[text width=\textwidth, font=\small, align=left] at (5.75,0.38) {$\theta_1$};
      \node[text width=\textwidth, font=\small, align=left] at (4.85,-0.36) {$M_J$};
      \node[text width=\textwidth, font=\small, align=left] at (5.75,-0.99) {$\theta_J$};
    \end{tikzpicture}
    \caption{Adaptive selection of sampling-reconstruction ($\mathcal{H}_{1.5}$, ours)}
    \label{fig:absc}
  \end{subfigure}

  \caption{%
  We propose the adaptive selection of sampling-reconstruction ($\mathcal{H}_{1.5}$, \textcolor{red}{c}). In Fourier compressed sensing, there were two classes of methods for finding the optimal sampling: joint optimization of sampling-reconstruction ($\mathcal{H}_1$, \textcolor{red}{a}) and adaptive sampling ($\mathcal{H}_2$, \textcolor{red}{b}). $\mathcal{H}_1$ has low potential as its mask ${M}$ is not adaptive to each data point. $\mathcal{H}_2$ poses a challenge in optimizing the mask generator ($\pi_\phi$) and then exhibits Pareto suboptimality, where a single $\theta$ is not optimal for multiple masks $M \in \mathcal{R}(\pi_\phi)$.
  In contrast, $\mathcal{H}_{1.5}$ is adaptive for input $k$ ($e_\psi$ in \textcolor{red}{c} selects the best $M$-$\theta$ pair), avoids the challenge of backpropagation to discrete space (red lines in \textcolor{red}{a, b, c}), and achieves Pareto optimality by dedicating each network $\theta_j$ exclusively to $M_j$. $\odot$ denotes the componentwise multiplication.}
  \label{fig:abs}
\end{figure}
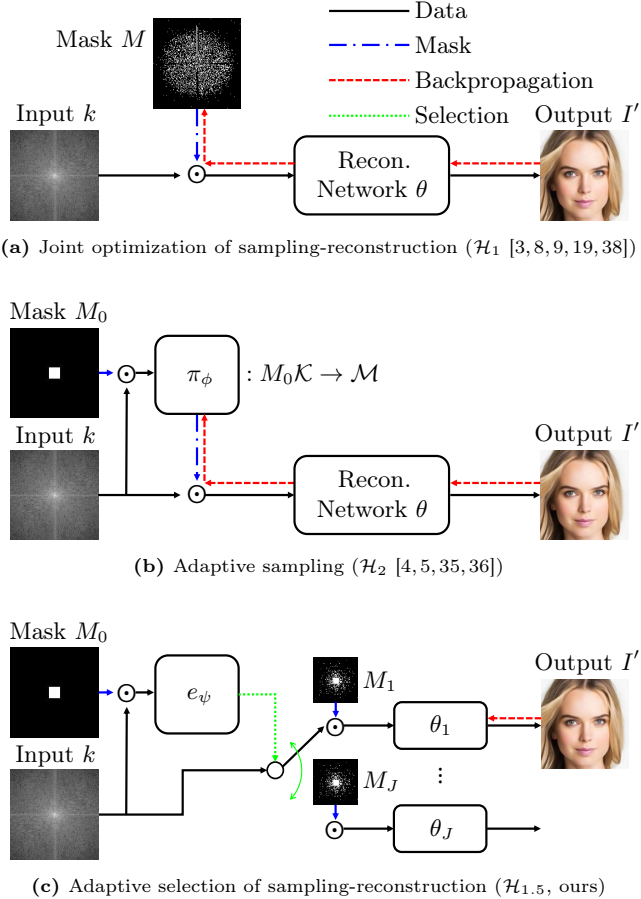

Compressed sensing (CS) has revolutionized the field of image acquisition, enabling the reconstruction of high-quality images from a reduced number of measurements. This remarkable feat is achieved by exploiting the sparsity of natural images (or medical images) in certain transform domains. The CS theory~\cite{candes2006cs,donoho2006compressed} guarantees that an image can be accurately recovered from a non-adaptive random sampling pattern, with much fewer samples than the Nyquist-Shannon sampling theorem requires, if the image has a sparse representation in that domain. Before the era of deep learning, CS used to refer to obtaining the final image through $l_1$-regularized reconstruction, \textit{i.e.}, solving Lasso~\cite{candes2006cs}. Recently, reconstruction has often been performed using deep neural networks trained on the data. In this paper, we focus on deep learning-based reconstructions.

Fourier compressed sensing (Fourier CS) refers to CS where the measurement is in the discrete Fourier transform (DFT) of an image. As electromagnetic waves are inherently wave-like, obtaining spatial information such as pixel values directly is not feasible. Instead, spatial information is acquired through DFT. Obtaining samples for every Fourier-transformed element can be costly. Unlike the CS theory that suggests random sampling is sufficient, the sampling results of Fourier CS, similar to the prior in natural images, concentrate a significant amount of energy in the low-frequency (LF) components~\cite{lustig2007sparse,wang2009variable}. Despite deviating from the random sampling principle of CS theory, Fourier CS achieves excellent image quality in many domains by extensively sampling LF components; hence it has been successfully applied to various electromagnetic imaging applications, including magnetic resonance imaging (MRI)~\cite{lustig2007sparse, lustig2008compressed} or radar~\cite{de2019compressed, hong2023advanced}.

\begin{table}[t!]
    \centering
    \caption{Adaptive selection of sampling-reconstruction ($\mathcal{H}_{1.5}$) alleviates the drawbacks of joint optimization of sampling-reconstruction ($\mathcal{H}_1$) and adaptive sampling ($\mathcal{H}_2$). 
    }
    \label{tab:abs}
    \setlength{\tabcolsep}{5pt}
    \begin{tabular}{lccc}      
         \toprule
         Methods & \begin{tabular}{c}Adaptive \\ to input $k$\end{tabular} & \begin{tabular}{c} Backprop to a\\ continuous space\end{tabular} & \begin{tabular}{c} Pareto \\ optimal $\theta$ \end{tabular} \\
         \midrule
         $\mathcal{H}_1$~\cite{bahadir2020deep, zhang2020extending, lazarus2019sparkling, chaithya2021learning, chaithya2022optimizing,wang2022b} & \xmark   & \xmark \tiny{(\cmark~\cite{wang2022b})}   & \cmark  \\
         \cmidrule{1-1}
         $\mathcal{H}_2$~\cite{ bakker2020experimental, yin2021end, yang2022l2sr, bakker2022learning,sanchez2022learning} &  \cmark & \xmark \tiny{(\cmark~\cite{sanchez2022learning})} & \xmark  \\
         \cmidrule{1-1}
         $\mathcal{H}_{1.5}$ (ours) & \cmark & \cmark & \cmark \\ \bottomrule
    \end{tabular}
\end{table}

However, finding an optimal sampling method that is both efficient and effective remains a challenge in Fourier CS. 
One approach is the \emph{joint optimization of sampling-reconstruction} (denoted by $\mathcal{H}_1$)~\cite{bahadir2020deep, zhang2020extending, lazarus2019sparkling, chaithya2021learning, chaithya2022optimizing, wang2022b}, where the parameters of both the sampling and reconstruction networks are jointly trained using the dataset, as depicted in \cref{fig:absa}. But this approach has two drawbacks, as described in \cref{tab:abs}. Firstly, it is not adaptive to each data point, \textit{i.e.}, the optimized sampling and the reconstruction parameter would not be the best pair for a specific input. Moreover, the reconstruction network is usually trained by backpropagation. To employ backpropagation, the parameters are expected to be defined within continuous spaces. This makes training the sampling mask not trivial, as it is defined in a discrete space. Most methods~\cite{bahadir2020deep, zhang2020extending, lazarus2019sparkling, chaithya2021learning, chaithya2022optimizing} just perform discrete optimization anyway using the straight-through estimator~\cite{bengio2013estimating, zhang2020extending}.  

The other approach is \emph{adaptive sampling} (denoted by $\mathcal{H}_2$)~\cite{zhang2019reducing, bakker2020experimental, pineda2020active, yin2021end, van2021active, yang2022l2sr, bakker2022learning}, which aims to generate the best sampling mask for each data point (or each image) based on the fact that a predetermined sampling mask may not be optimal for every situation.
Most adaptive sampling studies generate the optimal sampling mask based on the information from the initially measured LF components of each data point, which has the potential to achieve excellent results. Then, they usually have a single reconstruction network that is responsible for many optimal masks for all data points. Unfortunately, there are a couple of major issues in current adaptive sampling, as in \cref{tab:abs}. Similar to the joint optimization of sampling-reconstruction models, optimizing the mask generator is challenging due to the broad and discrete mask space, as depicted in \cref{fig:absb}. Secondly, the single reconstruction network for diverse sampling masks may not be Pareto optimal, which we called \emph{Pareto suboptimal reconstruction network}. Note that a similar issue can arise in the task of restoring various degradations (\textit{e.g.}, the performance of a blind denoising network trained on multiple noise levels is usually lower than that of an identical network trained only on the specific noise level used as the actual input~\cite{zhang2017beyond}).

In this paper, we propose a novel adaptive selection of the sampling-reconstruction framework for Fourier CS that alleviates the drawbacks of joint optimization of sampling-reconstruction and adaptive sampling. It is adaptive to each data point, avoids backpropagation to discrete spaces, and its reconstruction network is Pareto optimal. In the adaptive selection, we first sample LF components quickly and then leverage a super-resolution (SR) space generation model, to quantify the high-frequency (HF) Bayesian uncertainty. This approach ensures that HF components, which contain crucial details, are sampled more effectively, leading to improved reconstruction quality. 
The main contributions of our paper are as follows: 
\begin{itemize}[leftmargin=*]
\item Proposing a novel \emph{adaptive selection of sampling-reconstruction} framework for Fourier CS that alleviates the drawbacks of joint optimization of sampling-reconstruction and adaptive sampling with theoretical justification.
\item Designing the adaptive selection to efficiently quantify HF Bayesian uncertainty by leveraging an SR space generation model for determining sampling masks.
\item Demonstrating that our adaptive selection improves performance in multiple Fourier CS problems such as facial image restoration (up to 0.04 average gain in SSIM) and multi-coil MR reconstruction (up to 0.004 average gain in SSIM).

\end{itemize}

\section{Related Works}

\subsection{Fourier compressed sensing}
Fourier CS can be defined as the following regression problem.
Let us define the dataset $\mathcal{D}=\{(k^i,I^i)\}_{i=1}^{N}$ such that $k^1, k^2, \dots, k^N \in \mathcal{K} \subseteq \mathbb{C}^L$ are fully-sampled k-space data and $I^1, I^2, \dots, I^N \in \mathcal{I} \subseteq \mathbb{R}^L$ are the corresponding images, respectively. Let us define the mask space by $\mathcal{M} \subseteq \{0,1\}^{L \times L}$ whose element is a diagonal binary matrix (indicating acquired (1) and unacquired (0) grid points). %
Let $h(k; M, \theta):\mathcal{K} \times \mathcal{M} \times \Theta \rightarrow \mathcal{I}$ be a reconstruction function of $k$ for a sampling mask $M \in \mathcal{M}$ and a reconstruction network (\textit{e.g.}, U-Net~\cite{ronneberger2015u} or E2E-VarNet~\cite{sriram2020end}) parameterized by $\theta \in \Theta$. Then, this function $h$ can be used as a joint sampling-reconstruction model that optimizes both the sampling mask $M$ and the reconstruction network $\theta$. Specifically, for the given dataset $\mathcal{D}$, the model is optimized to minimize the following empirical risk \[\Hat{\mathcal{L}}[h(\mathcal{K};M,\theta)] = \frac{1}{N}\sum_{i=1}^{N} l(I^i,h(k^i;M,\theta)),\]
where $l$ is the loss function (\textit{e.g.}, $l(I,\hat{I}) = 1 - \mathrm{SSIM}(I,\hat{I})$).

\subsection{Joint optimization of sampling-reconstruction}\label{sec2.2}
One of the recent approaches to finding a good sampling mask is joint optimization of sampling-reconstruction~\cite{bahadir2020deep, zhang2020extending, lazarus2019sparkling, chaithya2021learning, chaithya2022optimizing, wang2022b}, which reconstructs the image with a non-adaptive mask $M \in \mathcal{M}$. They are defined as follows:
\begin{equation}\label{eq2.2.1}
    \mathcal{H}_1 = \{h(\cdot;M,\theta) | M \in \mathcal{M}, \theta \in \Theta\}.
\end{equation}
They jointly optimize $M$ and $\theta$ for a \emph{dataset}; however, $M$ is not adaptive to each data point. Whether using a tailored $M$ or not, the fundamental limitation of $\mathcal{H}_1$ is that the sampling mask is not optimal for each data point. %
Moreover, they exhibit highly varying results across different settings, as they require discrete optimization~\cite{bahadir2020deep, zhang2020extending, lazarus2019sparkling, chaithya2021learning, chaithya2022optimizing} or virtual data~\cite{wang2022b}, as shown in \cref{fig:absa} and \cref{tab:abs}. %

\subsection{Adaptive sampling}\label{sec:2.3}
Some recent works~\cite{bakker2020experimental, pineda2020active, yang2022l2sr,yin2021end, yang2022l2sr, bakker2022learning} employ adaptive mask, using a mask generator $\pi_\phi:M_0\mathcal{K} \rightarrow \mathcal{M}$, parameterized by $\phi \in \Phi$, as shown in \cref{fig:absb}. Here, $M_0 \in \mathcal{M}$ denotes a mask that samples only LF components. Adaptive sampling approaches minimize $\Hat{\mathcal{L}}[h]$ on
\begin{equation}\label{eq2.2.2}
    \mathcal{H}_2 = \{h(\cdot;\pi_\phi(\cdot),\theta) | \pi_\phi:M_0\mathcal{K} \rightarrow \mathcal{M}, \theta \in \Theta, \phi \in \Phi \}.
\end{equation}
Obviously, $\mathcal{H}_1 \subseteq \mathcal{H}_2$. That is, $\mathcal{H}_2$ has the greatest potential but is hard to train because of its complexity. 
Specifically, $\mathcal{H}_2$ faces two main issues: the difficulty of the mask generator ($\pi_\phi$) optimization, and Pareto suboptimality of $\theta$, due to the fact that a single reconstruction network is responsible for multiple masks.

Previous studies on $\mathcal{H}_2$ have used reinforcement learning~\cite{bakker2020experimental, pineda2020active, yang2022l2sr} or backpropagation~\cite{ bakker2020experimental, yin2021end, yang2022l2sr, bakker2022learning} using straight-through estimator~\cite{bengio2013estimating, zhang2020extending} to optimize mask generator $\pi_\phi$, but this is a complicated problem because the action space $\mathcal{M}$ is too broad and discrete. %
$\theta$ is Pareto suboptimal in adaptive sampling studies since there are multiple sampling masks $M$ while only one $\theta$ exists at inference time.
Due to these difficulties, most adaptive sampling studies in CS-MRI have been conducted in a clinically less relevant simple setting of single-coil~\cite{zhang2019reducing, bakker2020experimental, pineda2020active, yin2021end, van2021active, yang2022l2sr}. There was only one study conducted on realistic multi-coil setting~\cite{bakker2022learning}, but the final models of ~\cite{bakker2022learning} turned out to be non-adaptive, which is an unintended consequence.

To avoid optimization in discrete space, \cite{sanchez2022learning} proposed adaptive sampling using a conditional Generative Adversarial Network (cGAN). During sampling,  cGAN assesses the uncertainty of samples yet to be acquired. Subsequently, the user selects a sample with the highest uncertainty for acquisition (\textit{i.e.}, greedy algorithm). This process of quantifying and sampling is iteratively repeated. While this method benefits from backpropagation occurring only in continuous space (\cref{tab:abs}) it still faces the challenge of a single reconstruction network, having to perform reconstruction for all masks. Consequently, CS-MRI experiments were conducted using a simple single-coil setup.

\subsection{Super-resolution space generation}\label{sec2.3}

Super-resolution (SR) space generation~\cite{lugmayr2021ntire, lugmayr2022ntire} aims to create diverse high-resolution (HR) images that can be downsampled to the same low-resolution (LR) image (\textit{i.e.}, $q_\psi(I_\text{HR} | I_\text{LR}) \not\in \{ \delta_I | I\in \mathcal{I} \}$). For this purpose, a stochastic approach is used rather than a deterministic one. 
Conditional normalizing flow-based SR space generation methods~\cite{lugmayr2020srflow,Song_2022_CVPR,Hong_2023_ICCV} explicitly obtain $q_\psi(I_\text{HR} | I_\text{LR})$ using a diffeomorphic mapping $f_\psi:\mathcal{I} \rightarrow \mathcal{Z}$ and a simple base distribution $q_z$ (\textit{e.g.}, standard Gaussian), as $q_\psi(I_\text{HR} | I_\text{LR}) = q_z(f_{\psi}(I_\text{HR} ; I_\text{LR})) \lvert \mathrm{det} \frac{\partial f_{\psi}}{\partial I_\text{HR}} (I_\text{HR} ; I_\text{LR})\rvert.$
Since $f_\psi$ is invertible, $q_z$ and $f_\psi^{-1}$ can be used to directly sample $I_\text{HR}$ from $q_\psi$ (\textit{i.e.}, $z \sim q_z \Longrightarrow f_\psi^{-1}(z;I_\text{LR}) \sim q_\psi(\cdot | I_\text{LR})$). In this work, we trained and exploited a recent robust flow-based SR space generation method~\cite{Song_2022_CVPR}, with tuned hyperparameters \cite{Hong_2023_ICCV} for stability, to generate HR images from the corresponding LR image that is reconstructed from undersampled k-space data with mask $M_0$.

\section{Proposed methods}
\label{sec3.2}
In \cref{sec2.2} and \ref{sec:2.3}, we investigated the difficulty of optimizing the mask generator $\pi_\phi$, and Pareto optimal $\theta$ (for all masks). This section proposes a novel scheme, \emph{adaptive selection of sampling-reconstruction}, which does not encounter these problems. Using two Theorems \ref{thm1} and \ref{thm2}, we explain our adaptive selection ($\mathcal{H}_{1.5}$) is better than the joint optimization ($\mathcal{H}_1$) and the adaptive sampling ($\mathcal{H}_2$).  

\subsection{Adaptive selection of sampling-reconstruction}\label{sec3.2.1}
Our adaptive selection model $\mathcal{H}_{1.5}$ is defined as follows:
\begin{equation}\label{eq3.2.1.1}
    \mathcal{H}_{1.5} = \left\{\sum_{j=1}^J e_\psi(\cdot)_j h(\cdot;M_j,\theta_j) \middle|
    e_\psi: M_0\mathcal{K} \rightarrow \{e_j\}_{j=1}^J, M_j \in \mathcal{M}, \theta_j \in \Theta , \forall j \right\}   
\end{equation}
where $e_j$ is the $j$-th standard unit vector (\textit{i.e.}, one-hot vector). Each submodel $h(\cdot; M_j,\theta_j)$ contains mask $M_j$ and reconstruction network $\theta_j$ as a pair, which is Pareto optimal. At inference time, each data selects an appropriate submodel through the mask selector $e_\psi(\cdot)_j$, which takes input $M_0k$. This scheme is similar to a segmented regression problem that ensembles multiple submodels using one-hot encoding.
\begin{remark}\label{remark1}
If $\mathcal{H}_{1}$ is a linear regression, then $\mathcal{H}_{1.5}$ is a segmented linear regression. 
\end{remark}

We propose the following Theorems \ref{thm1} and \ref{thm2}. Theorem \ref{thm1} shows that $\mathcal{H}_{1.5}$ is better than $\mathcal{H}_1$ due to its adaptivity, and Theorem \ref{thm2} demonstrates that $\mathcal{H}_{1.5}$ is superior to $\mathcal{H}_{2}$ because $\mathcal{H}_{2}$ has poor Pareto optimality. %
\begin{theorem}[Adaptive selection is better than non-adaptive]\label{thm1}
    For a true risk $\mathcal{L}$,
    $\underset{{h \in \mathcal{H}_{1.5}}}{\inf} \mathcal{L}[h] \leq \underset{h \in \mathcal{H}_{1}}{\inf} \mathcal{L}[h]$.
\end{theorem}

\begin{theorem}[Adaptive selection is Pareto optimal]\label{thm2}
    For a true risk $\mathcal{L}$,
    $|\pi_\phi(M_0 \mathcal{K})| \leq J \Rightarrow$ 
    $\underset{{h \in \mathcal{H}_{1.5}}}{\inf} \mathcal{L}[h] \leq \underset{{h \in \mathcal{H}_{2}}}{\inf} \mathcal{L}[h]$.
\end{theorem}
 
\noindent Please see the supplementary material for the proofs. Theorem \ref{thm2} requires an assumption that optimizing $\pi_\phi$ is difficult (\textit{i.e.}, $|\pi_\phi(M_0 \mathcal{K})| \leq J$), which is justified in Section \ref{sec2.2} (\textit{e.g.}, The final model in \cite{bakker2022learning} converged to $|\pi_\phi(M_0 \mathcal{K})| \rightarrow 1$, which means non-adaptive).
Theorems \ref{thm1} and \ref{thm2} suggest that the proposed adaptive selection scheme ($\mathcal{H}_{1.5}$) may outperform both non-adaptive methods ($\mathcal{H}_{1}$) and adaptive sampling ($\mathcal{H}_{2}$). In Section \ref{sec3.2.2}, we describe the implementation of the scheme using the HF Bayesian uncertainty quantified by an SR space generation method~\cite{Song_2022_CVPR}.

\subsection{How to and what to adaptively select?}\label{sec3.2.2}
\subsubsection{(How to) proposed mask selector $e_\psi$:} 
The sample variance of a generative model to produce diverse samples can be utilized to quantify uncertainty for adaptive sampling~\cite{sanchez2022learning}. Specifically in Fourier CS, inspired by the idea of initially sampling LF components, we employ an SR space generation model~\cite{Song_2022_CVPR, Hong_2023_ICCV} as a HF uncertainty quantifier. 

The sample variance $v(M_0 k):=(\widehat{\mathrm{Var}}_{q_\psi}[k'_s])_{s=1}^S$ is an estimator of the mean square error in k-space domain, where $S$ is the number of the SR samples and $k'_s$ is the Fourier transform of the $s$-th sample. We make up the mask selector $e_\psi$ using $v(M_0 k)$. Specifically, at train time, we normalize $v(M_0 k)$ so that $u(v) := v / \lVert v \rVert_2 $ and then use the k-means++ clustering algorithm~\cite{arthur2007k} to $\{u(v(M_0 k^i))\}_{i=1}^N$ to create centroids $(c_j)_{j=1}^J$. At inference time, we select adaptive mask index $j$ by calculating the distance $u(v(M_0 k))$ and $(c_j)_{j=1}^J$. %

We also need to determine the number of the sampling-reconstruction pairs $J$. 
Thinking of \cref{remark1}, increasing $J$ doesn't always mean better average performance; while
increasing $J$ can help in robustly handling outliers. This trade-off can be organized as \cref{remark2}: 
\begin{remark}[Trade-off with the number of segments $J$]\label{remark2}
As $J$ increases, despite more training resources, the average performance reaches a plateau at some point, but it becomes more robust against outliers.
\end{remark}
The choice of $J$ depends on the user's needs; we defaulted to $J=3$. We delve into and validate \cref{remark2} in \cref{sec:5}.

\begin{figure}[t!]
    \par\bigskip
  \centering
  \begin{tikzpicture}[    
        every node/.style={anchor=east, inner sep=0pt, text width=\textwidth, align=left, font=\scriptsize},
      ] 
    \node[inner sep=0pt] (image) at (0,0){\includegraphics[width=\linewidth]{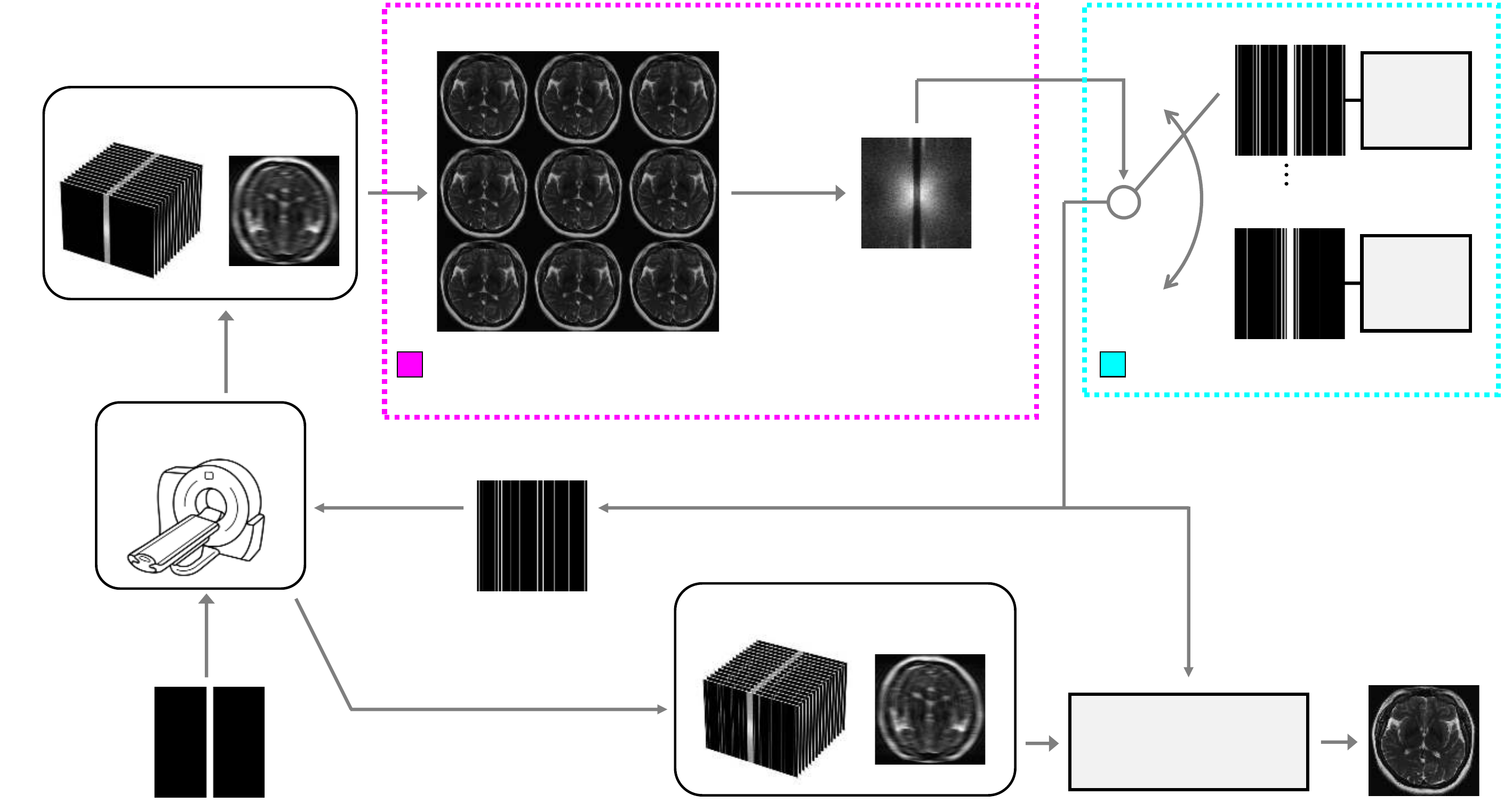}};
  
  \begin{pgfinterruptboundingbox}
  \begin{scope}[scale=1]
  \node[] at (0.62,2.3) {LF k-space data};
  \node[] at (3.55,3.0) {SR image samples};
  \node[] at (6.2,0.95) {\begin{tabular}{c}Sample variance \\ in k-space\end{tabular}};
  \node[] at (9.9,3.0) {Mask $M_1$};
  \node[] at (11,2.43) {\begin{tabular}{c}
       Recon  \\ $\theta_1$
  \end{tabular}};
  \node[] at (9.9,1.5) {Mask $M_J$};
  \node[] at (11,0.95) {\begin{tabular}{c}
       Recon  \\ $\theta_J$
  \end{tabular}};
  \node[] at (9.15,0.27) {\textbf{Adaptive selection}};
  \node[] at (3.5,0.22) {\textbf{HF Bayesian uncertainty \\ quantification ($e_\psi$)}};
  \node[] at (0.1, 0.4) {(\textcolor{red}{2}) Scan LF \\ \;\; k-space region};
  \node[] at (0.85 ,-0.22) {MRI scanner};
  \node[] at (4.2,-0.5) {\begin{tabular}{c}
       (\textcolor{red}{3}) HF component of $M_{j^\star}$  \\ (6.25\% samples)
  \end{tabular}}; 
  \node[] at (8.7, -0.5) {$(M_{j^\star},\theta_{j^\star})$ pair  };
  \node[] at (0.0,-1.9) {\begin{tabular}{c}
       (\textcolor{red}{1}) LF mask $M_0$  \\ (6.25\% samples)
  \end{tabular}}; 
  \node[] at (2.3,-3.0) {\begin{tabular}{l}
       (\textcolor{red}{4}) Additionally scan  \\ \;\; masked k-space region
  \end{tabular}}; 
  \node[] at (5.55 ,-1.7) {Masked k-space data};
  \node[] at (8.65,-2.75) {\begin{tabular}{c}
       Corresponding \\ recon $\theta_{j^\star}$
  \end{tabular}}; 
  \node[] at (10.2 ,-2.0) {\begin{tabular}{r}
       Reconstructed \\ image
  \end{tabular}};
  \end{scope}
  \end{pgfinterruptboundingbox}
  \end{tikzpicture}
  \par\bigskip
  \caption{Using the sample variance of SR space generation~\cite{Song_2022_CVPR,Hong_2023_ICCV} results, we can quantify HF Bayesian uncertainty (highlighted in \textcolor[RGB]{200,0,200}{magenta dotted box}). Then, we can adaptively select a sampling-reconstruction ($M-\theta$) pair (highlighted in \textcolor[RGB]{0,180,180}{cyan dotted box}). Here, we illustrate how our adaptive selection of sampling-reconstruction (\cref{alg:inference}) is employed in CS-MRI. In (\textcolor{red}{1})\&(\textcolor{red}{2}) in the figure, MRI scanner scans LF k-space region and adaptively selects mask $(M_{j^\star},\theta_{j^\star})$ pair using HF Bayesian uncertainty quantification. (\textcolor{red}{3})\&(\textcolor{red}{4}) After additionally scanned masked k-space region from $M_{j^\star}$, 
  reconstructed images are generated from masked k-space data using $\theta_{j^\star}$. See \cref{alg:inference} for details.}
  \label{fig:overview}
\end{figure}
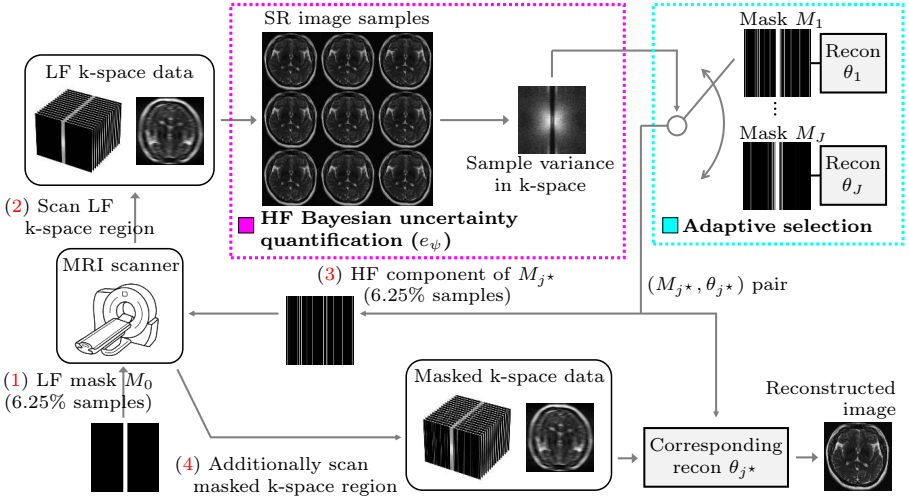
\begin{figure*}[t!]
    \centering
\begin{minipage}[t]{0.58\linewidth}
\begin{algorithm}[H]
\scriptsize
\caption{Training}\label{alg:train}
\begin{spacing}{1.1}
\begin{algorithmic}
\Require Training set $\{k^i\}_{i=1}^N$, initial sampling mask $M_0 \in \mathcal{M}$, trained SR space generation model $f_\psi : \mathcal{I} \rightarrow \mathcal{Z}$, the number of segments $J$, the number of SR generated images $S$, the number of total sampling points $N_M$, and the empirical risk $\hat{\mathcal{L}}$.
\Ensure Masks $(M_j)_{j=1}^J$, reconstruction parameters $(\theta_j)_{j=1}^J$, and centroids of uncertainty $(c_j)_{j=1}^J$
\For{$i=1$ to $N$}
    \For{$s=1$ to $S$}
        \State Sample $z^s \sim \mathcal{N}(0,\sigma_s^2)$
    \EndFor
    \State  $m^i \gets \frac{1}{S}\sum_{s'=1}^S f_\psi^{-1}(z^{s'};M_0 k^i)$
    \State $v^i \gets \frac{1}{S-1} \sum_{s=1}^S (f_\psi^{-1}(z^s;M_0 k^i) - m^i )^{\circ 2}$ %
    \State $u^i \gets v^i / \lVert v^i \rVert_2$ \Comment{Normalized $\widehat{\text{Var}}$}
\EndFor
\State $(c_j)_{j=1}^J \gets k\text{-}means\text{++}(\{u^i\}_{i=1}^N, J)$ \Comment{k-means++}
\For{$j=1$ to $J$}
    \State $M_j \gets M_0 + RejectionSampling(c_j, N_{M} - \mathrm{Tr}(M_0))$
    \State Train $\theta_j$ to minimize $\Hat{\mathcal{L}}[h(\cdot;M_j,\theta_j)]$
\EndFor
\end{algorithmic}
\end{spacing}
\end{algorithm}
\end{minipage}\hfill
\begin{minipage}[t]{0.39\linewidth}
\begin{algorithm}[H]
\scriptsize
\caption{Inference}\label{alg:inference}
\begin{spacing}{1.23}
\begin{algorithmic}
\Require k-space input $k$, initial sampling mask $M_0 \in \mathcal{M}$, trained SR space generation model $f_\psi : \mathcal{I} \rightarrow \mathcal{Z}$, the number of segments $J$, masks $(M_j)_{j=1}^J$, reconstruction network parameters $(\theta_j)_{j=1}^J$, and the centroids of uncertainty $(c_j)_{j=1}^J$
\Ensure Reconstructed image $I'$

\For{$s=1$ to $S$}
    \State Sample $z^s \sim \mathcal{N}(0,\sigma_s^2)$
\EndFor
\State  $m \gets \frac{1}{S}\sum_{s'=1}^S f_\psi^{-1}(z^{s'};M_0 k)$
\State $v \gets \frac{1}{S-1} \sum_{s=1}^S (f_\psi^{-1}(z^s;M_0 k) - m )^{\circ 2}$ %
\State $u \gets v / \lVert v \rVert_2$ \Comment{Normalized $\widehat{\text{Var}}$}

\State $j^\star \gets \argmin_j \, \lVert u-c_j \rVert_2$ \Comment{Selection}
\State $I' \gets h(k;M_{j^\star},\theta_{j^\star})$ \Comment{Recon.}
\end{algorithmic}
\end{spacing}
\end{algorithm}
\end{minipage}
\end{figure*}

\subsubsection{(What to) constructed sampling-reconstruction pairs $(M_j, \theta_j)_{j=1}^J$:} One might try to create $M_j$ from $c_j$ using just sorting, \textit{i.e.}, 
\begin{equation}\label{eq3.2.2.1}
    M_j = \argmax_M \, \lVert M c_j \rVert,
\end{equation}
based on the following proposition.%
\begin{proposition}[simplified version]\label{proposition1}
With mild assumptions, the sorted sample variance (\ref{eq3.2.2.1}) is the PSNR-maximizing mask.
\end{proposition}
However, this method is not the optimal approach for maximizing SSIM. 
In general, it is known that introducing randomness to the mask is effective in maximizing SSIM~\cite{wang2009variable}.
Therefore, we generate $M_j$ using rejection sampling proportional to $c_j$. Then, we train the dedicated $\theta_j$ for the corresponding $M_j$. Figure \ref{fig:overview} shows the overview of adaptive selection, clearly showing why the proposed adaptive selection is \emph{adaptive}. %
Algorithms \ref{alg:train} and \ref{alg:inference} provide detailed descriptions of the training and inference processes of our adaptive selection method, respectively.

\section{Experiments}\label{sec4}
We proposed the \emph{adaptive selection of sampling-reconstruction} scheme in \cref{sec3.2} to address the issues of $\mathcal{H}_1$ and $\mathcal{H}_2$. This section experimentally demonstrates that the proposed method performs well in various settings of Fourier CS.

\begin{figure}[t!]
\centering
    \setlength{\tabcolsep}{0pt}
\begin{tabularx}{\textwidth}{ccccc}
Ground truth & $\mathcal{H}_1$: VD~\cite{wang2009variable} & $\mathcal{H}_1$: LOUPE~\cite{bahadir2020deep} & $\mathcal{H}_2$: Policy~\cite{bakker2022learning} & $\mathcal{H}_{1.5}$ (ours) \\
\includegraphics[width=0.200\linewidth]{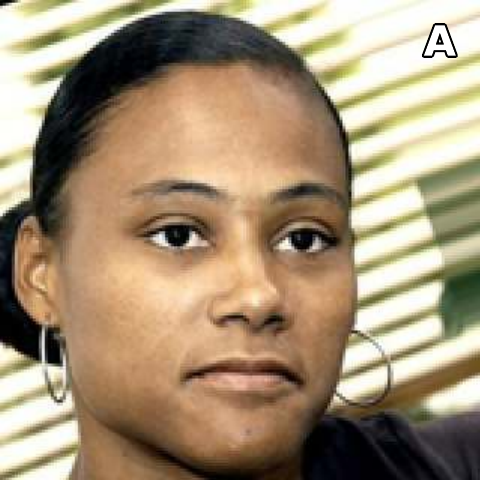} & \includegraphics[width=0.200\linewidth]{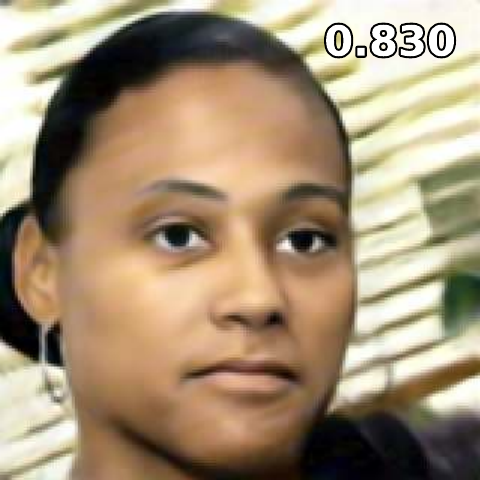} & \includegraphics[width=0.200\linewidth]{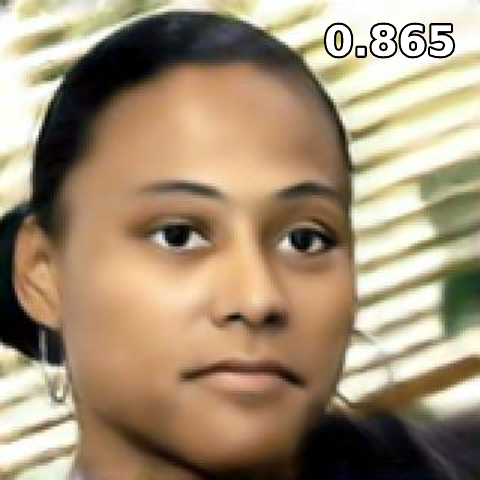} &  \includegraphics[width=0.200\linewidth]{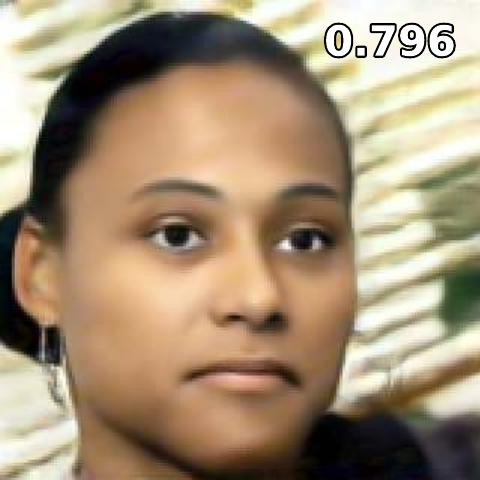} &  \includegraphics[width=0.200\linewidth]{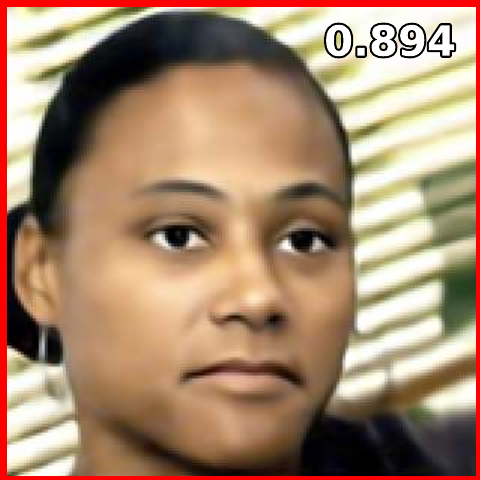}\\[-\dp\strutbox]
\includegraphics[width=0.200\linewidth]{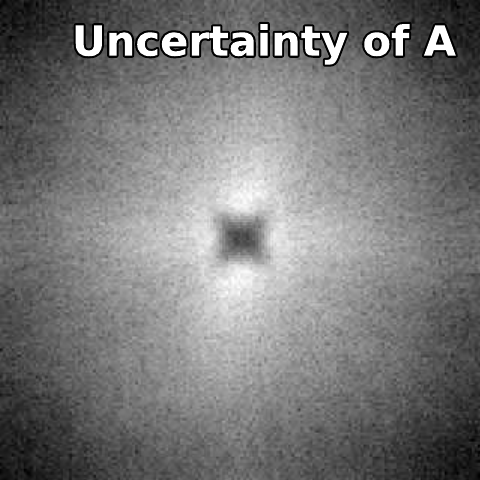} & \includegraphics[width=0.200\linewidth]{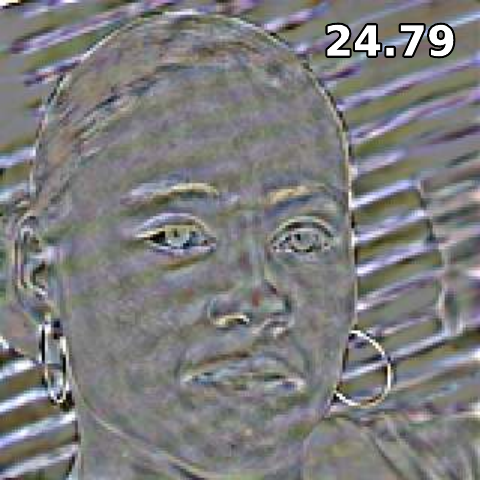} & \includegraphics[width=0.200\linewidth]{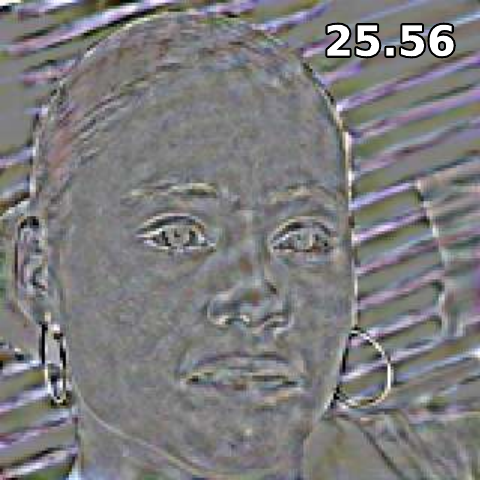} & \includegraphics[width=0.200\linewidth]{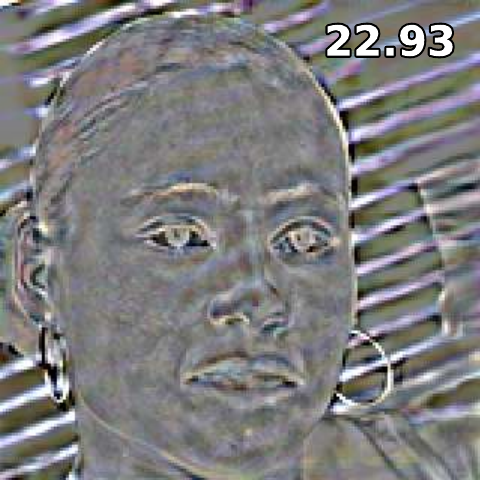} &  \includegraphics[width=0.200\linewidth]{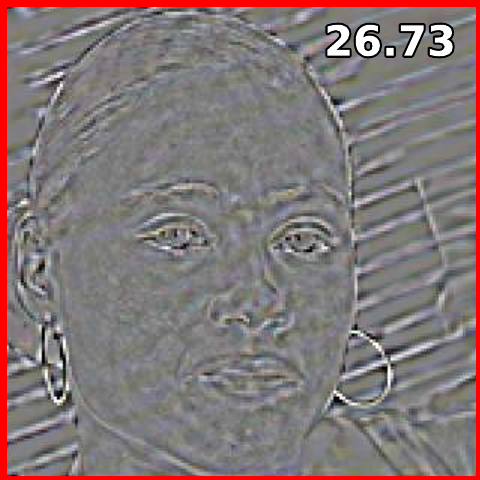}\\[-\dp\strutbox]
\includegraphics[width=0.200\linewidth]{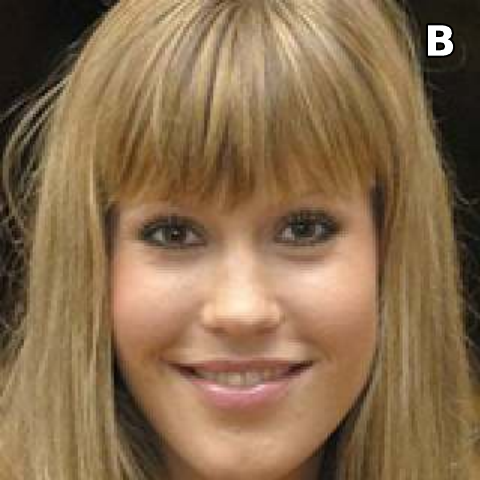} & \includegraphics[width=0.200\linewidth]{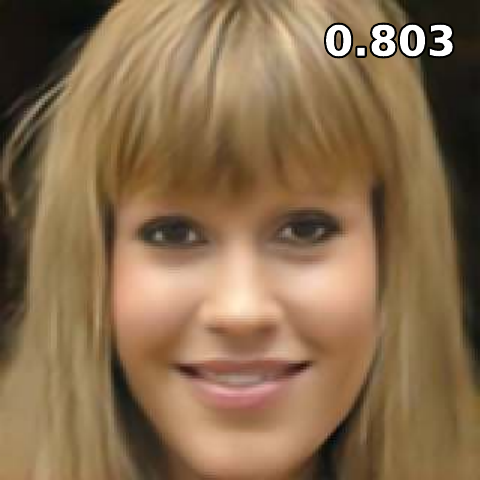} & \includegraphics[width=0.200\linewidth]{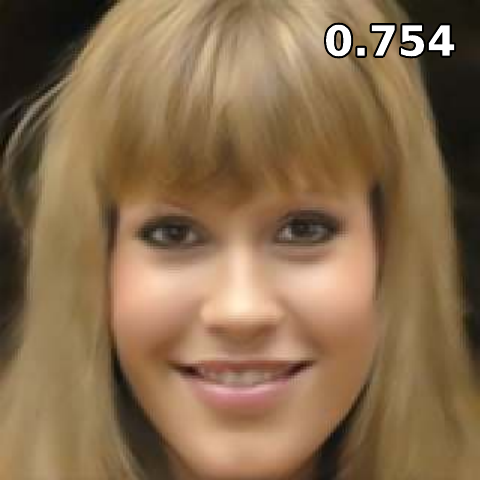} & \includegraphics[width=0.200\linewidth]{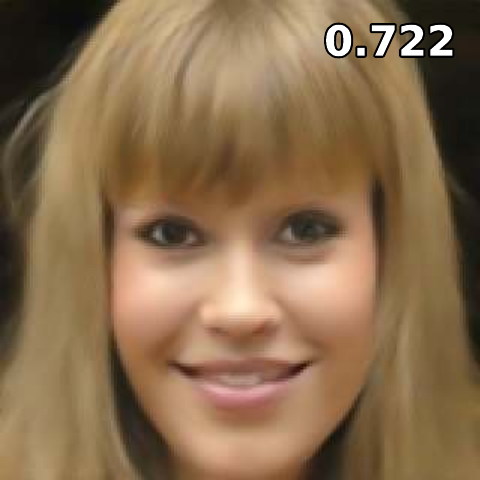} &  \includegraphics[width=0.200\linewidth]{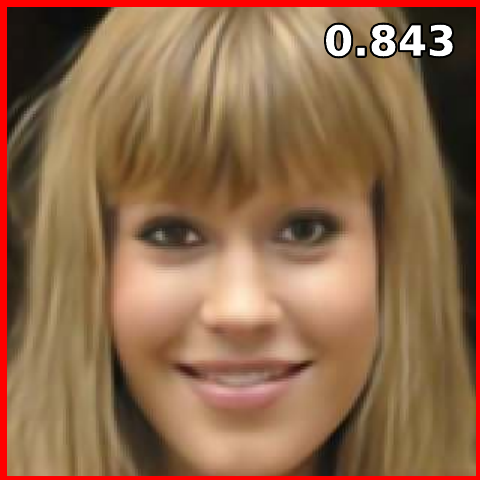} \\[-\dp\strutbox]
\includegraphics[width=0.200\linewidth]{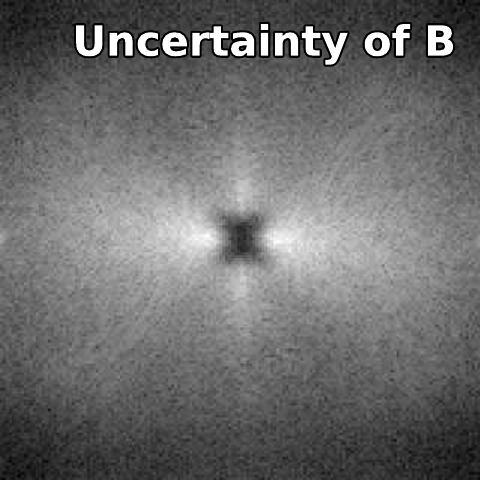} & \includegraphics[width=0.200\linewidth]{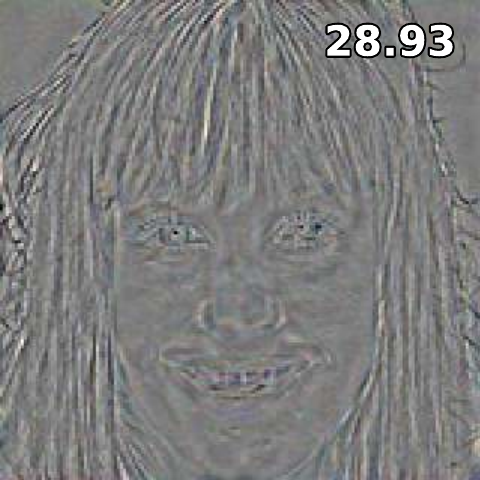} & \includegraphics[width=0.200\linewidth]{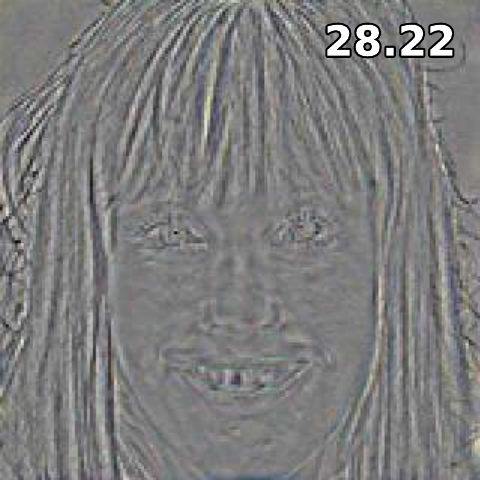} & \includegraphics[width=0.200\linewidth]{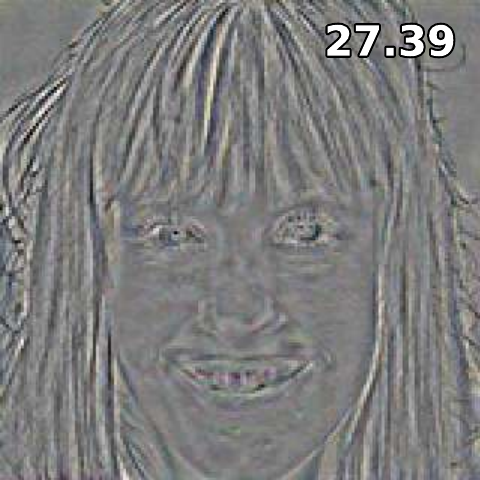} &  \includegraphics[width=0.200\linewidth]{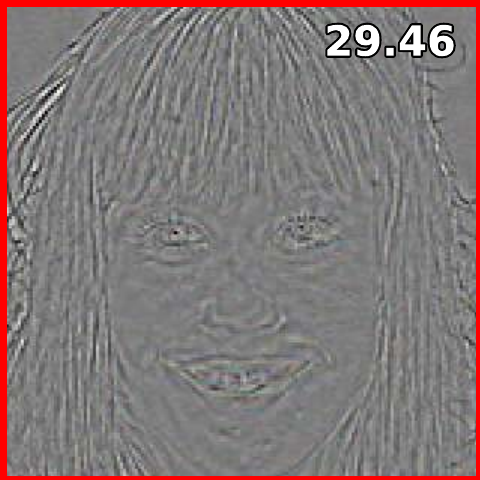} \\[-\dp\strutbox]

\end{tabularx}
  \caption{Our adaptive selection of sampling-reconstruction ($\mathcal{H}_{1.5}$) shows the strongest reconstruction performance (emphasized in red). Here, we show a qualitative comparison of reconstruction and error map at acceleration rate $16\times$ in the CelebA dataset~\cite{liu2015deep}. 
  For comparison, we also show the results of the variable density (VD) \cite{wang2009variable}, LOUPE~\cite{bahadir2020deep}, and policy-based adaptive sampling~\cite{bakker2022learning}.  SSIMs and PSNRs are included in the reconstructions and the error maps, respectively. %
}
  \label{fig:4.1}

\end{figure}

\begin{figure}[t!]
\small
    \setlength{\tabcolsep}{0pt}
\centering
\begin{tabularx}{0.99\textwidth}{cccccc}
\includegraphics[width=0.165\linewidth]{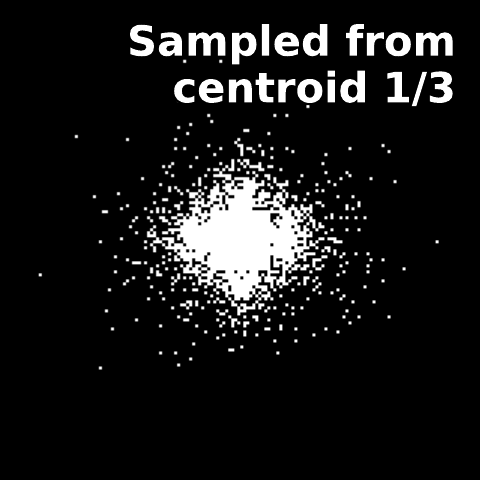} &
 \includegraphics[width=0.165\linewidth]{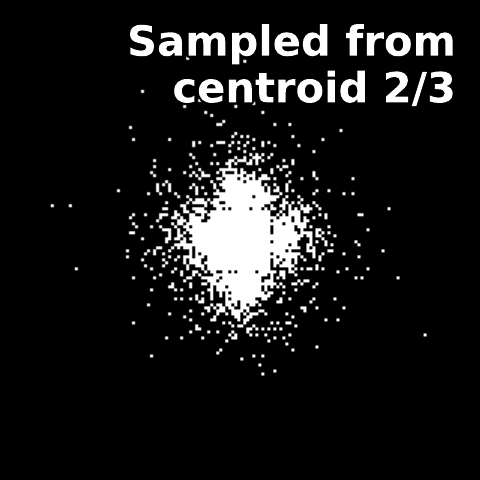} & \includegraphics[width=0.165\linewidth]{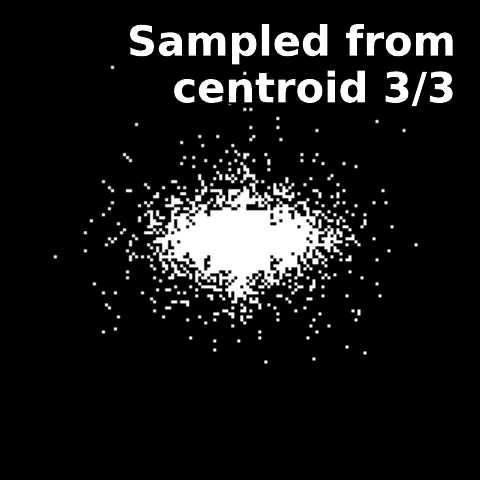} & \includegraphics[width=0.165\linewidth]{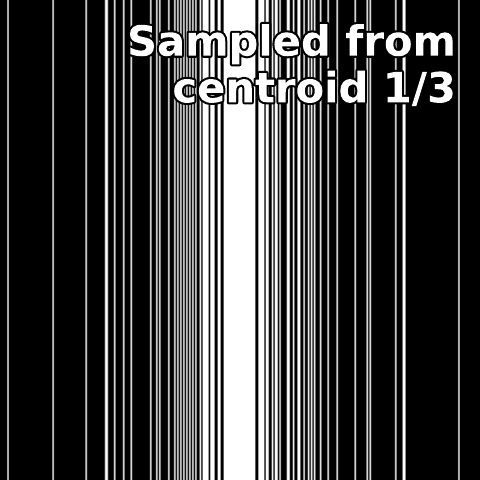} &
 \includegraphics[width=0.165\linewidth]{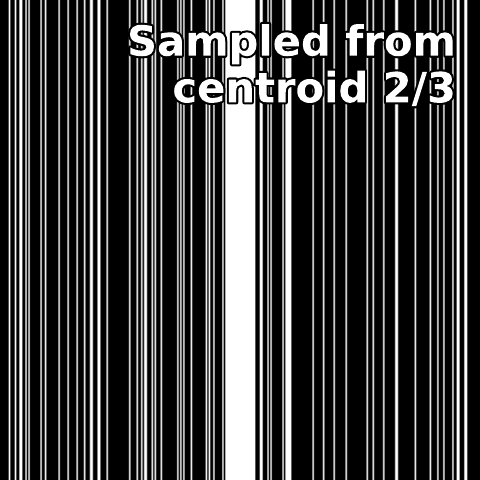} & \includegraphics[width=0.165\linewidth]{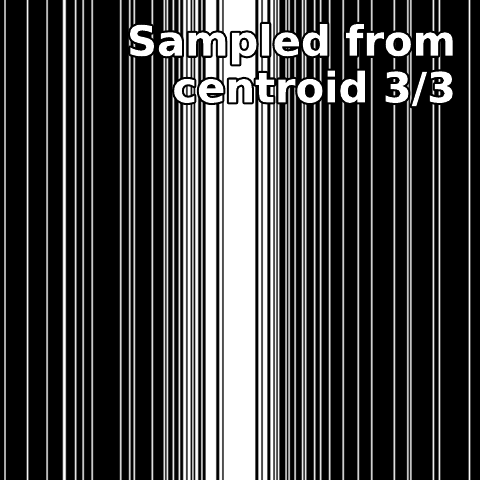}\\[-\dp\strutbox]
  \includegraphics[width=0.165\linewidth]{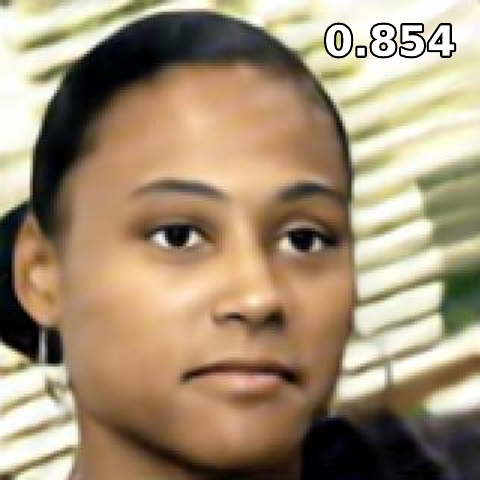} &
  \includegraphics[width=0.165\linewidth]{figs/face/ra2.pdf} & \includegraphics[width=0.165\linewidth]{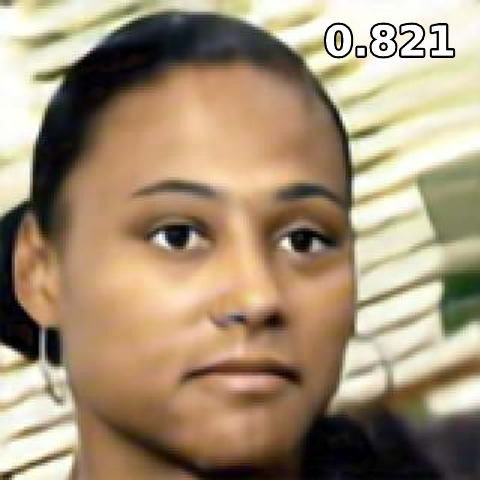} & \includegraphics[width=0.165\linewidth]{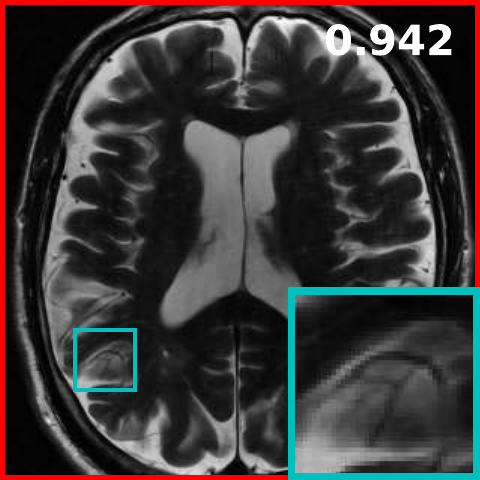} &
  \includegraphics[width=0.165\linewidth]{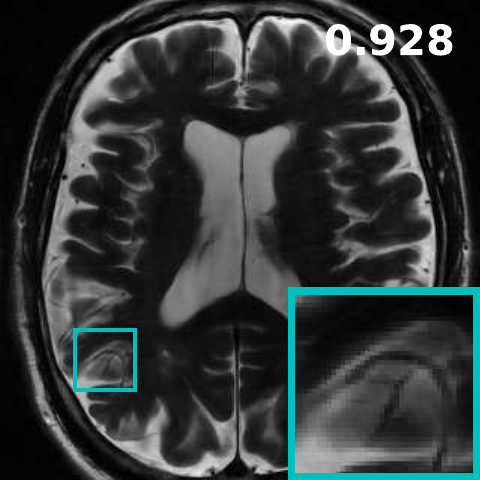} & \includegraphics[width=0.165\linewidth]{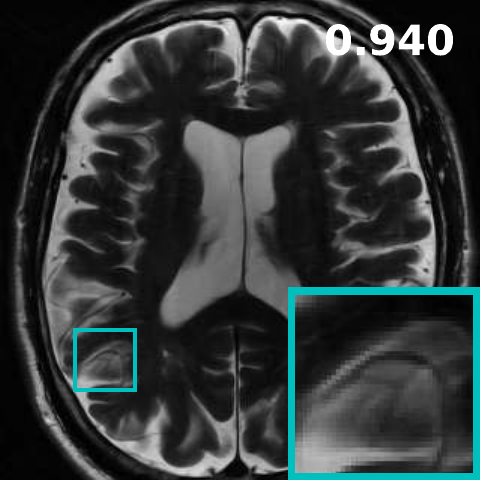}\\[-\dp\strutbox]
  \includegraphics[width=0.165\linewidth]{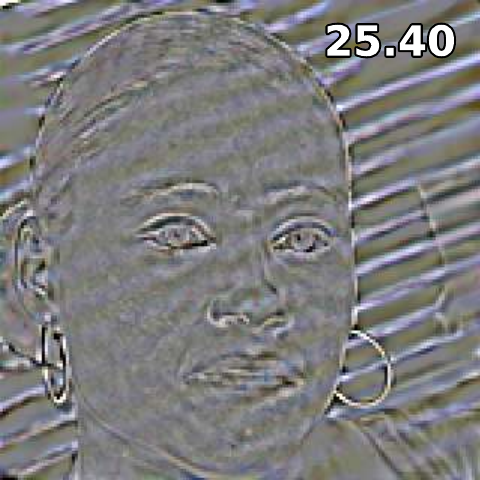} &
  \includegraphics[width=0.165\linewidth]{figs/face/ea2.pdf} & \includegraphics[width=0.165\linewidth]{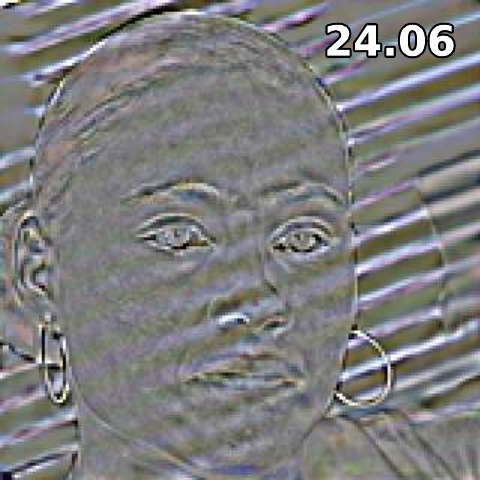} & \includegraphics[width=0.165\linewidth]{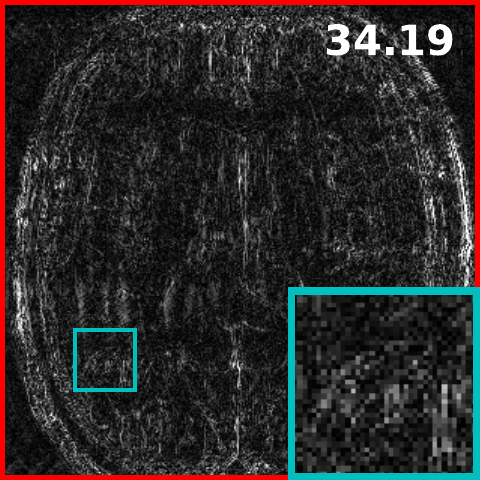} &
  \includegraphics[width=0.165\linewidth]{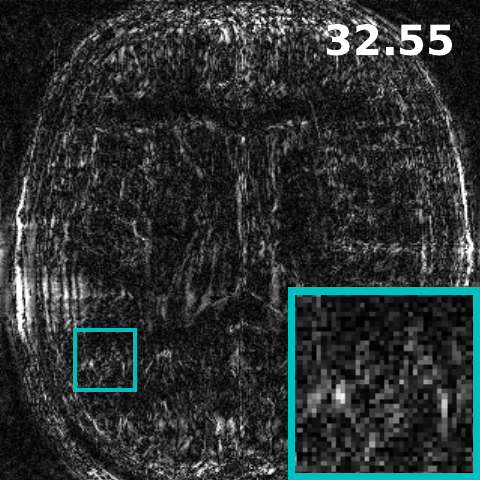} & \includegraphics[width=0.165\linewidth]{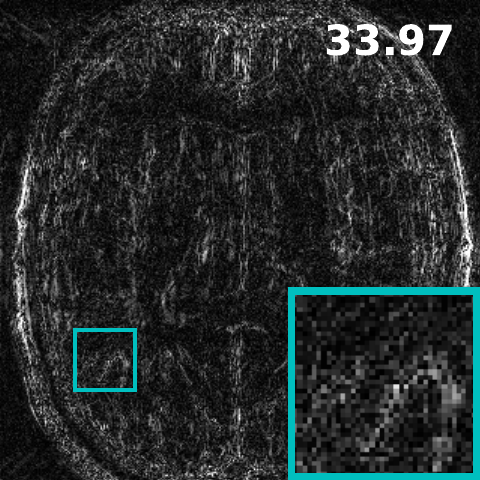}\\[-\dp\strutbox]
  \includegraphics[width=0.165\linewidth]{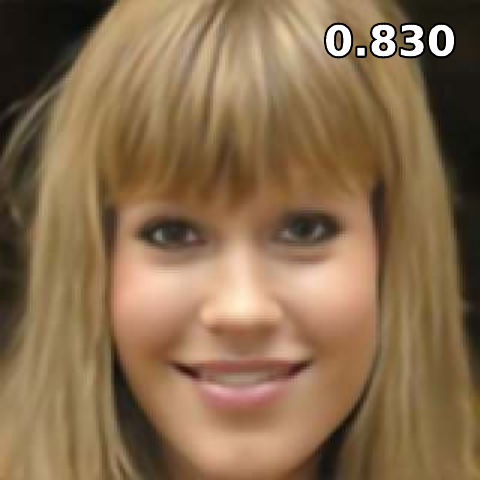} &
  \includegraphics[width=0.165\linewidth]{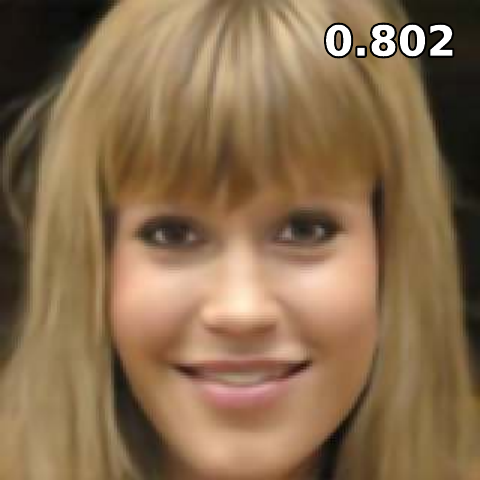} & \includegraphics[width=0.165\linewidth]{figs/face/rb3.pdf} & \includegraphics[width=0.165\linewidth]{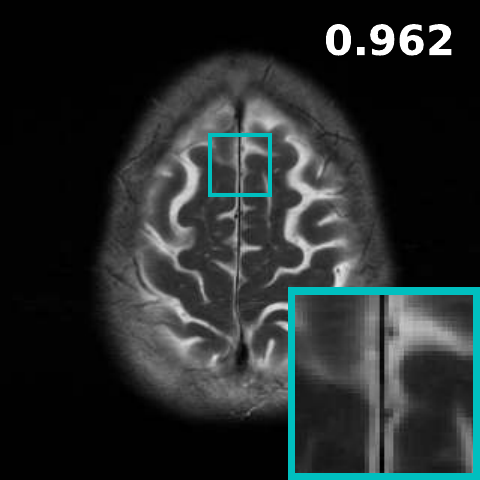} &
  \includegraphics[width=0.165\linewidth]{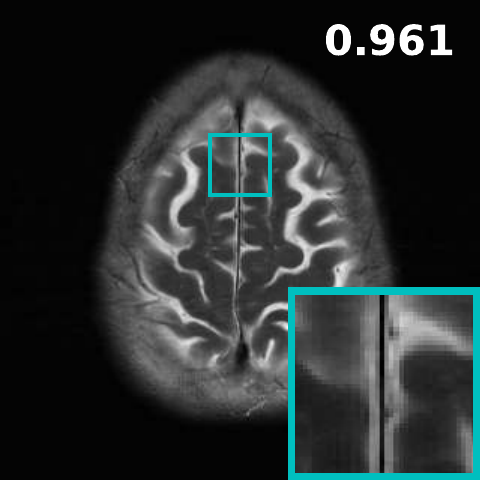} & \includegraphics[width=0.165\linewidth]{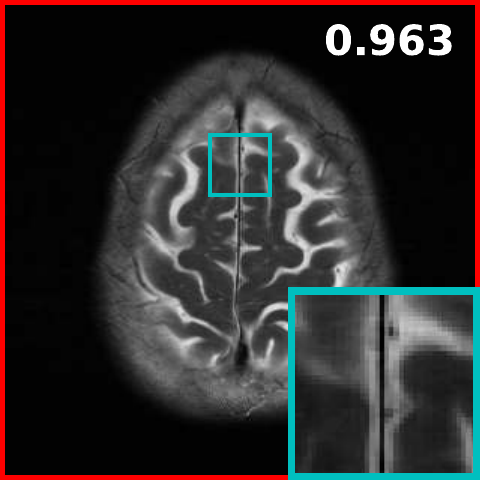}\\[-\dp\strutbox]
  \includegraphics[width=0.165\linewidth]{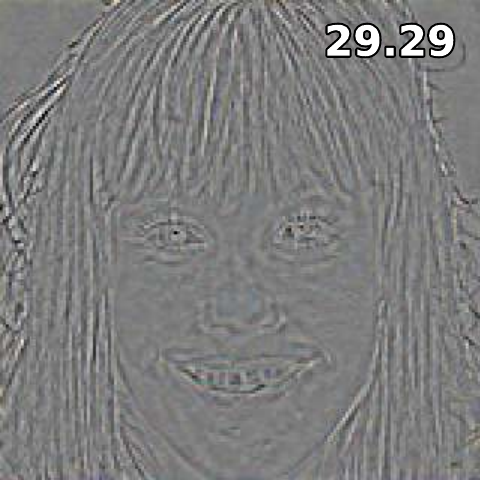} &
  \includegraphics[width=0.165\linewidth]{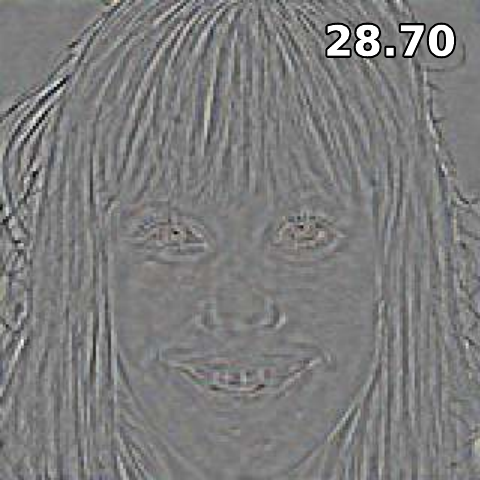} & \includegraphics[width=0.165\linewidth]{figs/face/eb3.pdf} & \includegraphics[width=0.165\linewidth]{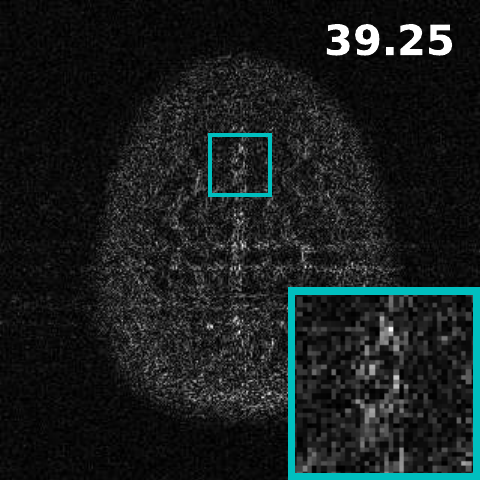} &
  \includegraphics[width=0.165\linewidth]{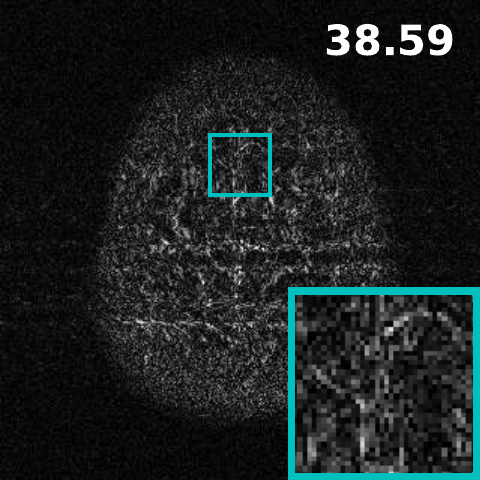} & \includegraphics[width=0.165\linewidth]{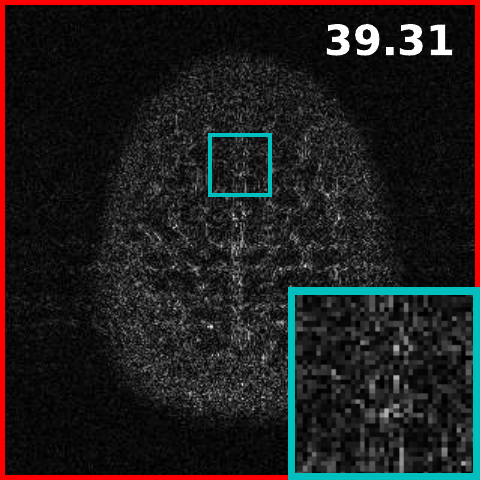}\\[-\dp\strutbox]
\end{tabularx}
  \caption{Our adaptive selection of sampling-reconstruction ($\mathcal{H}_{1.5}$) adaptively selects the best sampling-reconstruction pair based on the HF uncertainty of the input, leading to strong reconstruction performance. Here, we show a qualitative comparison of reconstruction and error map obtained using the mask-reconstruction pairs ($(M_j, \theta_j)_{j=1}^3$) generated from \cref{alg:train} at acceleration rate $16\times$ in the CelebA dataset~\cite{liu2015deep} and at acceleration rate $4\times$ in the fastMRI dataset~\cite{zbontar2018fastmri}. Our \cref{alg:inference} estimated the uncertainty of each image as in Figs. \ref{fig:4.1} and \ref{fig:4.2}, and then selected the appropriate mask $M_j$ (with $\theta_j$) as emphasized in red. For all images in this case, the selected $(M_j, \theta_j)$ resulted in the best reconstruction outcomes. SSIMs and PSNRs are included in the reconstructions and the error maps, respectively.}
  \label{fig:4.15}
\end{figure}

\begin{figure}[t!]
    \setlength{\tabcolsep}{0pt}
\centering
\begin{tabularx}{\textwidth}{ccccc}
Ground truth & $\mathcal{H}_1$: VD~\cite{wang2009variable} & $\mathcal{H}_1$: LOUPE~\cite{bahadir2020deep} & $\mathcal{H}_2$: Policy~\cite{bakker2022learning} & $\mathcal{H}_{1.5}$ (ours) \\
\includegraphics[width=0.200\linewidth]{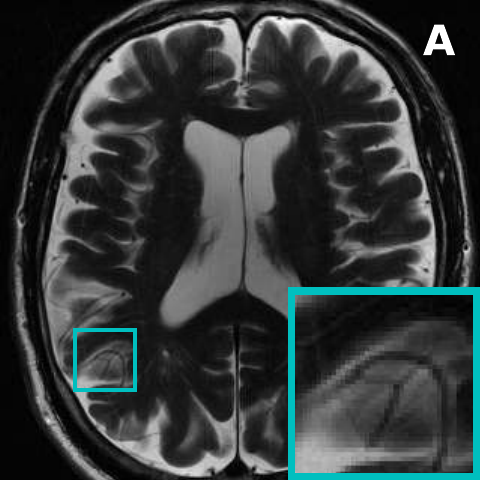} & \includegraphics[width=0.200\linewidth]{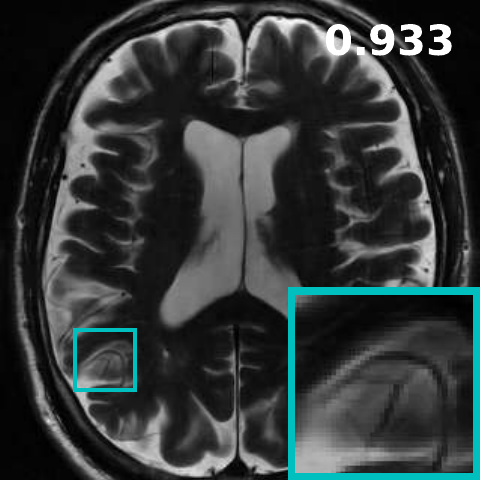} & \includegraphics[width=0.200\linewidth]{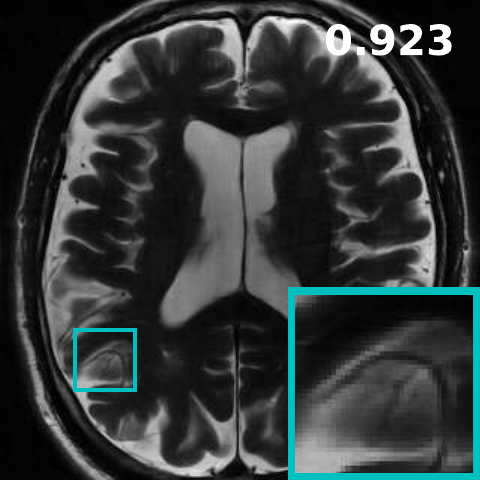} &  \includegraphics[width=0.200\linewidth]{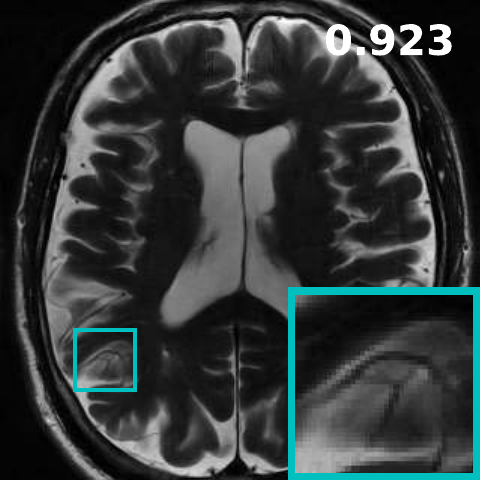} &  \includegraphics[width=0.200\linewidth]{figs/rebuttals/R1/ra1.pdf} \\[-\dp\strutbox]
\includegraphics[width=0.200\linewidth]{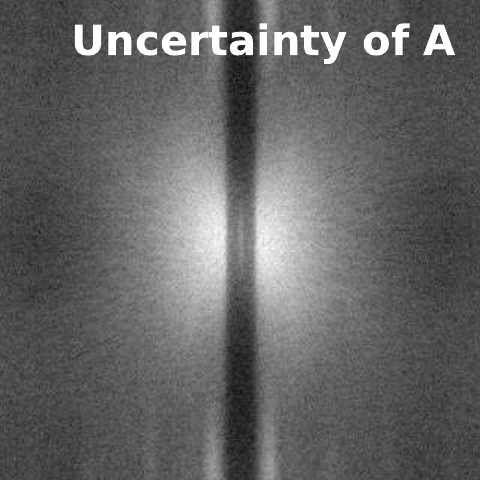} & \includegraphics[width=0.200\linewidth]{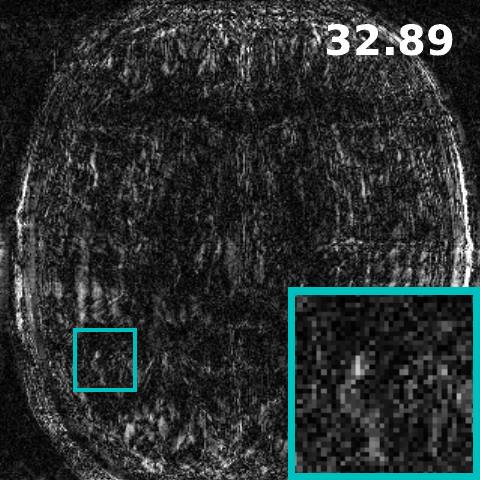} & \includegraphics[width=0.200\linewidth]{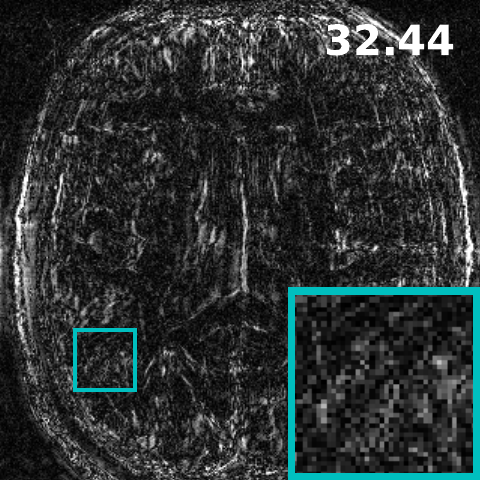} & \includegraphics[width=0.200\linewidth]{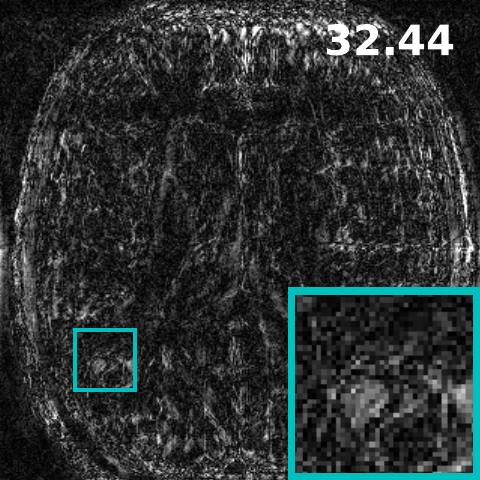} &  \includegraphics[width=0.200\linewidth]{figs/rebuttals/R1/ea1.pdf} \\[-\dp\strutbox]
\includegraphics[width=0.200\linewidth]{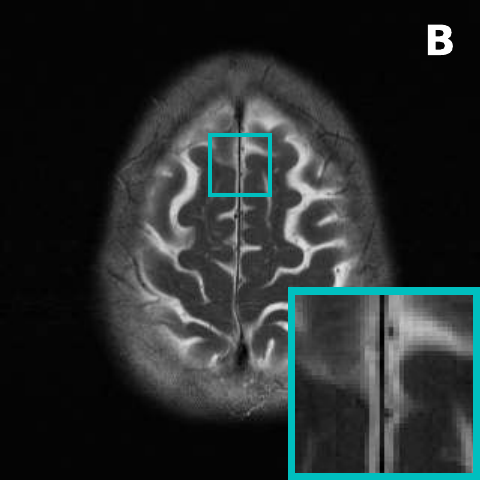} & \includegraphics[width=0.200\linewidth]{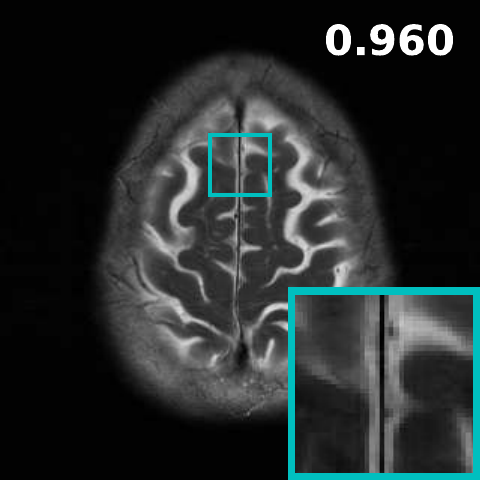} & \includegraphics[width=0.200\linewidth]{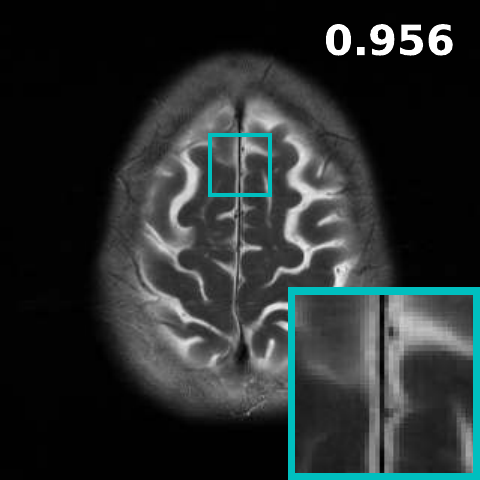} & \includegraphics[width=0.200\linewidth]{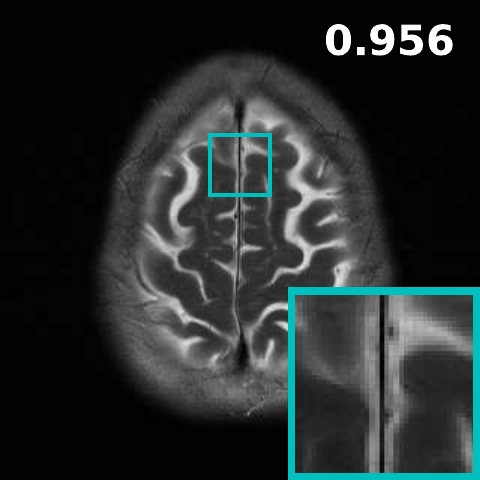} &  \includegraphics[width=0.200\linewidth]{figs/rebuttals/R1/rb3.pdf} \\[-\dp\strutbox]
\includegraphics[width=0.200\linewidth]{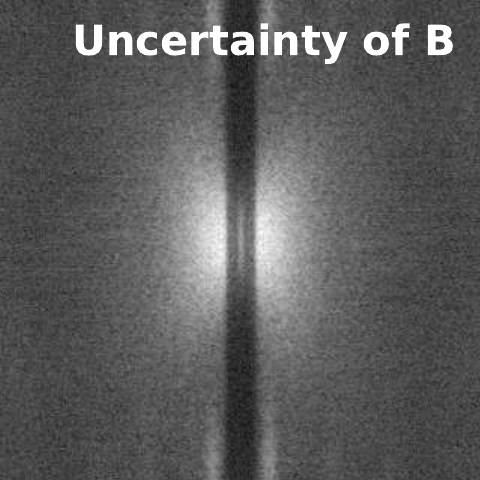} & \includegraphics[width=0.200\linewidth]{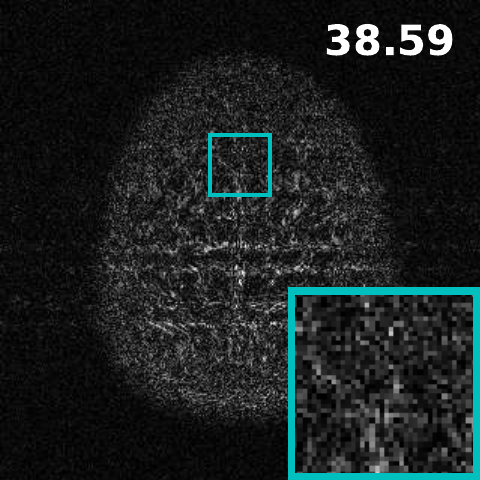} & \includegraphics[width=0.200\linewidth]{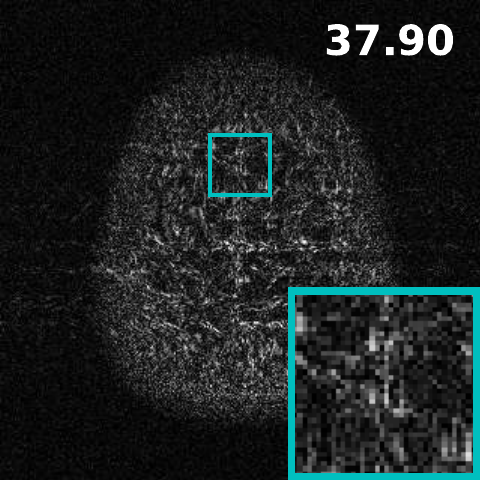} & \includegraphics[width=0.200\linewidth]{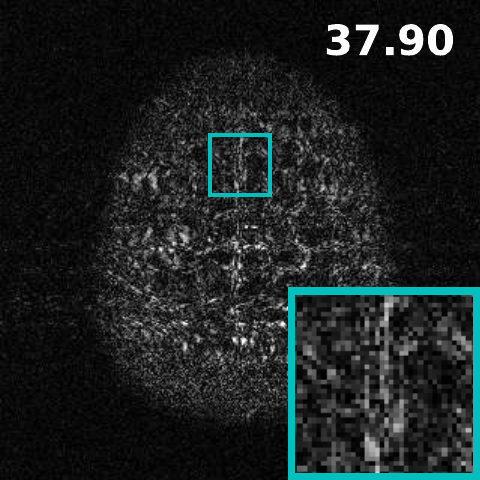} & \includegraphics[width=0.200\linewidth]{figs/rebuttals/R1/eb3.pdf} \\[-\dp\strutbox]
\end{tabularx}
  \caption{In a practical multi-coil CS-MRI 1D line sampling scenario, our adaptive selection of sampling-reconstruction shows the highest SSIM 
  (highlighted in red). Here, we show a qualitative comparison of reconstruction and error map at acceleration rate $4\times$ in the fastMRI dataset~\cite{zbontar2018fastmri}. 
  For comparison, we also show the results of the variable density (VD) \cite{wang2009variable}, LOUPE~\cite{bahadir2020deep}, and policy-based adaptive sampling~\cite{bakker2022learning}. SSIMs and PSNRs are included in the reconstructions and the error maps, respectively.
  }
  \label{fig:4.2}
\end{figure}

\subsection{Fourier CS face reconstruction}\label{sec4.1}
We performed Fourier compressed sensing on the CelebA dataset~\cite{liu2015deep}, which consists of $160 \times 160$ RGB human face images. Similar to LOUPE~\cite{bahadir2020deep}, the reconstruction network used a U-Net~\cite{ronneberger2015u} architecture, with 6 input channels and 3 output channels, because the input, zero-filling reconstruction, is complex.

\Cref{fig:4.1} shows a qualitative comparison of our method ($\mathcal{H}_{1.5}$) and other $\mathcal{H}_1$, $\mathcal{H}_2$ methods~\cite{wang2009variable, bahadir2020deep, bakker2022learning} at the acceleration rate $16\times$ with the metrics of SSIM and PSNR. 
Our final reconstruction results are superior to those of other methods, which supports Theorems \ref{thm1} and \ref{thm2}. 
In detail, our method adeptly selects sampling-reconstruction pairs using HF Bayesian uncertainty. Looking at the first column of \cref{fig:4.1}, in the case of A, the presence of horizontal stripes in the background results in a high uncertainty in the vertical direction, whereas in B, the elongated blonde hair leads to a high uncertainty in the horizontal HF components. In \cref{fig:4.15}, which shows the sampling masks (with the selection) and the corresponding reconstruction results of $\mathcal{H}_{1.5}$, our $e_\psi$ selected $M_2$, which has a shape similar to the uncertainty of A in the second column, obtained the highest SSIM for A. $M_2$ emphasizes in red in the error map, indicating effective suppression of artifacts caused by horizontal high-frequency components in the background, achieved by sampling more in the vertical direction. Similarly, $M_3$ in the third column, which has a shape similar to the uncertainty of B, achieved the highest SSIM for B. $M_3$ is highlighted in the error map, revealing reduced errors in the hair region of the subject due to increased sampling in the horizontal direction.

\subsection{Multi-coil CS-MRI reconstruction}\label{sec4.2}
We also performed Fourier compressed sensing on the fastMRI multi-coil brain dataset~\cite{zbontar2018fastmri}. We resized all slices to a size of $320 \times 320$. The number of coils was 16. Most implementations of $\mathcal{H}_1$ and $\mathcal{H}_2$ methods were based on the official fastMRI repository\footnote{https://github.com/facebookresearch/fastMRI}.  %
We experimented not only with 2D undersampling patterns, as described in \cref{sec4.1}, but also with 1D line subsampling used in actual MRI. For the latter, we modified the SR space generation model~\cite{Song_2022_CVPR} to achieve a $16\times$ SR only in the horizontal direction.

Figure \ref{fig:4.2} shows a qualitative comparison of our method ($\mathcal{H}_{1.5}$) and other $\mathcal{H}_1$, $\mathcal{H}_2$ methods~\cite{wang2009variable, bahadir2020deep, bakker2022learning} at $4\times$ 1D undersampling with the metrics of SSIM and PSNR. Our final reconstructions outperform other methods, supporting Theorems \ref{thm1} and \ref{thm2}. In the right half of \cref{fig:4.15}, our $e_\psi$ selected $(M_1, \theta_1)$, which samples more of the low-frequency components, obtained the best reconstruction result for A (second row). Besides, for input B (fourth row), $(M_3, \theta_3)$ generated the best reconstruction result (highlighted in red). Since $M_3$ samples the high-frequency components more, it made the clearest imaging of the longitudinal fissure, indicating the effectiveness of the proposed method (\textit{i.e.}, Algorithms \ref{alg:train} and \ref{alg:inference}).

\begin{table}[t]
\centering
\caption{Our adaptive selection of sampling-reconstruction ($\mathcal{H}_{1.5}$, Algorithms \ref{alg:train} and \ref{alg:inference}) shows the highestSSIM in Fourier CS in various settings (CelebA dataset~\cite{liu2015deep} w/ 2D sampling, fastMRI multi-coil dataset~\cite{zbontar2018fastmri} w/ 1D, 2D sampling).
}
\label{fig:4.2.2}
\begin{tabular}{ll|cccccc}
\toprule
\multicolumn{2}{c|}{SSIM$\uparrow$}           & \multicolumn{2}{c|}{CelebA} & \multicolumn{2}{c|}{CS-MRI 2D} & \multicolumn{2}{c}{CS-MRI 1D} \\
           \midrule
\multicolumn{2}{r|}{
\begin{tabular}{l@{       }r}
\multicolumn{1}{c}{Method \;\;\textbackslash} & \multicolumn{1}{r}{\; Accel.}
\end{tabular}}
  & \multicolumn{1}{p{1.4cm}}{\centering$8\times$}   & \multicolumn{1}{p{1.4cm}}{\centering$16\times$}   & \multicolumn{1}{p{1.4cm}}{\centering $4\times$}  & \multicolumn{1}{p{1.4cm}}{\centering$8\times$} & \multicolumn{1}{p{1.4cm}}{\centering$4\times$}  & \multicolumn{1}{p{1.4cm}}{\centering$8\times$}   \\
           \midrule
\multirow{4}{*}{$\mathcal{H}_1$}& Random  & 0.8378 & \cellcolor{tabthird}0.8684         & 0.9663     & 0.9506   & 0.9533          & 0.9255   \\
& VD~\cite{wang2009variable} &  \cellcolor{tabsecond}0.9073  & \cellcolor{tabsecond}0.8734         & \cellcolor{tabsecond}0.9698     & \cellcolor{tabsecond}0.9578  & \cellcolor{tabsecond}0.9603     & \cellcolor{tabsecond}0.9367     \\
& LOUPE~\cite{bahadir2020deep}  & \cellcolor{tabthird}0.8742  & 0.8673             & 0.9671          & 0.9525  & 0.9541     &  0.9218       \\
& Equispace (1D)  & -     & -        & -     & -    & \cellcolor{tabsecond}0.9603     &  \cellcolor{tabthird}0.9258        \\
\hline
$\mathcal{H}_2$ & Policy~\cite{bakker2022learning} & 0.8501 & 0.8394  & \cellcolor{tabsecond}0.9698 & \cellcolor{tabthird}0.9572 & 0.9569 & 0.9240\\
\hline
\multirow{1}{*}{$\mathcal{H}_{1.5}$} 
& \begin{tabular}{l}
Adaptive \\ selection (ours) 
\end{tabular} & \cellcolor{tabfirst}0.9405 & \cellcolor{tabfirst}0.8952         & \cellcolor{tabfirst}0.9704    & \cellcolor{tabfirst}0.9585 & \cellcolor{tabfirst}0.9624     & \cellcolor{tabfirst}0.9407    \\[-0.4\dp\strutbox]
\bottomrule
\end{tabular}
\end{table}

In \cref{fig:4.2.2}, we present the average SSIM of the proposed scheme in various accelerations and datasets. %
For comparison, LOUPE~\cite{bahadir2020deep} and policy-based method~\cite{bakker2022learning} are evaluated.
Two non-adaptive methods, uniformly random mask %
and sampling from VD~\cite{wang2009variable}, are also evaluated. For a 1D line sampling CS-MRI, equispaced masks are additionally evaluated.
As shown in \cref{fig:4.2.2}, our adaptive selection approach consistently achieves higher SSIM compared to other methods in all scenarios. For example in CelebA dataset at acceleration rate $8\times$, SSIM of our method (\colorbox{tabfirst}{0.9405}) is about 0.04 higher than the best of $\mathcal{H}_1$ (\colorbox{tabsecond}{0.9073}) and $\mathcal{H}_2$ (0.8501). In addition in a realistic setting, CS-MRI 1D at acceleration rate $8\times$, SSIM of our method (\colorbox{tabfirst}{0.9407}) is about 0.004 higher than the best of $\mathcal{H}_1$ (\colorbox{tabsecond}{0.9367}) and $\mathcal{H}_2$ (0.9240), which is a significant difference in MRI reconstruction problem~\cite{zhang2015denoising}.

\section{Discussion}\label{sec:5}

\subsubsection{Does the SR space generation model quantify the HF uncertainty well?}\label{sec4.1}
Since sample variance estimates MSE, evaluating SR space generation can be done by sorting the sample variance, as in \cref{proposition1}. After adaptive sampling and zero-filling for reconstruction, PSNR can be used as a metric for assessment.
We qualitatively (in the supplementary material) and quantitatively (in \cref{tab:5}) compare the adaptive sampling results in MRI~\cite{zbontar2018fastmri} at acceleration rate $8\times$. 
\Cref{tab:5} presents the average PSNR and SSIM of the proposed methods on the validation dataset. `Sorted-Self' refers to the zero-filling reconstruction results obtained when sorting its own HF Bayesian uncertainty to create a mask, while `Sorted-Another' randomly shuffles the masks among the data points. We additionally generate a mask sampled from VD~\cite{wang2009variable} for comparison. As a result, the `Sorted-Self' approach consistently achieves the highest PSNR and SSIM. These results show that the SR space generation effectively quantifies HF Bayesian uncertainty.%

\begin{table}[t]
    \small
    \centering
    \caption{Quantitative comparison of PSNR and SSIM with \emph{zero-filling}. `Sorted-Self' achieved the highest PSNR, which supports our \cref{proposition1}. Thus, we can assert that the SR space generation model effectively quantifies the HF uncertainty.}%
    \label{tab:5}
    \setlength{\tabcolsep}{5pt}
    \begin{tabular}{c|ccc}
    \toprule
              PSNR / SSIM  & Sorted-Self        & Sorted-Another & VD~\cite{wang2009variable} \\
    \midrule
    $4\times$     & \textbf{37.15 / 0.939}    & 36.36 / 0.922                                                                                                                    & 33.33 / 0.854 \\
    $8\times$     & \textbf{34.79 / 0.910}    & 34.24 / 0.894                                                                                                                    & 32.16 / 0.834 \\
    \bottomrule
    \end{tabular}
\end{table}

\begin{figure}[t!]
    \centering
    \begin{subfigure}[b]{0.45\textwidth}
        \centering
    \begin{tikzpicture}
    \begin{axis}[
        xlabel={$J$, number of segments},
        ylabel={SSIM margin \scriptsize{w.r.t.} \small{$J\!=\!1$}},
        legend style={
            anchor=south,
            legend columns=2,
            cells={anchor=west},
            font=\scriptsize,
            rounded corners=2pt,
            draw=none,
        },
        grid=both,
        grid style={dashed,gray!30},
        xmin=1, xmax=4,
        ymin=0.000, ymax=0.004,
        xtick={1,2,3,4},
        ytick={0.0000,0.0005,0.0010,0.0015,0.0020,0.0025,0.0030,0.0035,0.0040},
        tick label style={font=\scriptsize},
        width=5cm,
        height=4.5cm,
    ]
    
    \addplot[color=red,mark=square] coordinates {
        (1,0.0)
        (2,0.9583-0.9579)
        (3,0.9585-0.9579)
        (4,0.9586-0.9579)
    };
    
    \addplot[color=blue,mark=*] coordinates {
        (1,0.0)
        (2,0.9269-0.9249)
        (3,0.9268-0.9249)
        (4,0.9273-0.9249)
    };
    
    \addplot[color=green,mark=triangle] coordinates {
        (1,0.0)
        (2,0.9112-0.9089)
        (3,0.9109-0.9089)
        (4,0.9116-0.9089)
    };
    
    \end{axis}
    \end{tikzpicture}
        \captionsetup{margin={-30pt, 0pt}}
        \caption{CS-MRI 2D 8$\times$}
        \label{fig:trade-off:a}
    \end{subfigure}
    \hfill
    \begin{subfigure}[b]{0.54\textwidth}
        \centering
    \begin{tikzpicture}
    \begin{axis}[
        xlabel={$J$, number of segments},
        legend style={
            at={(1.35,0.2)},
            anchor=south,
            legend columns=1,
            cells={anchor=west},
            rounded corners=2pt,
            draw=none,
        },
        grid=both,
        grid style={dashed,gray!30},
        xmin=1, xmax=4,
        ymin=0.000, ymax=0.004,
        xtick={1,2,3,4},
        ytick={0.0000,0.0005,0.0010,0.0015,0.0020,0.0025,0.0030,0.0035,0.0040},
        tick label style={font=\scriptsize},
        width=5cm,
        height=4.5cm,
    ]
    
    \addplot[color=red,mark=square] coordinates {
        (1,0.0)
        (2,0.9406-0.9389)
        (3,0.9407-0.9389)
        (4,0.9408-0.9389)
    };
    
    \addplot[color=blue,mark=*] coordinates {
        (1,0.0)
        (2,0.8990-0.8968)
        (3,0.8992-0.8968)
        (4,0.8999-0.8968)
    };
    
    \addplot[color=green,mark=triangle] coordinates {
        (1,0.0)
        (2,0.8810-0.8806)
        (3,0.8839-0.8806)
        (4,0.8845-0.8806)
    };
    
    \legend{$\overline{\text{SSIM}}$,$\overline{\scriptstyle{\text{SSIM}}}^{\text{ low}}_{10\%}$, $\overline{\scriptstyle{\text{SSIM}}}^{\text{ low}}_{5\%}$}
    \end{axis}
    \end{tikzpicture}
        \captionsetup{margin={45pt, 0pt}}
        \caption{CS-MRI 1D 8$\times$}
        \label{fig:trade-off:b}
    \end{subfigure}
    \caption{Trade-off with the number of segments $J$. The average of the lowest \{5\%, 10\%, 100\%\} of SSIM values according to the number of segments $J$ in CS-MRI (a) 2D 8$\times$ and (b) 1D 8$\times$ experiment, respectively. All SSIM values were shown as margin with respect to $J=1$. $\overline{\text{SSIM}}$ reaches a plateau after $J=2$, but $\overline{\scriptstyle{\text{SSIM}}}^{\text{ low}}_{10\%}$ and $\overline{\scriptstyle{\text{SSIM}}}^{\text{ low}}_{5\%}$ become more higher in $J=2,3$ or $4$. These results support our \cref{remark2}.}\label{fig:trade-off}
\end{figure}
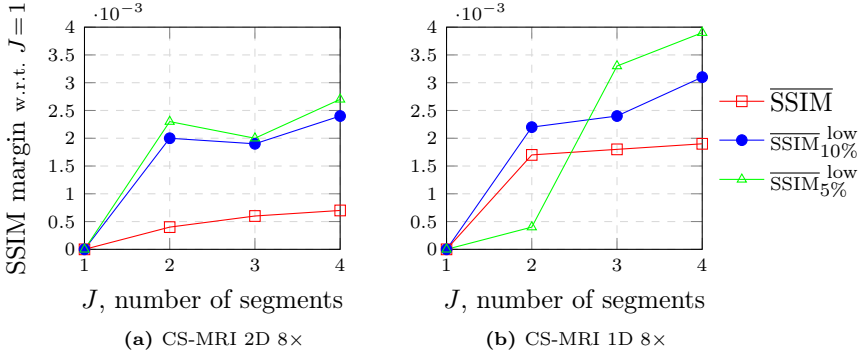

\subsubsection{Effect of the number of segments $J$}
Here, we validate \cref{remark2} by conducting an ablation study using the dataset employed in our experiments.
\Cref{fig:trade-off} shows the average of the lowest 5\%, 10\%, and 100\% of SSIM values ($\overline{\scriptstyle{\text{SSIM}}}^{\text{ low}}_{5\%}$, $\overline{\scriptstyle{\text{SSIM}}}^{\text{ low}}_{10\%}$, and $\overline{\text{SSIM}}$) for all $J=1,\dots,4$ in the CS-MRI 8$\times$ experiments (2D and 1D). 
The fact that the increase in $\overline{\text{SSIM}}$ is less noticeable when transitioning from $J=2$ to $J=3$ or $J=4$ compared to the transition from $J=1$ to $J=2$ supports the first part of \cref{remark2}, ``As $J$ increases, despite more training resources,  the average performance reaches a plateau at some point.'' Additionally, the relatively significant increase in $\overline{\scriptstyle{\text{SSIM}}}^{\text{ low}}_{10\%}$ and $\overline{\scriptstyle{\text{SSIM}}}^{\text{ low}}_{5\%}$ when transitioning from $J=2$ to $J=3$ or from $J=3$ to $J=4$ supports the latter part of \cref{remark2}, ``As $J$ increases, it becomes more robust against outliers.'' Therefore, users can choose $J$ considering this trade-off.

\subsubsection{Other sampling methods}
We employed the rejection sampling to introduce randomness to the mask~\cite{wang2009variable}. To check the effectiveness of the rejection sampling, we compare it with `kmeans-Sorted' method (\textit{i.e.,} applying `Self-Sorted' in \cref{tab:5} to the k-means centroids). In \cref{tab:r2}, SSIM in rejection sampling is much higher than `kmeans-Sorted' in MRI datasets, supporting the effectiveness of rejection sampling.

For more discussions such as runtimes, see the supplementary material.

\begin{table}[t!]
\caption{Comparison of rejection sampling and `kmeans-Sorted'.}\label{tab:r2}
\small
\centering
\setlength{\tabcolsep}{5pt}
\begin{tabular}{l|c|c|c|c}
    \toprule
     SSIM in CS-MRI $J\!=\!3$ & 2D 4$\times$ & 2D 8$\times$ & 1D 4$\times$ & 1D 8$\times$\\ \midrule
     Rejection sampling (ours) &     \textbf{0.9704} & \textbf{0.9585}  &   \textbf{0.9624} & \textbf{0.9407}\\ 
     kmeans-Sorted & 0.9612 & 0.9478 & 0.9493 & 0.9167 \\
     \bottomrule
\end{tabular}
\end{table}

\section{Conclusion}
We have presented an adaptive selection of sampling-reconstruction $\mathcal{H}_{1.5}$ framework for Fourier CS. Our method uses an SR space generation model to quantify the high-frequency Bayesian uncertainty of each input; hence is adaptive compared to the existing joint optimization of sampling-reconstruction ($\mathcal{H}_1$) (\cref{thm1}). Since our method has a dedicated reconstruction network for each sampling mask, unlike adaptive sampling ($\mathcal{H}_2$), our method does not suffer from the Pareto suboptimality (\cref{thm2}). 
The proposed method improved SSIM in various Fourier CS experiments, such as CS of facial images and CS-MRI in a practical multi-coil setting.

\section*{Acknowledgements}
This work was supported by Institute of Information \& communications Technology Planning \& Evaluation (IITP) grant funded by the Korea government(MSIT) [NO.RS-2021-II211343, Artificial Intelligence Graduate School Program (Seoul National University)], the National Research Foundation of Korea(NRF) grant funded by the Korea government(MSIT) (No. NRF-2022R1A4A1030579), Creative-Pioneering Researchers Program through Seoul National University and AI-Bio Research Grant through Seoul National University. Also, the authors acknowledged the financial support from the BK21 FOUR program of the Education and Research Program for Future ICT Pioneers, Seoul National University.

\appendix

\renewcommand{\thesection}{S\arabic{section}}
\renewcommand{\thefigure}{S\arabic{figure}}
\renewcommand{\thetable}{S\arabic{table}}
\renewcommand{\theequation}{S\arabic{equation}}
\renewcommand{\thetheorem}{S\arabic{theorem}}
\renewcommand{\thelemma}{S\arabic{lemma}}
\renewcommand{\theassumption}{S\arabic{assumption}}

\setcounter{section}{0}
\setcounter{figure}{0}
\setcounter{table}{0}
\setcounter{equation}{0}
\setcounter{theorem}{0}
\setcounter{lemma}{0}
\setcounter{assumption}{0}

\section{Proofs and Clarifications}
\subsection{Proofs of theorems}
\begin{proof}[Proof of Theorem 1]
    Showing $\mathcal{H}_{1} \subseteq \mathcal{H}_{1.5}^J$ is sufficient.
    
    Let $h^* := h(\cdot;M^*,\theta^*) \in \mathcal{H}_{1}$. 
    Then, setting $e_\psi(\cdot) = e_1$ and $(M_1, \theta_1) = (M^*,\theta^*)$ makes $h^* = \sum_{j=1}^J e_\psi(\cdot)_j h(\cdot;M_j,\theta_j) \in \mathcal{H}_{1.5}^J$. 
\end{proof}
\begin{proof}[Proof of Theorem 2]
    Showing $\mathcal{H}_{2} \subseteq \mathcal{H}_{1.5}^J$ is sufficient. 
    
    Let $h^* := h(\cdot;\pi_\phi(\cdot),\theta^*) \in \mathcal{H}_{2}$. 
    Since $|\pi_\phi(M_0 \mathcal{K})| \leq J$, let $\pi_\phi(M_0 \mathcal{K}) = \{M_1^*, M_2^*, \dots, M_{J}^* \}$. 
    Then, setting $e_\psi(\pi_\phi^{-1}(M_j^*)) = e_j$  and $(M_j, \theta_j) = (M_j^*,\theta^*)$ for all $j \in [1,J]$ makes $h^* = \sum_{j=1}^J e_\psi(\cdot)_j h(\cdot;M_j,\theta_j) \in \mathcal{H}_{1.5}^J$. 
\end{proof}

\subsection{Mathematical notations}
\begin{itemize}[leftmargin=*]
\item In our paper, $p$ represents the true probability density function, and $q_\theta$ represents the probability density function modeled by a parameter $\theta$.
\item In Section S2.1, $\delta$ represents the Dirac delta distribution, and $\hat{k}$ represents the intermediate k-space output of the reconstruction network.
\item In Algorithm 1, $\mathrm{Tr}$ represents the trace of a matrix, and $\widehat{\mathrm{Var}}$ refers to the sample variance (matrix).
\item In Proposition 1, $\mathbf{1}$ is the indicator function.
\item In Algorithms 1 and 2, $(\cdot)^{\circ 2}$ represents the Hadamard power.
\item In Algorithm 1, $RejectionSampling(c_j, n)$ to sample a total of $n$ items without duplication in proportion to $c_j$.
\end{itemize}

\section{Notes on High-frequency Bayesian uncertainty quantification} %
\label{sec3.1}

To quantify HF Bayesian uncertainty, we employed the sample variance of SR space generation. Although a similar idea has been proposed in adaptive sampling~\cite{sanchez2022learning}, we provide notes on how this is achievable in our setting.

\subsection{Analysis for mean squared error in k-space}\label{sec3.1.1}
Let $q_\theta(\cdot | k, M)$ be a probability distribution of the k-space output of the reconstruction network parameterized by $\theta$, conditioned on $k \in \mathcal{K}$ and $M \in \mathcal{M}$.
Then the mean squared error of the output $ \mathrm{MSE}_{q_\theta}[k'|k, M] := \int_{k'} \lVert k'-k \rVert_2^2 dq_\theta(k'|k,M)$ is expressed as follows: 
\begin{equation}\label{eq3.1.1.1}
\begin{aligned}
    \mathrm{MSE}_{q_\theta}&[k'|k, M] = \int_{k'} \left(\lVert Mk'-Mk \rVert_2^2 \right.\\ 
    & \left. + \lVert (E-M)k'-(E-M)k \rVert_2^2 \right) dq_\theta(k'|k,M),%
\end{aligned}
\end{equation}
where $E \in \mathcal{M}$ is the identity matrix (i.e., all-sampling mask). One might think of minimizing the projected generalized Stein's unbiased risk estimator~\cite{abu2022image} instead of MSE; however, it does not estimate the error of $(E-M)k$ that we are interested in.

As most reconstruction networks are designed to receive input in the form of $Mk$, estimating $(E-M)k$ is considerably more challenging than estimating $Mk$. As a result, we can rely on the following Assumption \ref{ass1}.
\begin{assumption}\label{ass1}
$0 < \exists \epsilon \ll 1$ s.t. $\mathbb{E}_{k' \sim q_\theta}[\lVert M(k'-k) \rVert_2^2|k, M] < \epsilon \mathbb{E}_{k' \sim q_\theta}[\lVert (E-M)(k'-k) \rVert_2^2|k, M]$.
\end{assumption}
\noindent By Assumption \ref{ass1}, we have
\begin{equation}\label{eq3.1.1.4}
    \mathbb{E}_{k' \sim q_\theta}[\lVert k'-k \rVert_2^2|k, M]
    < (1+\epsilon) \int_{k'} \lVert (E-M)k'-(E-M)k \rVert_2^2 dq_\theta(k'|k,M).
\end{equation}
One could attempt to find the optimizer $M$ that minimizes the tight upper bound of MSE obtained through \cref{eq3.1.1.4}. However, this task is infeasible for three reasons. Firstly, $(E-M)k$ is out of observation on input. Secondly, since the output of the reconstruction network is deterministic (i.e., $q_\theta(k'|k,M) = \delta(k'-\hat{k})$), it is impossible to estimate $(E-M)k$ from $(E-M)k'$. Thirdly, the process needs joint optimization of $M$-$\theta$, which is difficult.
Section \ref{sec3.1.2} addresses these three issues.

\subsection{Uncertainty quantification via super-resolution space generation}\label{sec3.1.2}
We can feasibly estimate the upper bound of MSE (i.e., HF Bayesian uncertainty) in k-space by leveraging SR space generation.
Using the discrete Fourier transform, we can define the SR space generation model so that its conditional input and output are defined in k-space (i.e., $q_\psi(\cdot|\cdot): \mathcal{K}\times\mathcal{K} \rightarrow \mathbb{R}$).
Since the downsampling in $\mathcal{I}$ is almost identical to masking with $M_0$ (i.e., masking only LF components) in $\mathcal{K}$, we can represent the SR space generation model as $q_\psi(k'|M_0k)$. Replacing $q_\theta$ in \cref{eq3.1.1.4} with $q_\psi$ can be expressed as an expectation over $z$ as follows:
\begin{equation}\label{eq3.1.2.1}
    \int_{k'}  \lVert (E-M)(k'-k) \rVert_2^2 dq_\psi(k'|M_0k) 
     = \mathbb{E}_{z\sim q_z} [\lVert(E-M)(f_\psi^{-1}(z;M_0 k)-k)\rVert_2^2].
\end{equation}
Unlike $q_\theta$, diverse sampling is possible from $q_\psi$. Therefore, by sampling, the right-hand side of \cref{eq3.1.2.1} can be estimated in the form of sample variance of $(E-M)k'$ as follows:
\begin{equation}\label{eq3.1.2.2}
    \mathrm{Tr}(\widehat{\mathrm{Var}}_{q_\psi}[(E-M)k']) %
    = \frac{1}{S-1} \sum_{s=1}^S \left\lVert (E-M)(f_\psi^{-1}(z^s;M_0 k) - m) \right\rVert_2^2,
\end{equation}
where $m= \frac{1}{S}\sum_{s'=1}^S f_\psi^{-1}(z^{s'};M_0 k)$, and $z^s  {\sim_{\text{i.i.d.}}} q_z$ for $s=1,\dots,S$. Note that if $f_\psi^{-1}(z;M_0 k)$ is an unbiased estimator for $k$ (i.e., $\int_{k'} k' dq_\psi(k'|M_0k)=k$), then \cref{eq3.1.2.2} is an unbiased estimator for \cref{eq3.1.2.1}.

In summary, with simplified expressions, the equations used in this section can be related under each of the following conditions as follows:
\begin{equation}\label{eq3.1.2.3}
\begin{aligned}
    & \mathrm{Tr}(\widehat{\mathrm{Var}}_{q_\psi}[(E-M)k']) \overset{\text{(A)}}{\longrightarrow} \mathrm{MSE}_{q_\psi}[(E-M)k'] \\
    & \overset{\text{(B)}}{\longrightarrow} \mathrm{MSE}_{q_\theta}[(E-M)k'] \overset{\text{(C)}}{\longrightarrow} \mathrm{MSE}_{q_\theta}[k'], 
\end{aligned}
\end{equation}
(A) holds if $f_\psi^{-1}(z;M_0 k)$ is an unbiased, uncorrelated estimator for $k$, (B) holds if both models are well-trained, i.e., if $q_\psi \simeq p$ and $q_\theta \simeq p$, and (C) holds if Assumption \ref{ass1} holds. 

We don't necessarily need the minimum-variance unbiased estimator of MSE, but it is important to have the correct tendency. Thus, it is acceptable not to satisfy the challenging conditions of \cref{eq3.1.2.3}. 
In case of CS-MRI, Parseval's theorem guarantees that the MSE in k-space equals the MSE in the coil-by-coil (i.e., just before doing root-sum-square) image space. Therefore, we can obtain a mask that maximizes PSNR as follows:
\begin{equation}\label{eq3.1.2.4}
\begin{aligned}
    & \underset{M}{\argmax} \, \mathrm{PSNR}_{q_\theta}[I'] 
    = \underset{M}{\argmin} \, \mathrm{MSE}_{q_\theta}[I'] \simeq \underset{M}{\argmin} \, \mathrm{MSE}_{q_\theta}[k']\\
    & \simeq \underset{M}{\argmin} \, \mathrm{Tr}(\widehat{\mathrm{Var}}_{q_\psi}[(E - M)k']) = \underset{M}{\argmax} \, M(\widehat{\mathrm{Var}}_{q_\psi}[ k'_l])_{l=1}^L.
\end{aligned}
\end{equation}
\Cref{eq3.1.2.4} implies that selecting k-space variances in descending order as a mask is an adaptive sampling method that maximizes PSNR.
We express the conclusion of this subsection as the following Proposition \ref{proposition1}:
\begin{proposition}[Sorted sample variance is the PSNR-maximizing mask]\label{proposition1}
    $\argmax_M \, \mathrm{PSNR}_{q_\theta}[I'] \simeq M_0 + \mathrm{diag}(\mathbf{1}(\widehat{\mathrm{Var}}_{q_\psi}[ k'_l] \geq V_{th})_{l=1}^L)$, where $V_{th}$ is determined by the acceleration rate $r$ of $M$. %
\end{proposition}
\noindent We refer to $(\widehat{\mathrm{Var}}_{q_\psi}[ k'_l])_{l=1}^L$ as the \emph{HF Bayesian uncertainty} because $q_\psi$ is the posterior probability given $M_0 k$.

\section{More Discussions}
\subsubsection{Processing delay}
One might argue that the time required for generating SR space could be a disadvantage. To deal with that, we can employ multislice imaging, which is the default mode on modern clinical MRI scanners. We can acquire ACS lines for all slices first, and then subsequentially acquire high-frequency components later. The detailed acquisition order is as follows: 
\begin{equation*}
\small
\begin{aligned}
& Slice_1 ACS_1, Slice_2 ACS_1, \dots , Slice_N ACS_1, \\
& Slice_1  ACS_2, Slice_2 ACS_2, \dots , Slice_N ACS_2,\\
& \vdots\\
& Slice_1 ACS_{Last}, Slice_2 ACS_{Last}, \dots , Slice_N ACS_{Last},\\
& Slice_1 HF_1, Slice_2 HF_1, \dots, Slice_N HF_1,\\
& Slice_1 HF_2, Slice_2 HF_2, \dots, Slice_N HF_2,\\
& \vdots
\end{aligned}
\end{equation*}
where $Slice_i ACS_j$ is $j$-th ACS line of $i$-th slice, and $Slice_i HF_j$ is $j$-th non-ACS line acquisition (by Algorithm 2) of $i$-th slice. This scenario allows SR space generation to be performed during the ACS acquisition of other slices, potentially further minimizing possible time wastage. Our runtime of SR-space generation is 0.4 s / slice; 
real-time is possible.

\subsubsection{Output of the reconstruction network}
Typically, the output of a reconstruction network is in $\mathcal{I}$. However, in the case of the E2E-VarNet~\cite{sriram2020end} used in CS-MRI, there exists an intermediate output $\hat{k} \in \mathcal{K}$, and the image (i.e., the final output) is obtained by the root sum of squares (i.e., $\mathrm{RSS}(\hat{k})$. Note that $\mathrm{RSS}$ is not learnable. Therefore, for simplicity, we consider $\hat{k}$ as the final output of the reconstruction network.
\subsubsection{Why training a neural net classifier is difficult}
One na\"ive approach is to train a neural net classifier, $e_\psi$, that minimizes the cross-entropy 
\begin{equation}
    -\frac{1}{N}\sum_{i=1}^{N}\sum_{j=1}^{J}y^i_j \log e_\psi(M_0 k^i)_j,
\end{equation}
where $y^i = e_{\arg\min_{j} l(I^i,h(k^i;M_{j},\theta_{j}))}$. However, this approach is not straightforward because it assumes that $k$ (specifically, $M_0 k$) comes from the same distribution in the dataset, which means that classifying $k$ is difficult. Moreover, $\{y_i\}_{i=1}^N$ can have a class imbalance, which further complicates the training of the classifier.
\subsubsection{Employing SR Bayesian uncertainty generator}
There is a method to directly estimate uncertainty along with the SR image~\cite{kar2021fast}. One can try to obtain HF Bayesian uncertainty by applying the inverse discrete Fourier transform (IDFT) to the SR Bayesian uncertainty. However, this approach is inappropriate because the component-wise variance of the k-space is not the IDFT of the variance of the image, as in the following Theorem \ref{thmS1}.
\begin{theorem}\label{thmS1}
    Let $F \in \mathbb{C}^{L\times L}$ be the $L$-point DFT matrix, and $L>1$. Then, there exist an integer $i \in [1,L]$ and a random variable $I' \in \mathcal{I} \subseteq \mathbb{R}^L$ such that
    $\mathrm{Var}[F^H(I')]_{ii} \neq (F^H (\mathrm{Var}[I']_{ll})_{l=1}^L)_i$.
\end{theorem}
\begin{proof}
    Let's denote the $i$-th column of $F$ as $f_i$. Let $\mathrm{Var}[I]=E$ and $i=2$. Then we have 
    \begin{equation}\label{eqS1}
    \mathrm{Var}[F^H(I')]_{ii} = (F^H \mathrm{Var}[I'] F)_{ii} = (F^H F)_{ii} = 1,        
    \end{equation}
    and
    \begin{equation}
    (F^H (\mathrm{Var}[I']_{ll})_{l=1}^L)_i = f_i^H (\mathrm{Var}[I']_{ll})_{l=1}^L = f_i^H 1 = 0.   
    \end{equation}
\end{proof}
As in \eqref{eqS1}, $\mathrm{Var}[F^H(I')] = (F^H \mathrm{Var}[I'] F)$, but SR Bayesian uncertainty is an estimator of component-wise variance $(\mathrm{Var}[I']_{ll})_{l=1}^L$ instead of a variance matrix $\mathrm{Var}[I']$. Note that $(\mathrm{Var}[I']_{ll})_{l=1}^L$ is the diagonal entries of $\mathrm{Var}[I']$. Therefore, we cannot directly obtain HF Bayesian uncertainty in k-space using SR Bayesian uncertainty. To compute the desired HF Bayesian uncertainty, we need to rely on SR space generation and compute the sample variance, as our proposed method.

\subsubsection{Other SR space generations} Flow-based SR models have demonstrated good diversity and robustness in the recent NTIRE challenge~\cite{Song_2022_CVPR, lugmayr2022ntire}, so we chose to use a flow-based model~\cite{Hong_2023_ICCV} for generating the SR space. Other SR space generation models would be also possible, such as diffusion models~\cite{li2022srdiff}, or dropout method in conditional generative adversarial networks~\cite{deshpande2017learning}.

\subsubsection{Limitations} Our proposed method relies on several assumptions. 
In reality, SR space generation may have inaccuracies, and therefore, HF Bayesian uncertainty may also contain errors. 
Nevertheless, despite these uncertainties, the reconstruction results can be trusted because the SR space generation is only used for the mask generation and determination process, not for the reconstruction process.

\section{Experiment details}
\subsection{Dataset and Implementation}
We use CelebA~\cite{liu2015deep} dataset for CS of facial images. Among the dataset, CelebA 1-182,340 is the training set, while CelebA 182,341-202,600 serve as the validation set. For k-means clustering of the HF uncertainty, CelebA 1-18,234 is used. There size is $(160 \times 160)$, so LF regions for initial sampling mask $M_0$ are set to center $(20 \times 20)$ region. For CS-MRI experiment, NYU fastMRI multi-coil brain dataset~\cite{zbontar2018fastmri} is used. Among the dataset, we only use T2 weighted images which contain 16 multi-coil for data consistency. We also 20\% randomly sampled volumes from the dataset, which results in 172 train volumes and 47 validation volumes. Each slice in the volumes is treated independently in our proposed method, so a total of 2,730 training slices and 752 validation slices are used as our dataset. For each coil image of the slices, we further center crop the images to the $(320 \times 320)$ region in the image domain, and apply Fourier transform to generate a fully sampled k-space dataset. The ground truth of the reconstructed image is created by applying root-sum-of-squares (RSS) on center-cropped coil images. Low-frequency regions for initial sampling mask $M_0$ are set to center $(40 \times 40)$ region for 2D, and autocalibration lines $(20 \times 320)$ for 1D sampling setting, respectively.

\subsection{Model description}
\subsubsection{SR space generation}
\paragraph{Architecture} We utilized the latest flow-based SR space generation model proposed by \cite{Hong_2023_ICCV}. This model, similar to the winning model FS-NCSR~\cite{Song_2022_CVPR} in the NTIRE 2022 challenge on learning the super-resolution space~\cite{lugmayr2022ntire}, is a robust approach that suppresses unintended severe artifacts present in FS-NCSR~\cite{Song_2022_CVPR}. To achieve this, \cite{Hong_2023_ICCV} adjusted the bounds of the coefficients of the affine coupling layer~\cite{DBLP:journals/corr/KingmaW13, DBLP:conf/iclr/DinhSB17}. The original model accepts RGB images as conditional inputs and performs $4\times$ or $8\times$ SR space generation. Thus we used the original model without modifications for CS of facial images. For CS-MRI experiment, We modified it to accommodate MR images with a single channel as the conditional input. Since the size of the HR images was $320 \times 320 \times 1$, the size of the conditional inputs (i.e., LR images) was adjusted to $40 \times 40 \times 1$ (for 2D) or $20 \times 320$ (for 1D), while the remaining parameters remained consistent with \cite{Hong_2023_ICCV}.

\paragraph{Training} 
For facial images, we used the pretrained model of \cite{Hong_2023_ICCV}. For CS-MRI, we utilized the Adam optimizer~\cite{kingma2014adam} with parameters $(\beta_1,\beta_2) = (0.9, 0.999)$ and an initial learning rate of $2\times 10^{-4}$ to train the network. The learning rate was reduced by half at 50\%, 75\%, 90\%, and 95\% of the total iterations. The training was performed for 10,000 iterations with a batch size of 16. As we modified the input channel size, we did not utilize any pre-trained model. The training process was executed on a single NVIDIA A100 GPU, taking approximately 1 day to complete.

\subsubsection{Reconstruction of facial images}
\label{sub:S4.2.2}
\paragraph{Architecture} Our implementation of U-Net~\cite{ronneberger2015u} features an input layer with 6 channels (since the zero-filling reconstruction is complex), followed by a down-sampling path comprising four convolutional blocks, each performing a $3 \times 3$ convolution, batch normalization, and ReLU activation. The number of output channels of the first convolution layer is set to 32. After the down-sampling layers, a central convolutional block captures high-level features. The up-sampling path then consists of four transpose convolutional blocks followed by corresponding convolutional blocks, resulting in a total of four up-sampling layers. The final layer employs a 1x1 convolution operation, reducing the number of channels to 3. Overall, the U-Net architecture follows the classic structure with skip connections. The outermost skip connection adds the absolute value of the zero-filling reconstruction to the output, which is a similar technique used in LOUPE~\cite{bahadir2020deep}.

Our implementation of $\mathcal{H}_1$ : LOUPE is based on ~\cite{bahadir2020deep}. We implemented reconstruction network with U-Net which is same as $\mathcal{H}_{1.5}^J$. For under sampling pattern optimization network, we used non-differentiable threshold operation with a sigmoid function, which slope value of $s=200$ for $\sigma_s$ and slope value of $t=10$ for $\sigma_t$. 

For implementation of $\mathcal{H}_2$ : Policy, we replaced probabilistic mask in ~\cite{bahadir2020deep} with U-Net. Policy network takes LF component in kspace domain as an input and outputs probabilistic mask. The architecture of policy network is same as the reconstruction network, except an output layer is 1 channel and the number of output channels of the first convolution layer is set to 16.

\paragraph{Training}
We utilized the Adam optimizer~\cite{kingma2014adam} with parameters $(\beta_1,\beta_2) = (0.9, 0.999)$ and an initial learning rate of $1\times 10^{-3}$ to train the network. All models are trained to optimize L1-regularization loss. The training was performed for 10 iterations with a batch size of 128. The training process was executed on a single NVIDIA GeForce RTX 3090Ti GPU, taking approximately 4 hours to complete.

\subsubsection{Reconstruction of MR images}
\paragraph{Architecture} We used simplified version of E2E-VarNet~\cite{sriram2020end} for DL-based reconstruction network in our adaptive selection scheme. To be specific, we use 5 cascades architecture rather than 12 cascades in the original architecture. In addition, we downsized the number of pooling layers for U-Net and the number of pooling layers for sense estimation to 3, which is 4 in the original. The remaining parameters remained consistent with \cite{sriram2020end}.

Our implementation of $\mathcal{H}_1$ : LOUPE and $\mathcal{H}_2$ : Policy is based on \cite{bakker2022learning}. We implemented reconstruction network with E2E-VarNet which is same as $\mathcal{H}_{1.5}^J$. In addition, we used non-differentiable threshold operation with a sigmoid function, which slope value of $s=200$ for $\sigma_s$ and slope value of $t=10$ for $\sigma_t$. For implementation of $\mathcal{H}_2$ : Policy, we implemented policy network with U-Net as in ~\cite{bahadir2020deep} to output probabilistic mask from LF component in kspace domain. The architecture of policy network is same as U-Net in \labelcref{sub:S4.2.2}, except an input layer is 2 channels and an output layer is 1 channel.

\paragraph{Training}
All networks are trained with the Adam optimizer~\cite{kingma2014adam} with parameters $(\beta_1,\beta_2) = (0.9, 0.999)$ and an initial learning rate of $1\times 10^{-3}$. The models are trained to optimize SSIM. The training is performed for 50 iterations with a batch size of 4, and the learning rate is reduced by 10\% on epoch 40. The training process was executed on a single NVIDIA GeForce RTX 3090Ti GPU, taking approximately 8 hours to complete.

\subsection{Sampling masks}
We generated masks that sampled from variable density probability map with $a_F = 1.5$~\cite{wang2009variable}.

\subsection{Evaluation}
For quantitative evaluations of the reconstruction, we assess two metrics, peak-to-signal ratio (PSNR) and structural similarity index measure (SSIM).
PSNR is defined as:
\begin{equation}
    \mathrm{{PSNR}} = 20 \log_{10}(\mathrm{{MAX}}) - 10 \log_{10}(\mathrm{{MSE}}),
\end{equation}
where $\mathrm{MAX}$ represents the maximum possible value that a pixel can have, and $\mathrm{MSE}$ refers to the mean-squared error of the reconstruction output. This also explains the first equality in equation (8).

SSIM quantifies the structural similarity between two images. It takes into account luminance, contrast, and structural information. The SSIM index is calculated using the following formula:
\begin{equation}
    \mathrm{{SSIM}}(\hat{I}, I) = \frac{{(2 \mu_{\hat{I}} \mu_I + c_1)(2 \sigma_{\hat{I}I} + c_2)}}{{(\mu_{\hat{I}}^2 + \mu_I^2 + c_1)(\sigma_{\hat{I}}^2 + \sigma_I^2 + c_2)}},
\end{equation}
where $\hat{I}$ and $I$ represent the reconstructed and original images, respectively.
$\mu_{\hat{I}}$ and $\mu_I$ denote the mean values of $\hat{I}$ and $I$,
$\sigma_{\hat{I}}$ and $\sigma_I$ represent the standard deviations of  $\hat{I}$ and $I$, and $\sigma_{\hat{I}I}$ represents the covariance between  $\hat{I}$ and $I$. The constants $c_1$ and $c_2$ are used to ensure stability in the division when the denominator approaches zero. The specific values of these constants were obtained from the official fastMRI repository~\cite{zbontar2018fastmri}.

\section{Additional results}
\begin{figure*}[h!]
\scriptsize
\centering
    \setlength{\tabcolsep}{0pt}
\begin{subfigure}{0.95\textwidth}
\begin{tabularx}{\linewidth}{cccccc}
Ground truth & $\mathcal{H}_1$: Random & $\mathcal{H}_1$: VD~\cite{wang2009variable} & $\mathcal{H}_1$:LOUPE\!\cite{bahadir2020deep} & $\mathcal{H}_2$: Policy~\cite{bakker2022learning} & $\mathcal{H}_{1.5}^J$ (ours) \\
\includegraphics[width=0.166\linewidth]{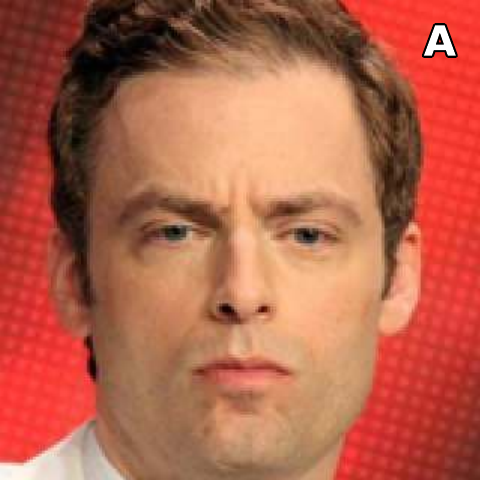} & \includegraphics[width=0.166\linewidth]{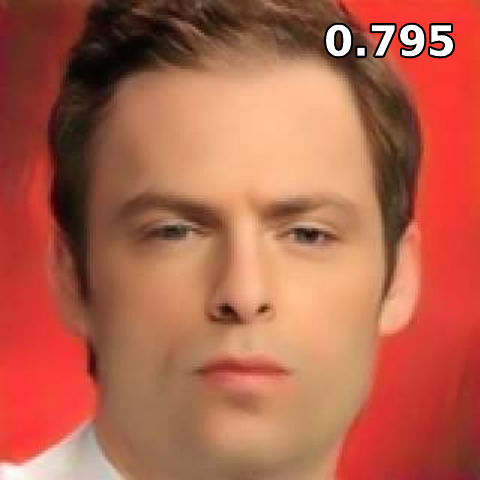} & \includegraphics[width=0.166\linewidth]{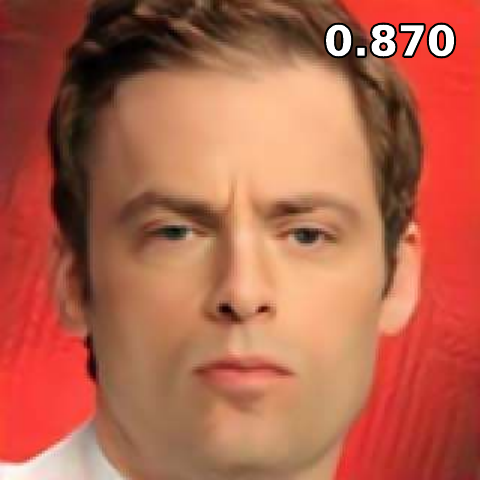} & \includegraphics[width=0.166\linewidth]{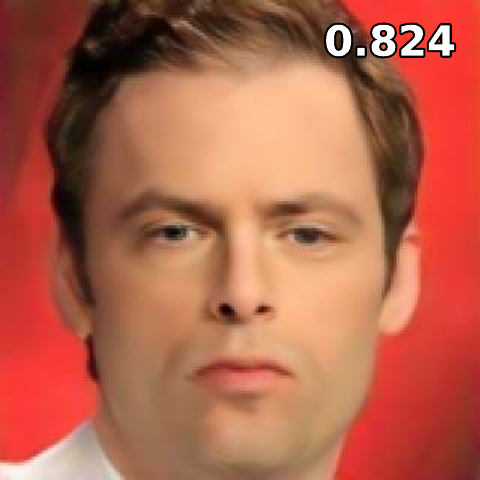} &  \includegraphics[width=0.166\linewidth]{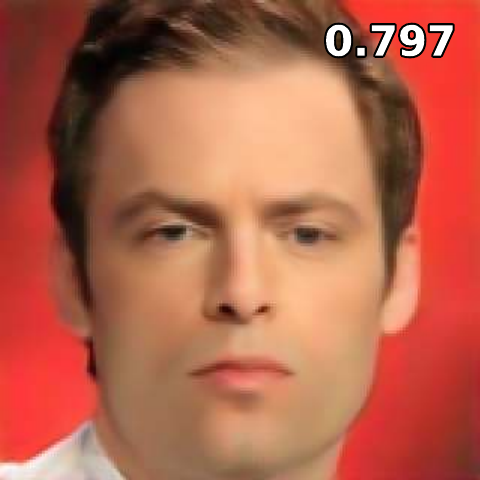} &  \includegraphics[width=0.166\linewidth]{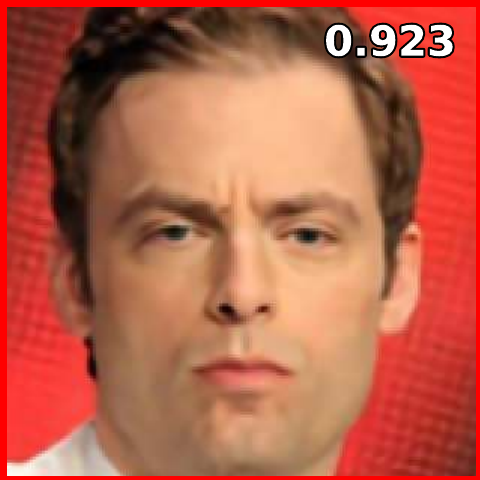}\\[-\dp\strutbox]
\includegraphics[width=0.166\linewidth]{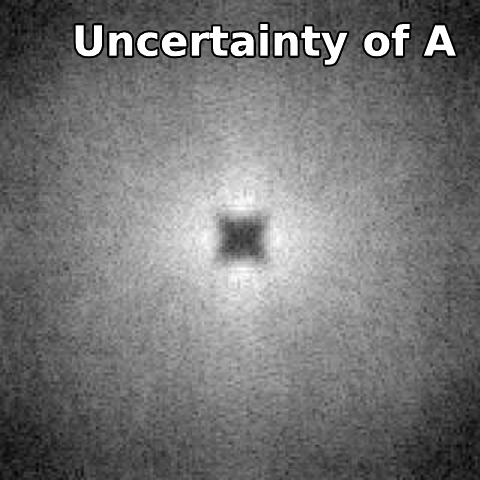} & \includegraphics[width=0.166\linewidth]{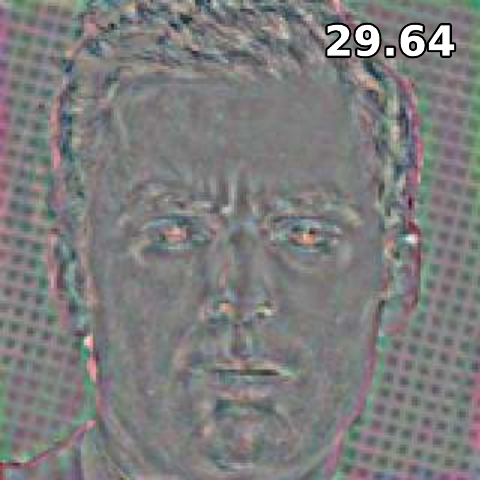} & \includegraphics[width=0.166\linewidth]{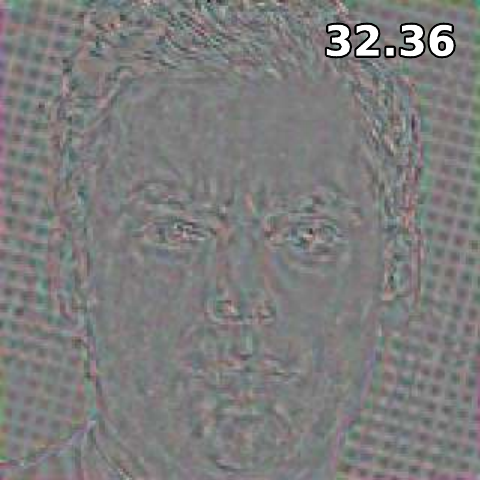} & \includegraphics[width=0.166\linewidth]{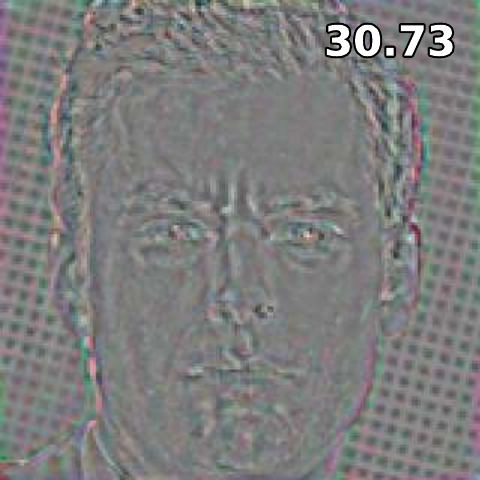} & \includegraphics[width=0.166\linewidth]{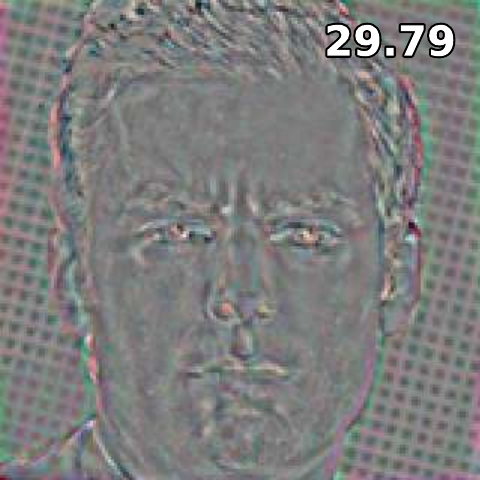} &  \includegraphics[width=0.166\linewidth]{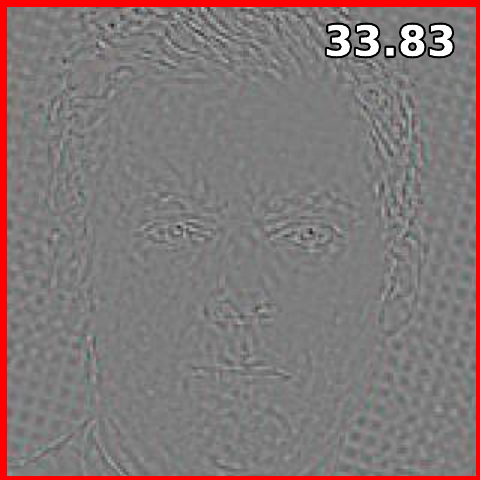}\\[-\dp\strutbox]
\includegraphics[width=0.166\linewidth]{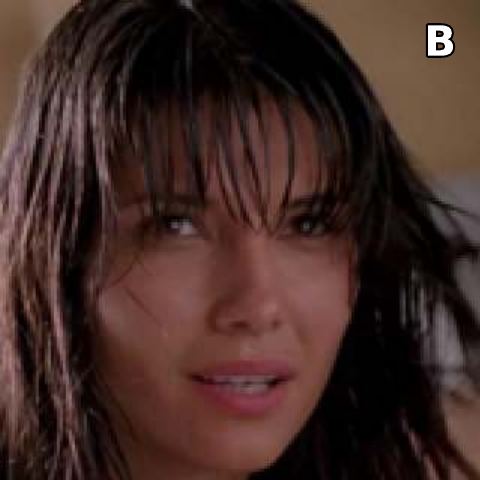} & \includegraphics[width=0.166\linewidth]{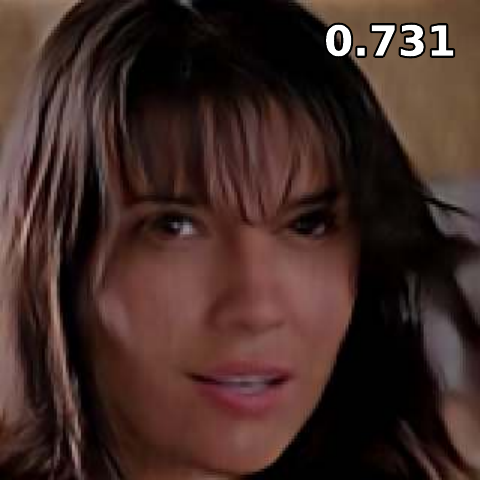} & \includegraphics[width=0.166\linewidth]{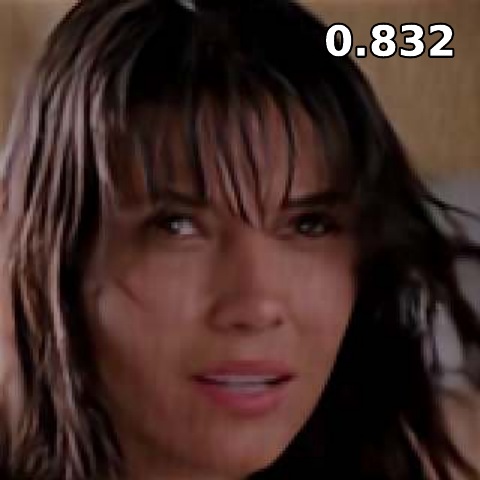} & \includegraphics[width=0.166\linewidth]{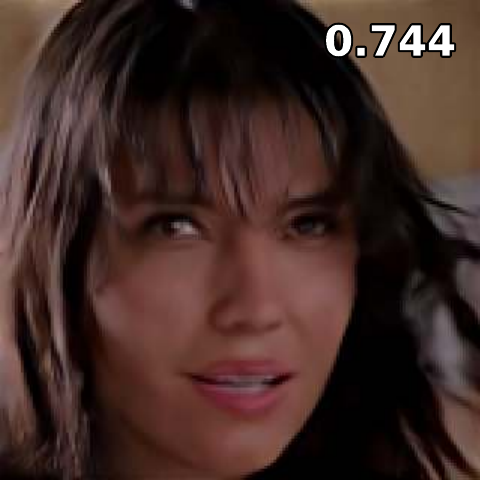} & \includegraphics[width=0.166\linewidth]{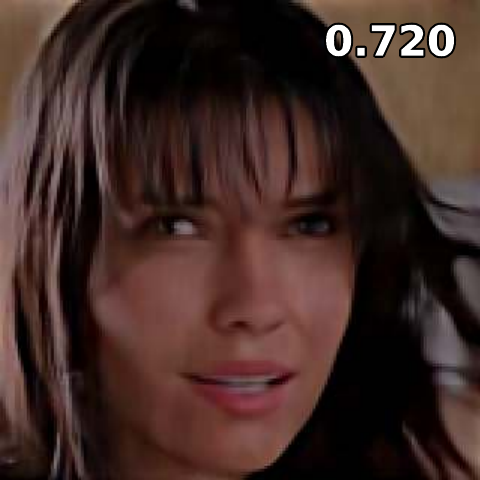} &  \includegraphics[width=0.166\linewidth]{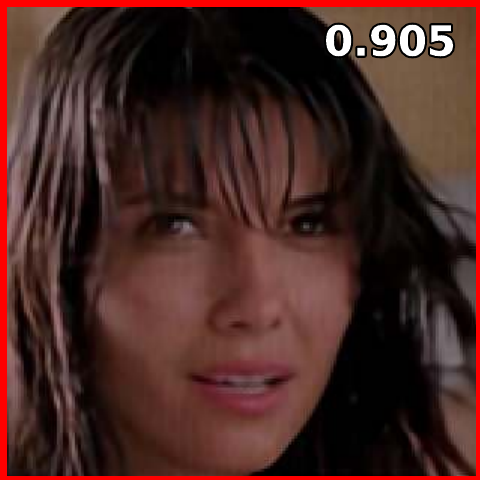} \\[-\dp\strutbox]
\includegraphics[width=0.166\linewidth]{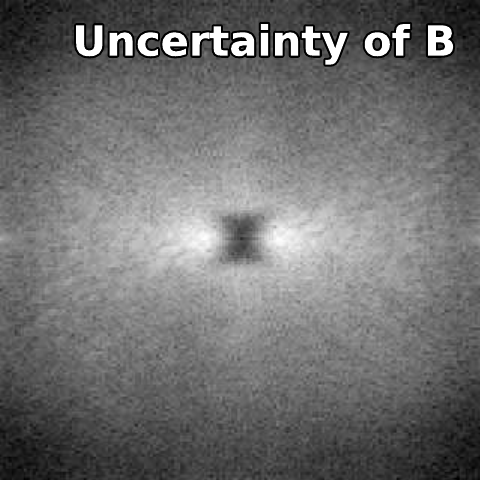} & \includegraphics[width=0.166\linewidth]{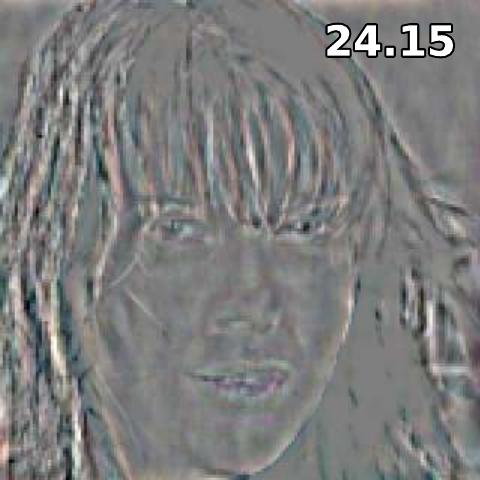} & \includegraphics[width=0.166\linewidth]{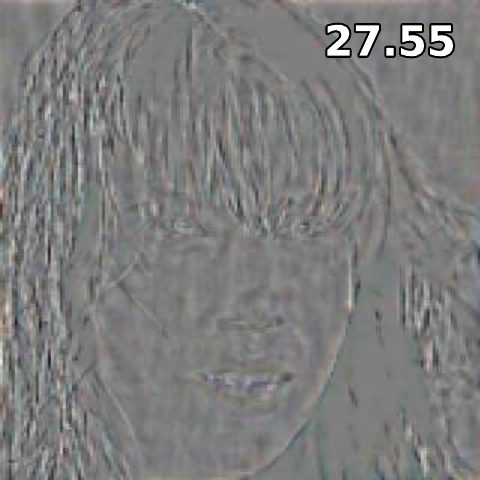} & \includegraphics[width=0.166\linewidth]{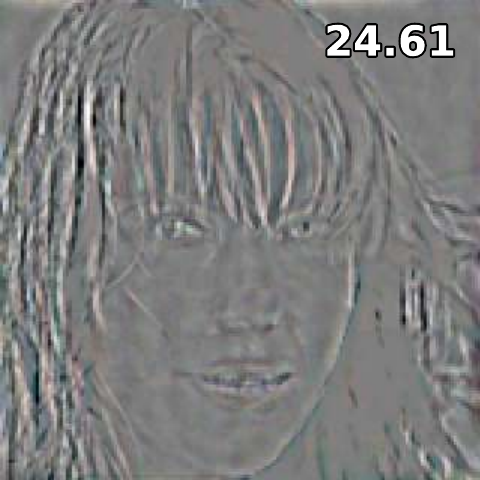} & \includegraphics[width=0.166\linewidth]{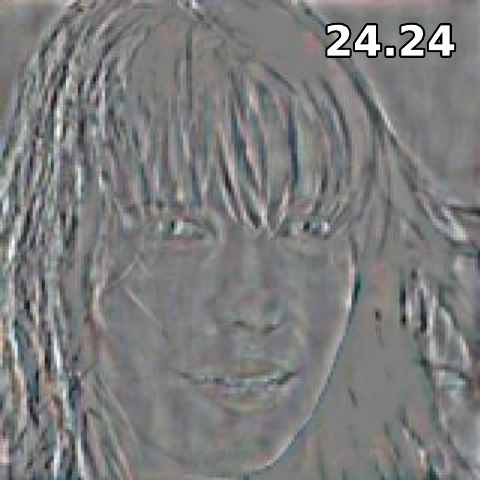} &  \includegraphics[width=0.166\linewidth]{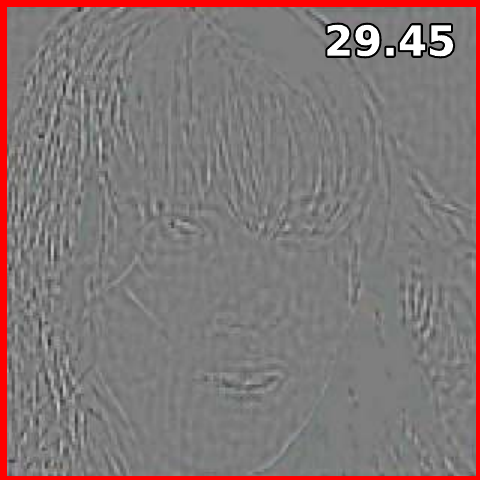} \\[-\dp\strutbox]
\end{tabularx}
\caption{Comparison of various methods.}
\end{subfigure}
\par\bigskip
\begin{subfigure}{0.95\textwidth}
\centering
\begin{tabularx}{0.833\linewidth}{ccccc}
    \includegraphics[width=0.166\linewidth]{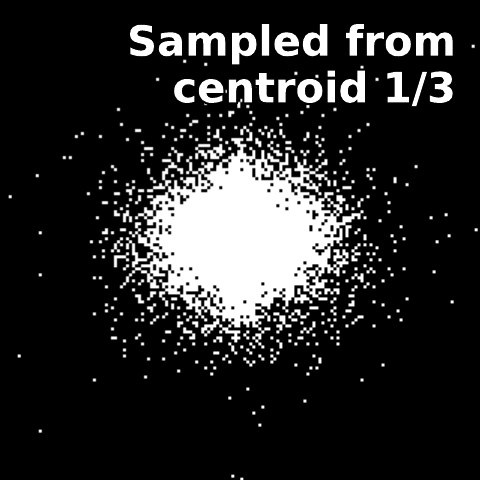} & \includegraphics[width=0.166\linewidth]{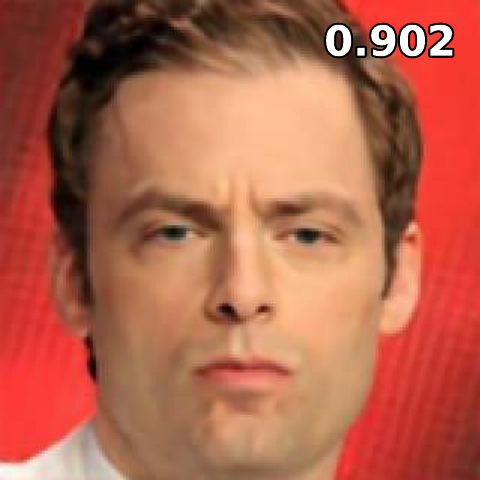} & \includegraphics[width=0.166\linewidth]{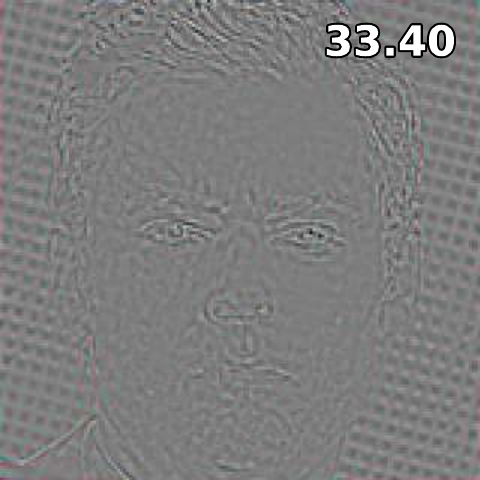} & \includegraphics[width=0.166\linewidth]{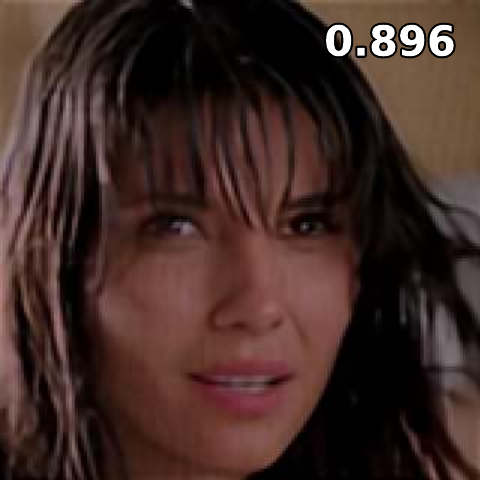} & \includegraphics[width=0.166\linewidth]{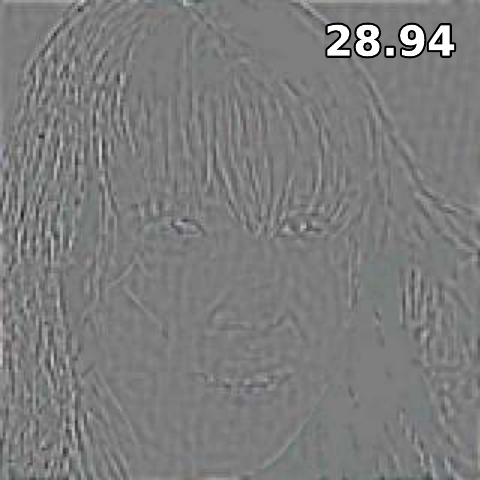} \\[-\dp\strutbox]
    \includegraphics[width=0.166\linewidth]{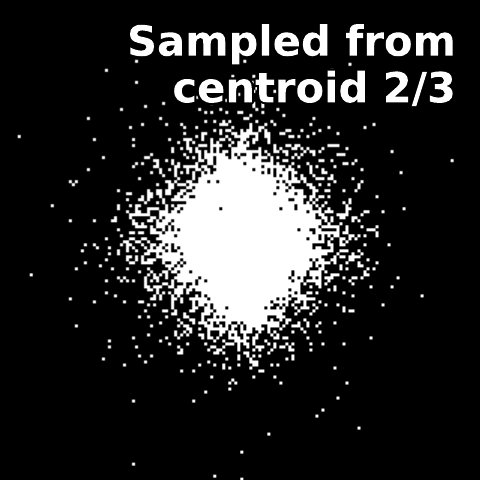} & \includegraphics[width=0.166\linewidth]{figs/face_8x/ra2.pdf} & \includegraphics[width=0.166\linewidth]{figs/face_8x/ea2.pdf} & \includegraphics[width=0.166\linewidth]{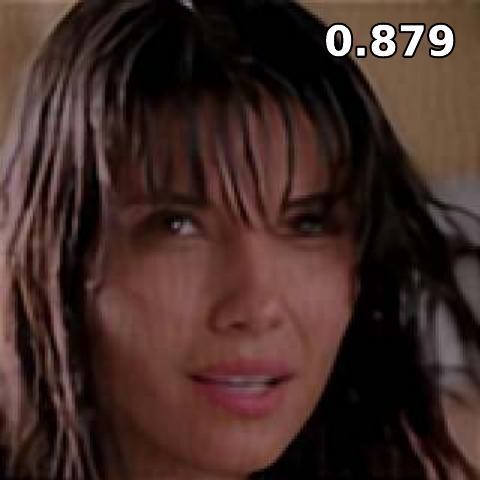} & \includegraphics[width=0.166\linewidth]{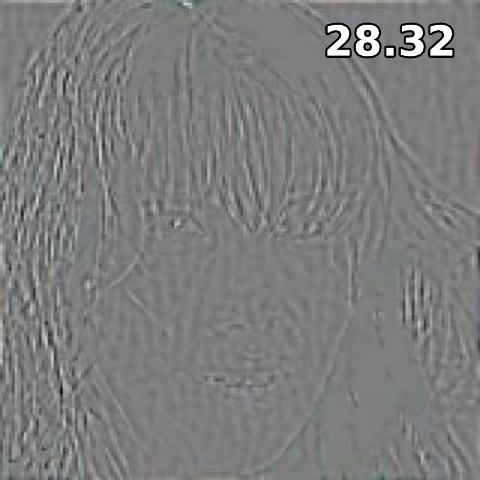}  \\[-\dp\strutbox]
    \includegraphics[width=0.166\linewidth]{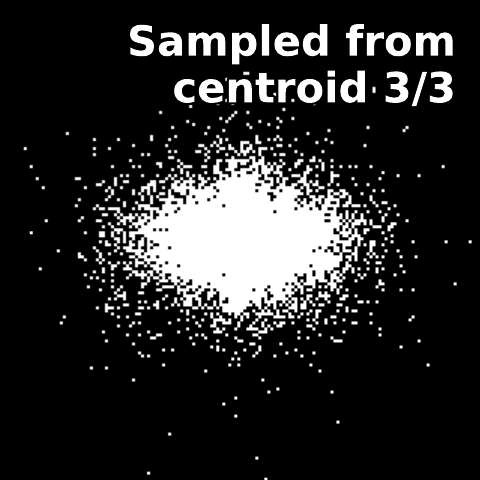} & \includegraphics[width=0.166\linewidth]{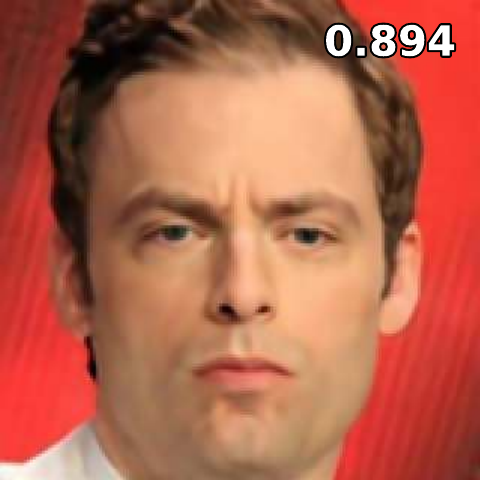} & \includegraphics[width=0.166\linewidth]{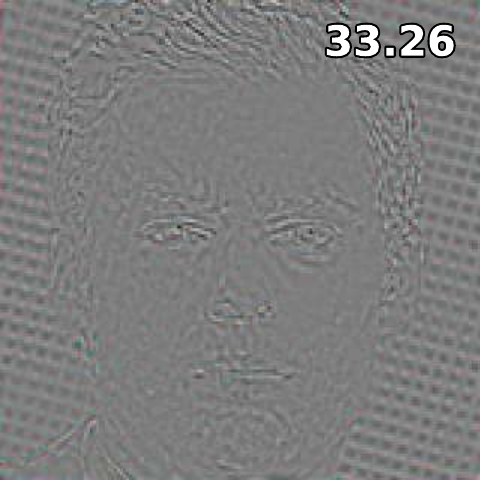} & \includegraphics[width=0.166\linewidth]{figs/face_8x/rb3.pdf} & \includegraphics[width=0.166\linewidth]{figs/face_8x/eb3.pdf}  \\[-\dp\strutbox]
\end{tabularx}
\caption{Comparison of the sampling-reconstruction pairs $(M_j,\theta_j)_{j=1}^3$ in our method $\mathcal{H}_{1.5}^J$.}
\end{subfigure}
  \caption{
  Qualitative comparison of reconstruction and error map (a) of various methods, (b) obtained using the mask-reconstruction pairs ($(M_j, \theta_j)_{j=1}^3$) generated from Algorithm 1, at acceleration rate $8\times$ in the CelebA dataset~\cite{liu2015deep}. 
  For comparison, we also show the results of the variable density (VD) \cite{wang2009variable}, LOUPE~\cite{bahadir2020deep}, and policy-based adaptive sampling~\cite{bakker2022learning} in (a).  SSIMs and PSNRs are included in the reconstructions and the error maps, respectively. 
   The images highlighted in \textcolor{red}{red} in (b) demonstrate that our Algorithm 2 estimated the HF Bayesian uncertainty of each image and selected an appropriate sampling-reconstruction pair $(M_j,\theta_j)$ using the uncertainty. The images thus selected become the final output of our method $\mathcal{H}_{1.5}^J$, as emphasized in \textcolor{red}{red} in (a).
}
  \label{sfig:5.1}

\end{figure*}

\begin{figure*}[h!]
\scriptsize
\centering
    \setlength{\tabcolsep}{0pt}
\begin{subfigure}{0.95\textwidth}
\begin{tabularx}{\linewidth}{cccccc}
Ground truth & $\mathcal{H}_1$: Random & $\mathcal{H}_1$: VD~\cite{wang2009variable} & $\mathcal{H}_1$:LOUPE\!\cite{bahadir2020deep} & $\mathcal{H}_2$: Policy~\cite{bakker2022learning} & $\mathcal{H}_{1.5}^J$ (ours) \\
\includegraphics[width=0.166\linewidth]{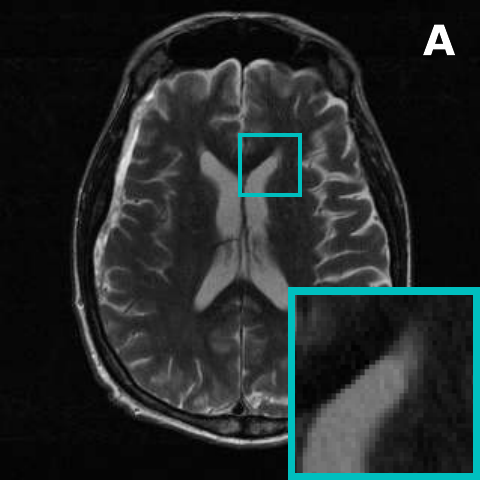} & \includegraphics[width=0.166\linewidth]{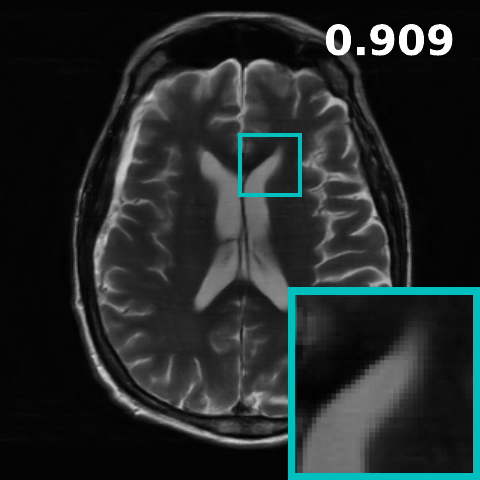} & \includegraphics[width=0.166\linewidth]{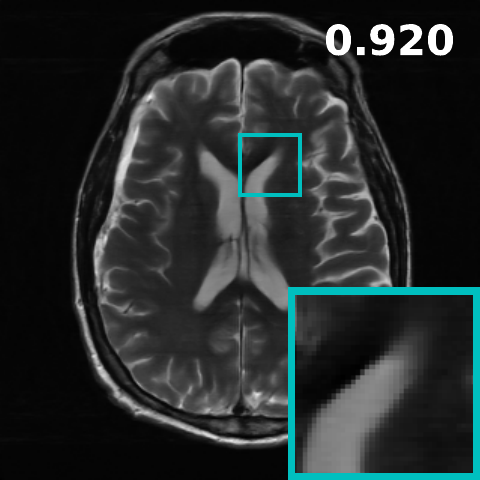} & \includegraphics[width=0.166\linewidth]{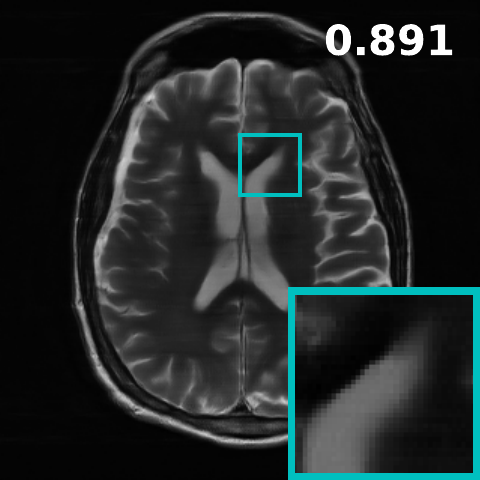} &  \includegraphics[width=0.166\linewidth]{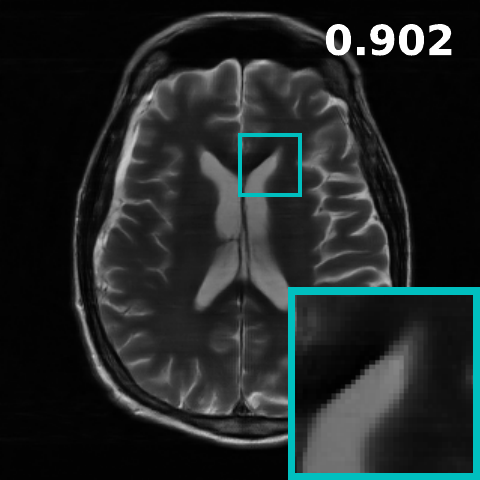} &  \includegraphics[width=0.166\linewidth]{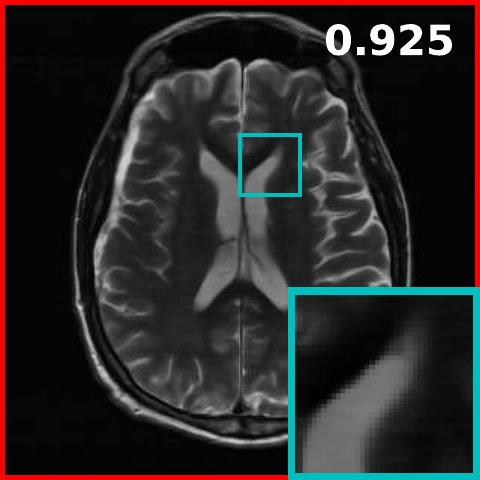}\\[-\dp\strutbox]
\includegraphics[width=0.166\linewidth]{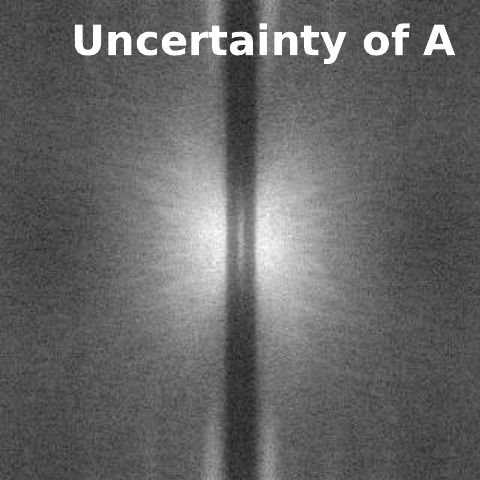} & \includegraphics[width=0.166\linewidth]{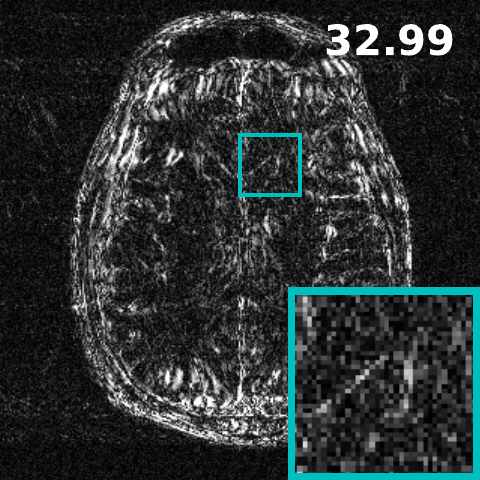} & \includegraphics[width=0.166\linewidth]{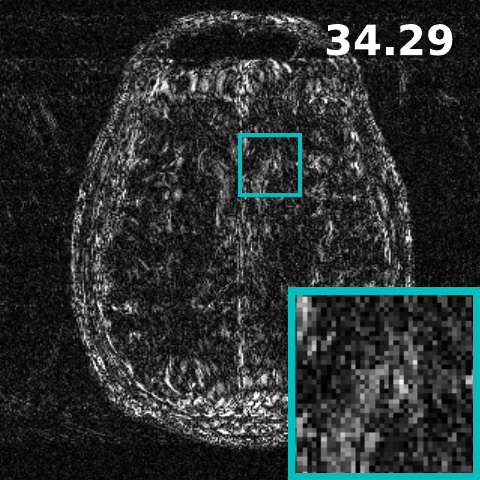} & \includegraphics[width=0.166\linewidth]{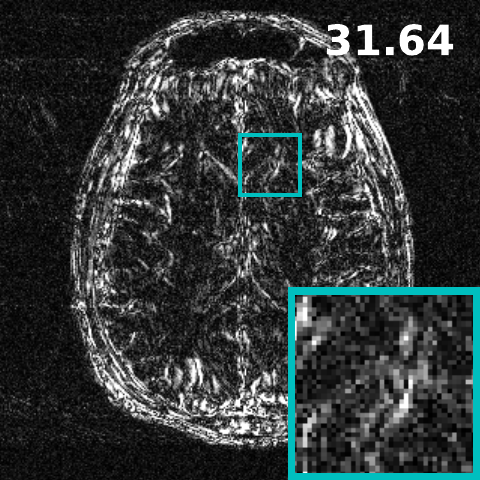} & \includegraphics[width=0.166\linewidth]{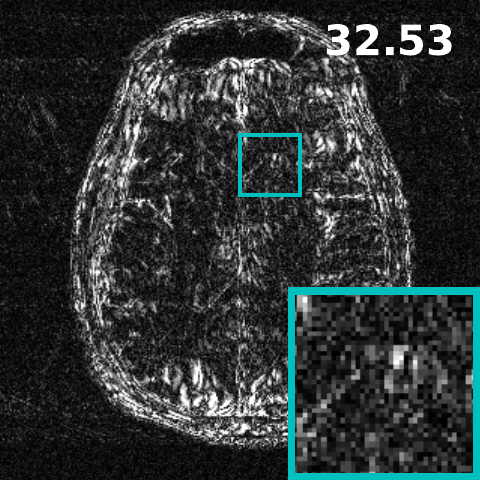} &  \includegraphics[width=0.166\linewidth]{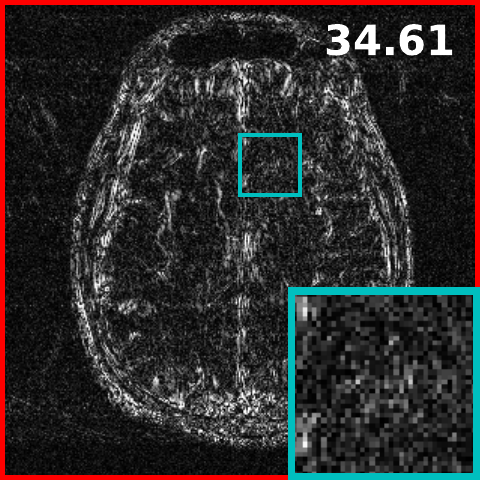}\\[-\dp\strutbox]
\includegraphics[width=0.166\linewidth]{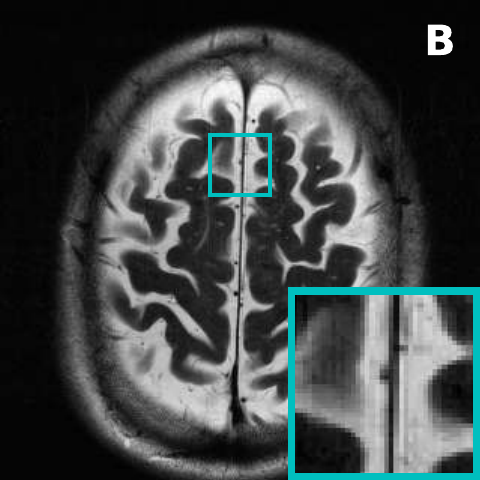} & \includegraphics[width=0.166\linewidth]{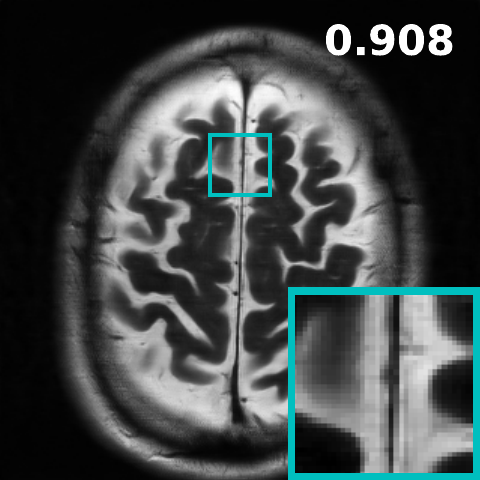} & \includegraphics[width=0.166\linewidth]{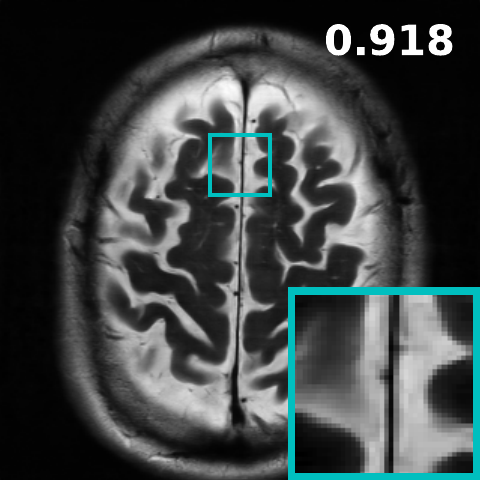} & \includegraphics[width=0.166\linewidth]{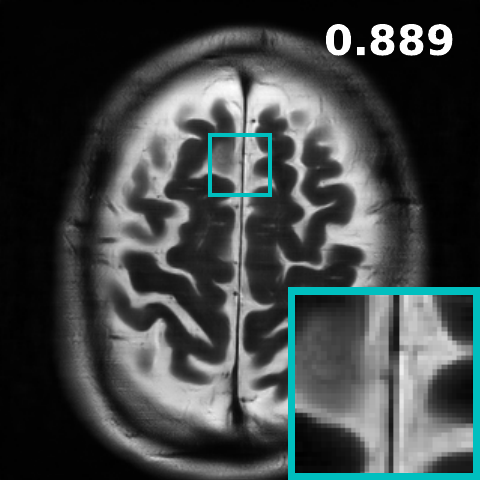} & \includegraphics[width=0.166\linewidth]{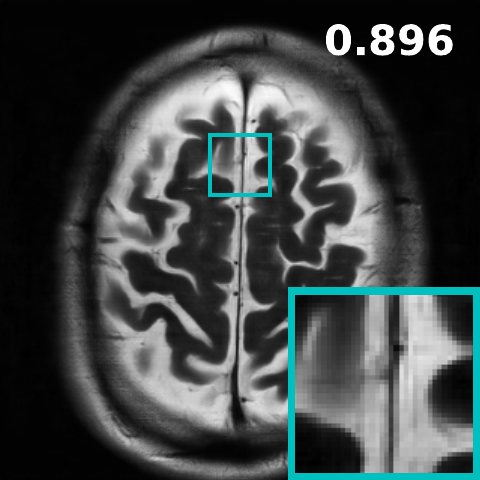} &  \includegraphics[width=0.166\linewidth]{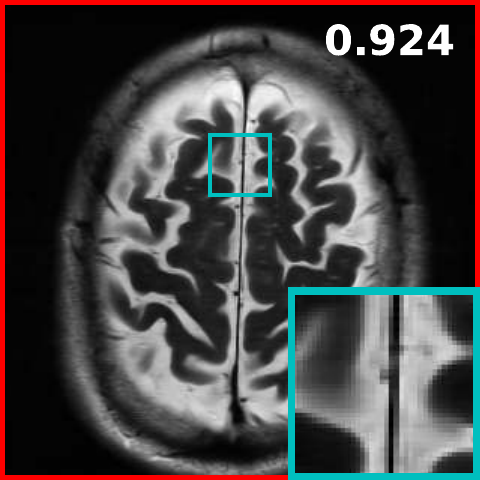} \\[-\dp\strutbox]
\includegraphics[width=0.166\linewidth]{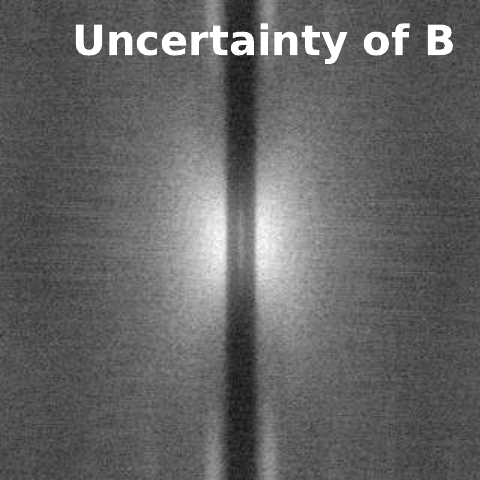} & \includegraphics[width=0.166\linewidth]{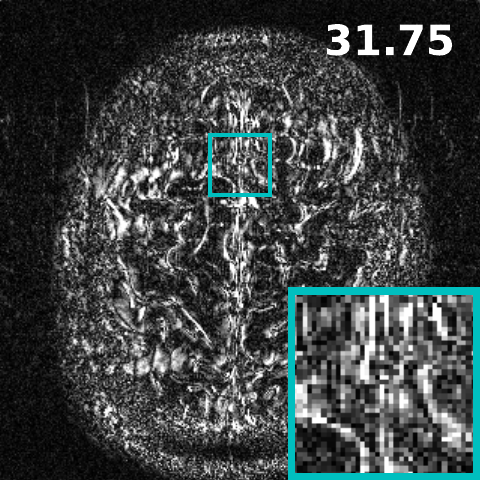} & \includegraphics[width=0.166\linewidth]{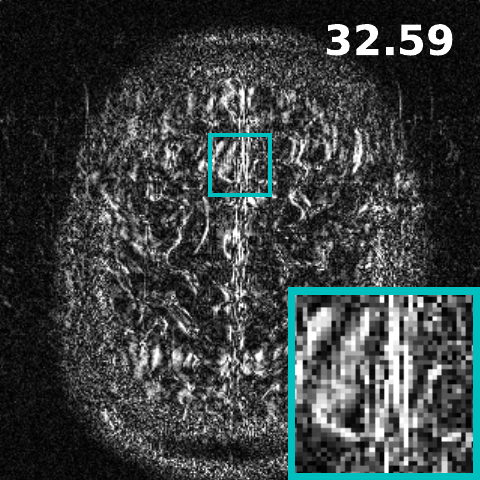} & \includegraphics[width=0.166\linewidth]{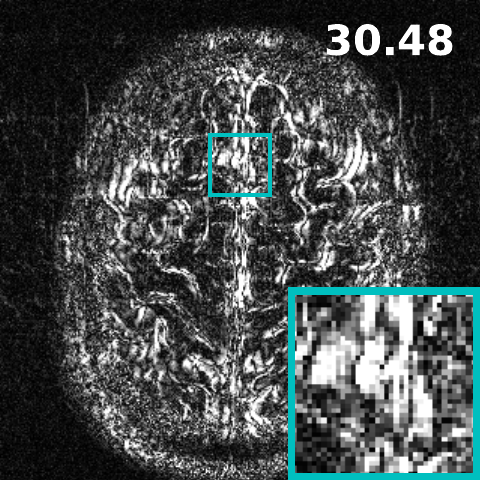} & \includegraphics[width=0.166\linewidth]{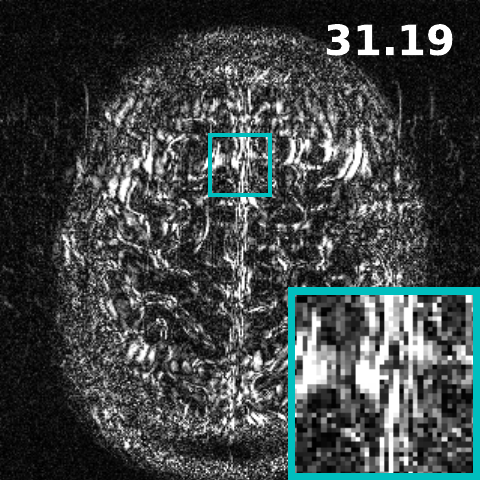} &  \includegraphics[width=0.166\linewidth]{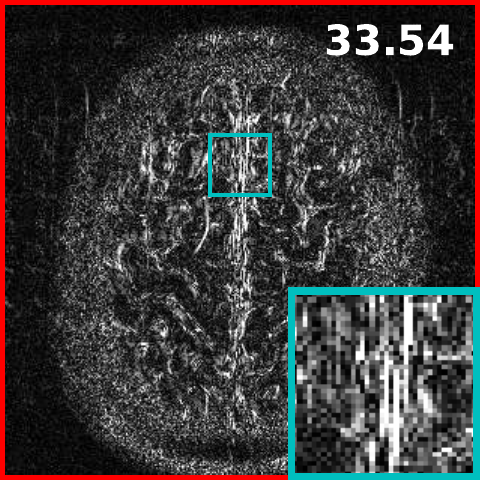} \\[-\dp\strutbox]
\end{tabularx}
\caption{Comparison of various methods.}
\end{subfigure}
\par\bigskip
\begin{subfigure}{0.95\textwidth}
\centering
\begin{tabularx}{0.833\linewidth}{ccccc}
    \includegraphics[width=0.166\linewidth]{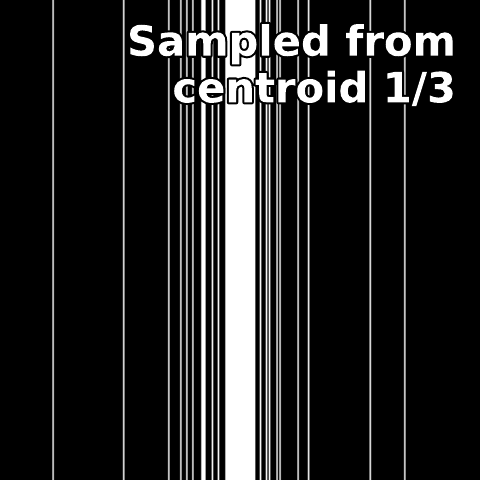} & \includegraphics[width=0.166\linewidth]{figs/mr_1d8x/ra1.pdf} & \includegraphics[width=0.166\linewidth]{figs/mr_1d8x/ea1.pdf} & \includegraphics[width=0.166\linewidth]{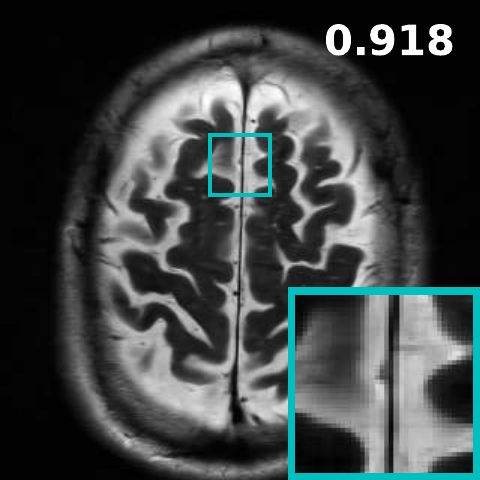} & \includegraphics[width=0.166\linewidth]{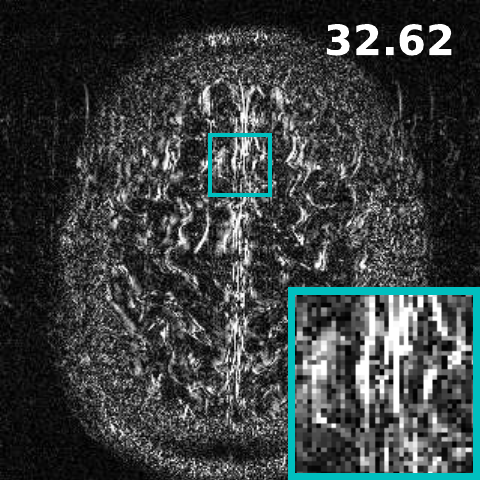} \\[-\dp\strutbox]
    \includegraphics[width=0.166\linewidth]{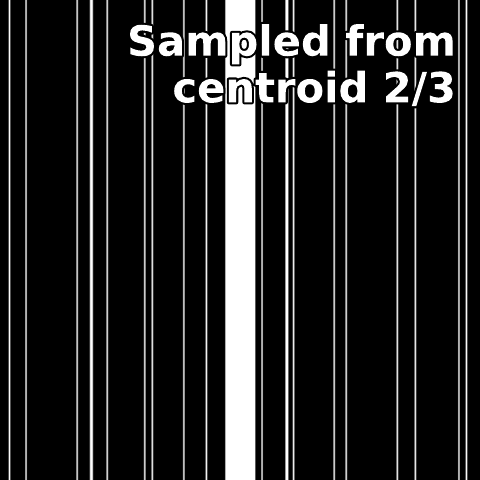} & \includegraphics[width=0.166\linewidth]{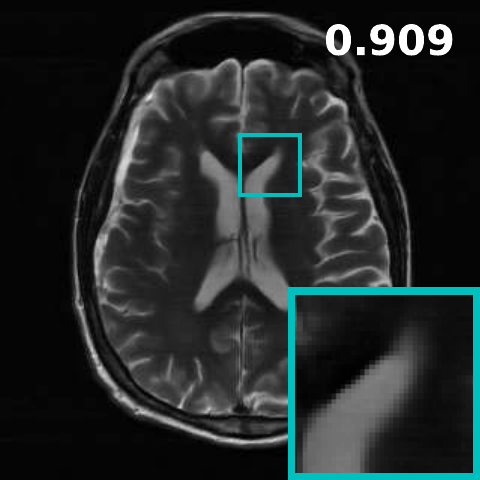} & \includegraphics[width=0.166\linewidth]{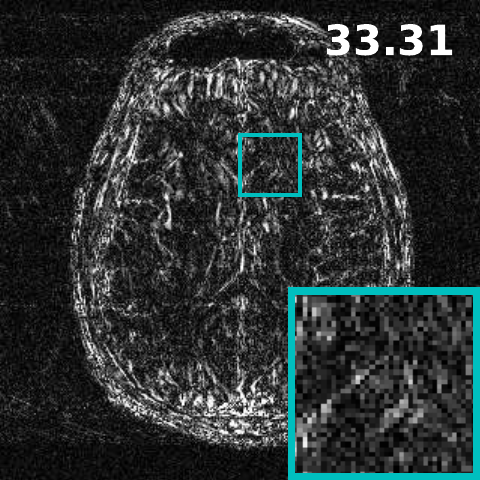} & \includegraphics[width=0.166\linewidth]{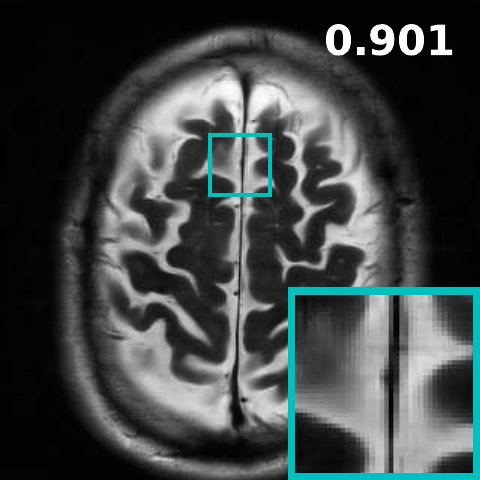} & \includegraphics[width=0.166\linewidth]{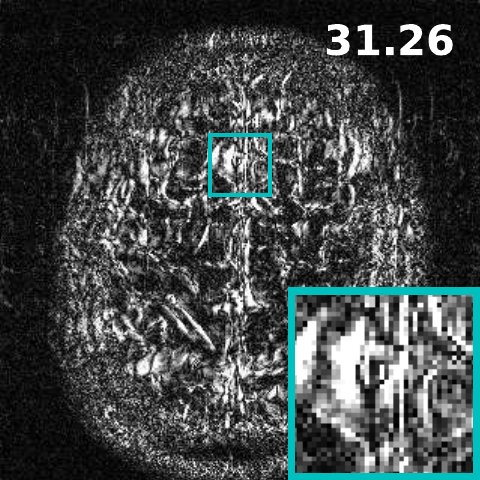}  \\[-\dp\strutbox]
    \includegraphics[width=0.166\linewidth]{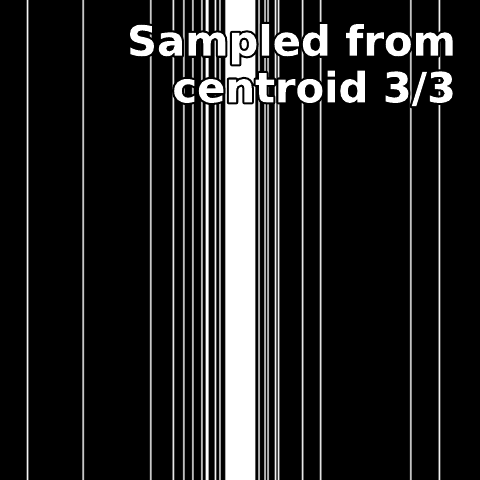} & \includegraphics[width=0.166\linewidth]{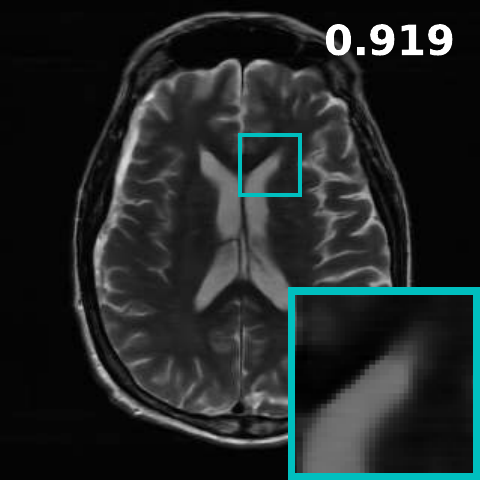} & \includegraphics[width=0.166\linewidth]{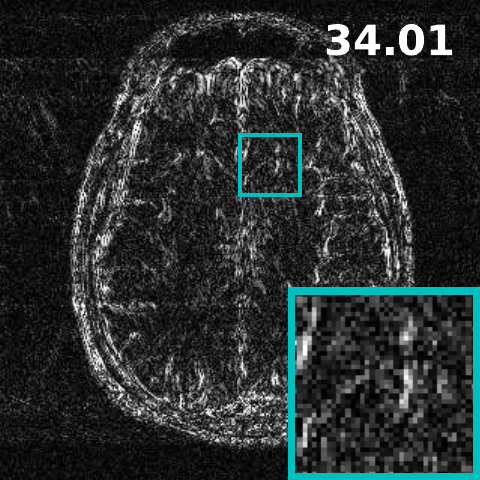} & \includegraphics[width=0.166\linewidth]{figs/mr_1d8x/rb3.pdf} & \includegraphics[width=0.166\linewidth]{figs/mr_1d8x/eb3.pdf}  \\[-\dp\strutbox]
\end{tabularx}
\caption{Comparison of the sampling-reconstruction pairs $(M_j,\theta_j)_{j=1}^3$ in our method $\mathcal{H}_{1.5}^J$.}
\end{subfigure}
  \caption{  Qualitative comparison of reconstruction and error map (a) of various methods, (b) obtained using the mask-reconstruction pairs ($(M_j, \theta_j)_{j=1}^3$) generated from Algorithm 1, at acceleration rate $8\times$ in the fastMRI dataset~\cite{zbontar2018fastmri}. 
  For comparison, we also show the results of the variable density (VD) \cite{wang2009variable}, LOUPE~\cite{bahadir2020deep}, and policy-based adaptive sampling~\cite{bakker2022learning} in (a).  SSIMs and PSNRs are included in the reconstructions and the error maps, respectively. 
  The images highlighted in \textcolor{red}{red} in (b) demonstrate that our Algorithm 2 estimated the HF Bayesian uncertainty of each image and selected an appropriate sampling-reconstruction pair $(M_j,\theta_j)$ using the uncertainty. The images thus selected become the final output of our method $\mathcal{H}_{1.5}^J$, as emphasized in \textcolor{red}{red} in (a).
}
  \label{sfig:5.2}

\end{figure*}

\begin{figure*}[h!]
\scriptsize
\centering
    \setlength{\tabcolsep}{0pt}
\begin{subfigure}{0.95\textwidth}
\begin{tabularx}{\linewidth}{cccccc}
Ground truth & $\mathcal{H}_1$: Random & $\mathcal{H}_1$: VD~\cite{wang2009variable} & $\mathcal{H}_1$:LOUPE\!\cite{bahadir2020deep} & $\mathcal{H}_2$: Policy~\cite{bakker2022learning} & $\mathcal{H}_{1.5}^J$ (ours) \\
\includegraphics[width=0.166\linewidth]{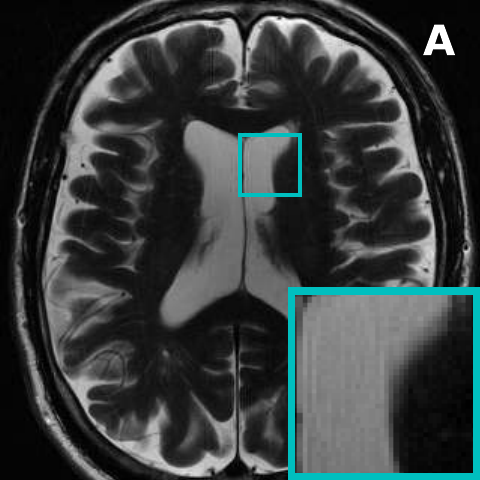} & \includegraphics[width=0.166\linewidth]{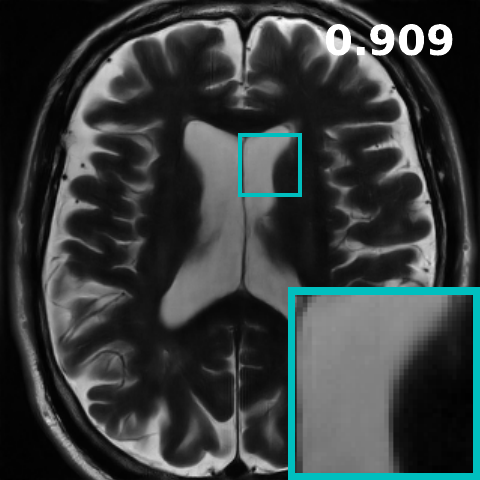} & \includegraphics[width=0.166\linewidth]{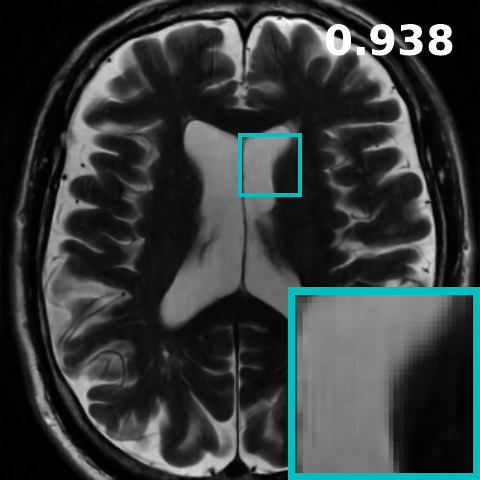} & \includegraphics[width=0.166\linewidth]{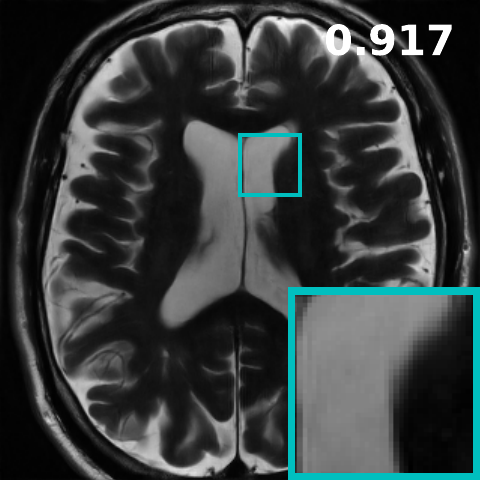} &  \includegraphics[width=0.166\linewidth]{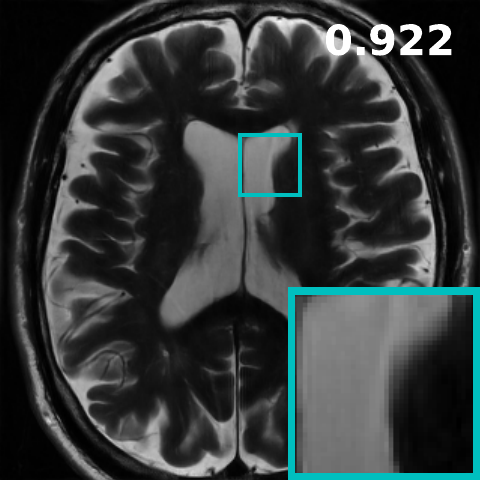} &  \includegraphics[width=0.166\linewidth]{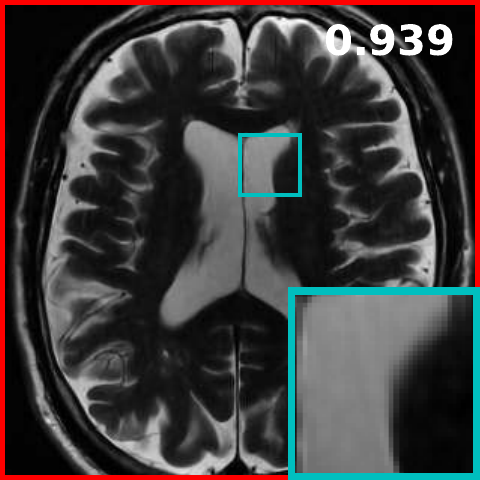}\\[-\dp\strutbox]
\includegraphics[width=0.166\linewidth]{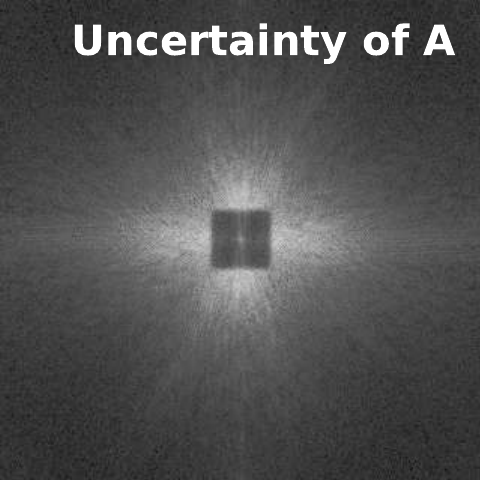} & \includegraphics[width=0.166\linewidth]{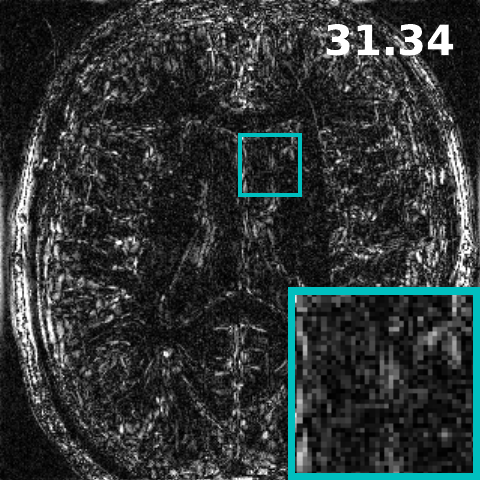} & \includegraphics[width=0.166\linewidth]{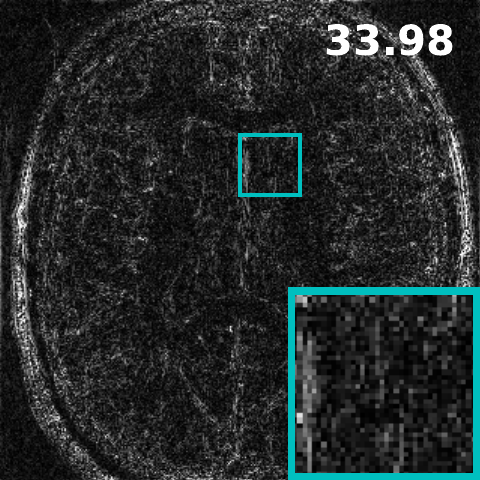} & \includegraphics[width=0.166\linewidth]{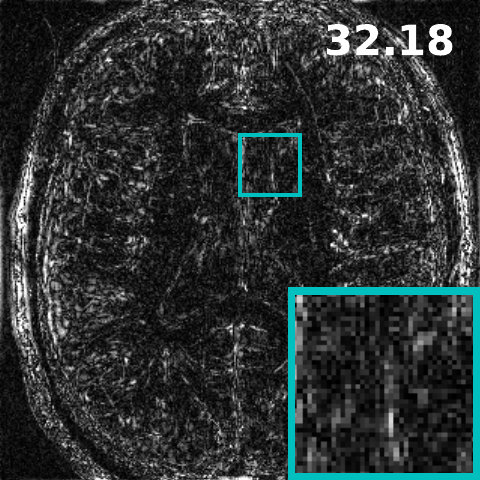} & \includegraphics[width=0.166\linewidth]{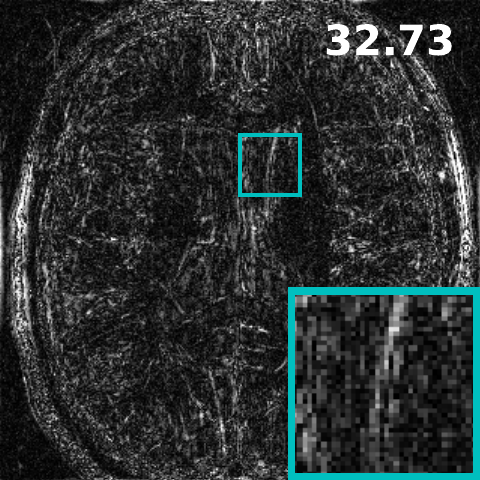} &  \includegraphics[width=0.166\linewidth]{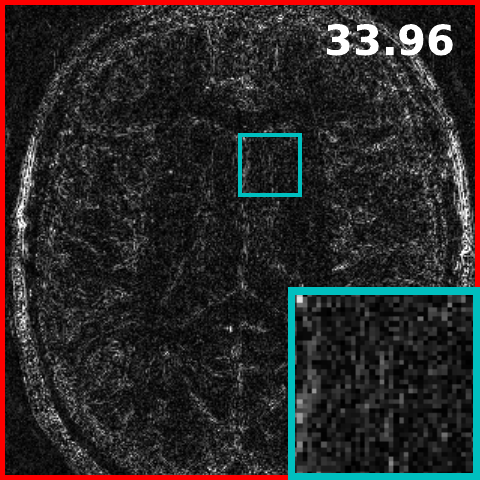}\\[-\dp\strutbox]
\includegraphics[width=0.166\linewidth]{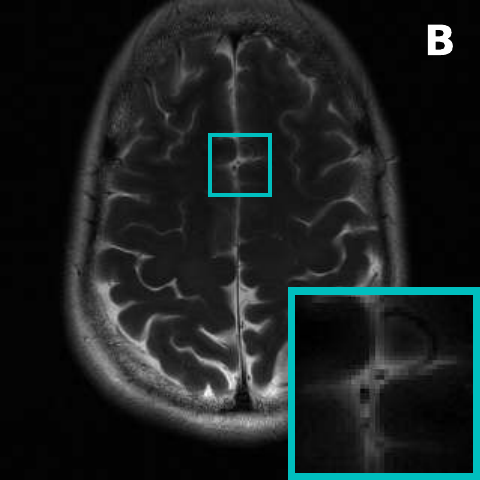} & \includegraphics[width=0.166\linewidth]{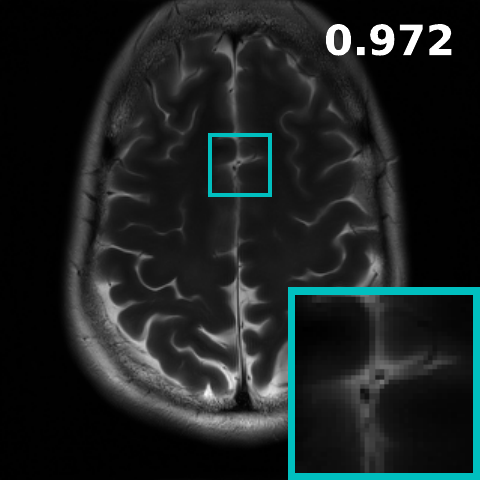} & \includegraphics[width=0.166\linewidth]{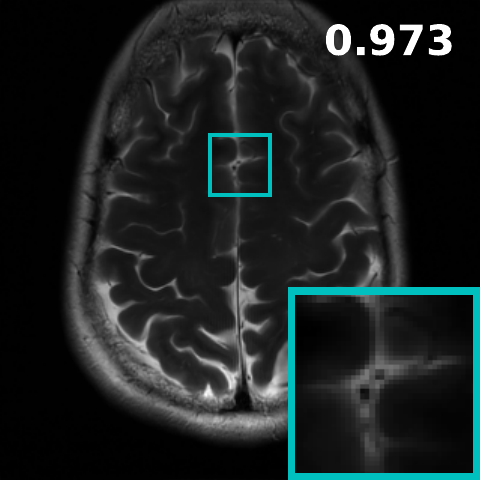} & \includegraphics[width=0.166\linewidth]{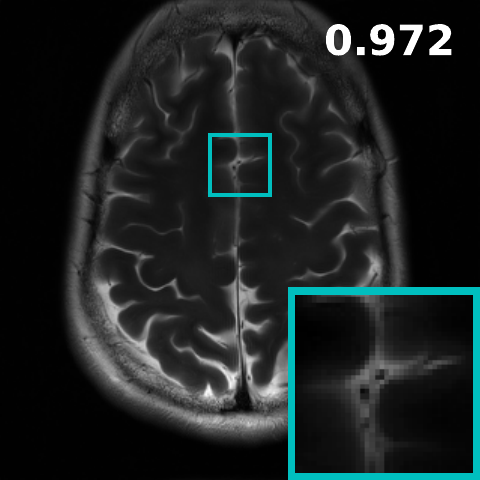} & \includegraphics[width=0.166\linewidth]{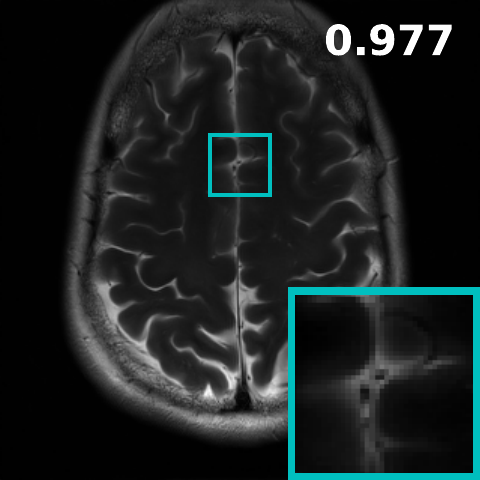} &  \includegraphics[width=0.166\linewidth]{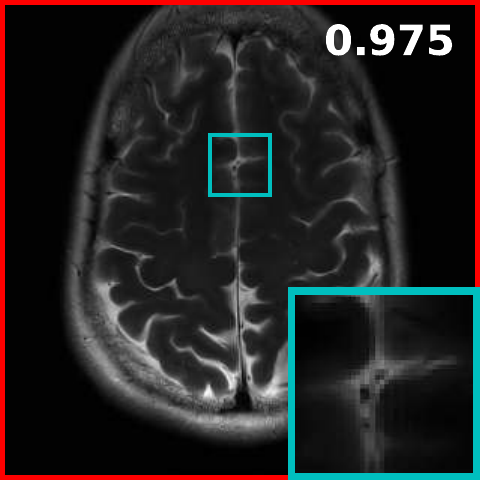} \\[-\dp\strutbox]
\includegraphics[width=0.166\linewidth]{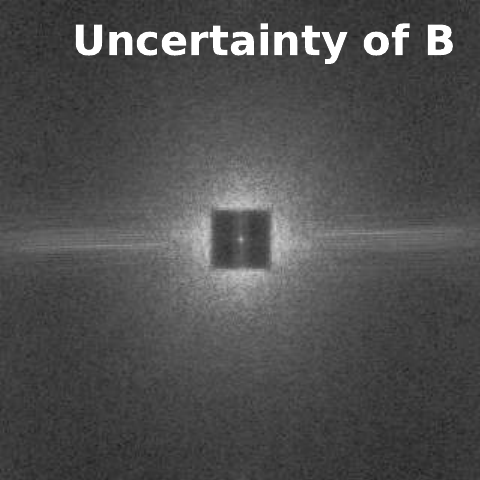} & \includegraphics[width=0.166\linewidth]{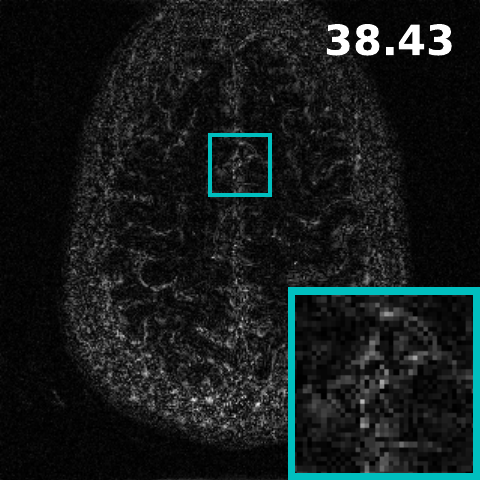} & \includegraphics[width=0.166\linewidth]{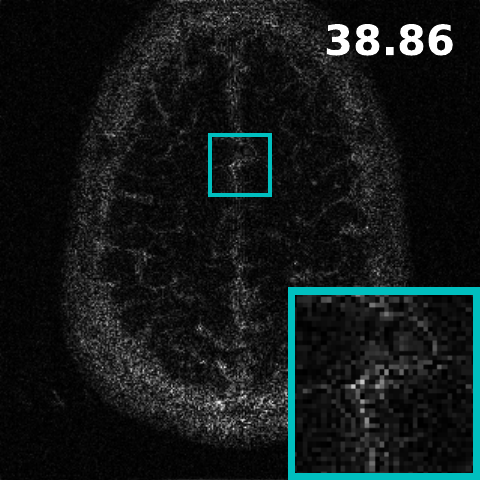} & \includegraphics[width=0.166\linewidth]{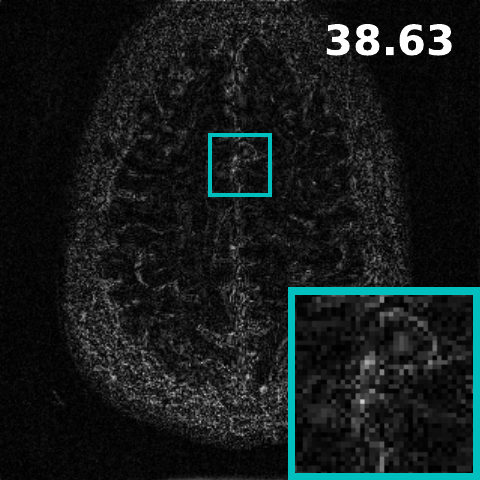} & \includegraphics[width=0.166\linewidth]{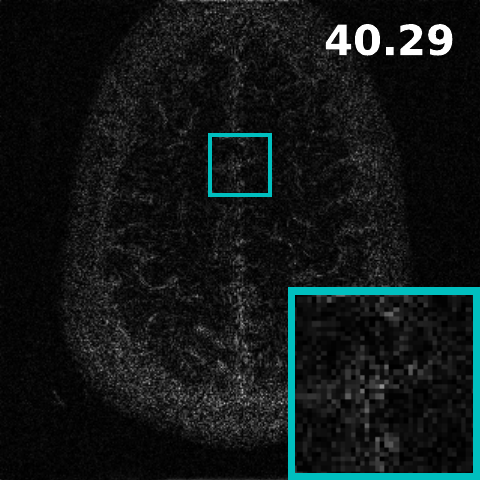} &  \includegraphics[width=0.166\linewidth]{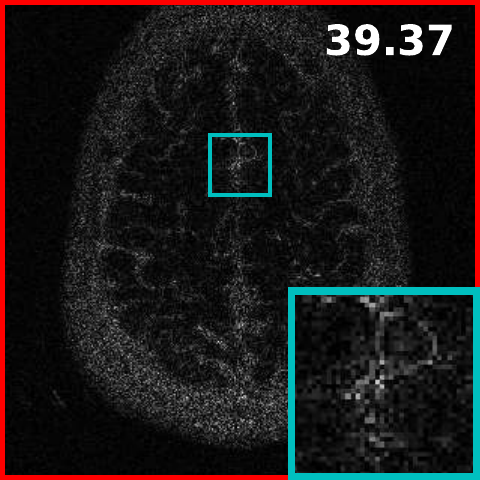} \\[-\dp\strutbox]
\end{tabularx}
\caption{Comparison of various methods.}
\end{subfigure}
\par\bigskip
\begin{subfigure}{0.95\textwidth}
\centering
\begin{tabularx}{0.833\linewidth}{ccccc}
    \includegraphics[width=0.166\linewidth]{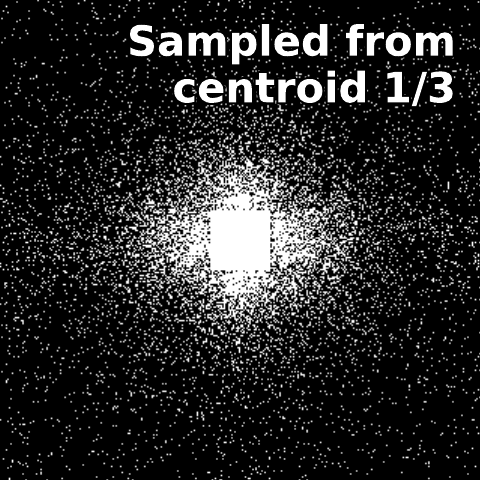} & \includegraphics[width=0.166\linewidth]{figs/mr_2d8x/ra1.pdf} & \includegraphics[width=0.166\linewidth]{figs/mr_2d8x/ea1.pdf} & \includegraphics[width=0.166\linewidth]{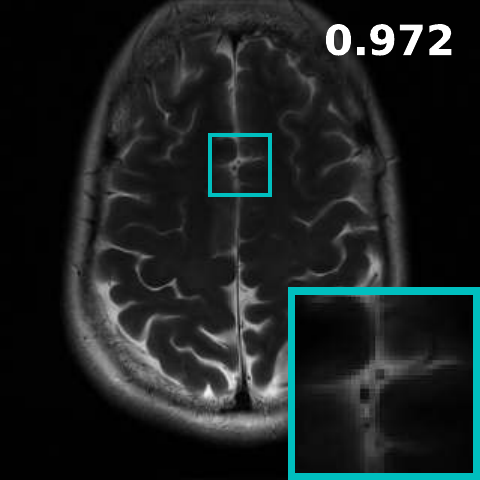} & \includegraphics[width=0.166\linewidth]{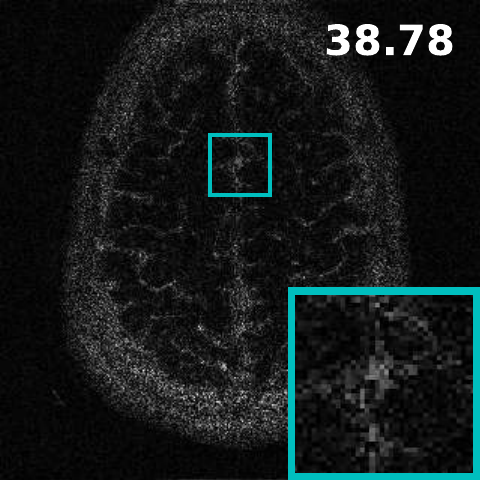} \\[-\dp\strutbox]
    \includegraphics[width=0.166\linewidth]{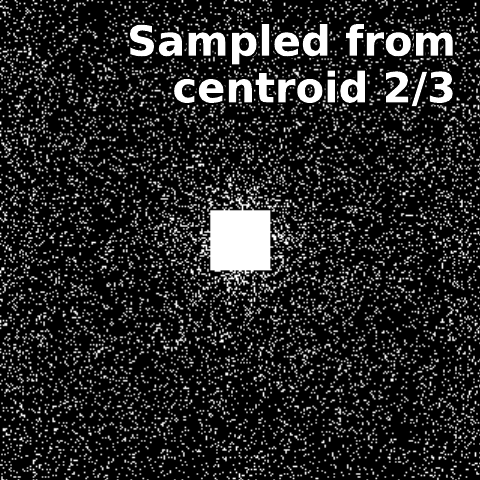} & \includegraphics[width=0.166\linewidth]{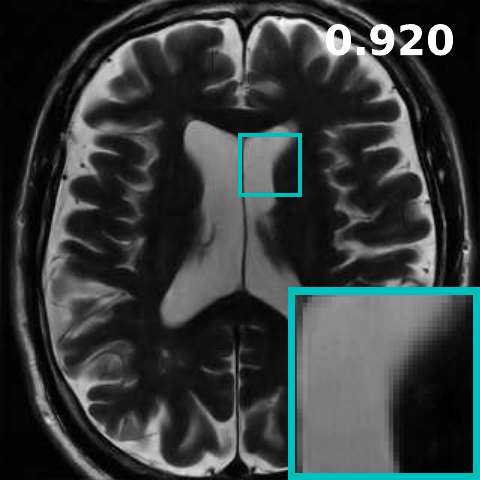} & \includegraphics[width=0.166\linewidth]{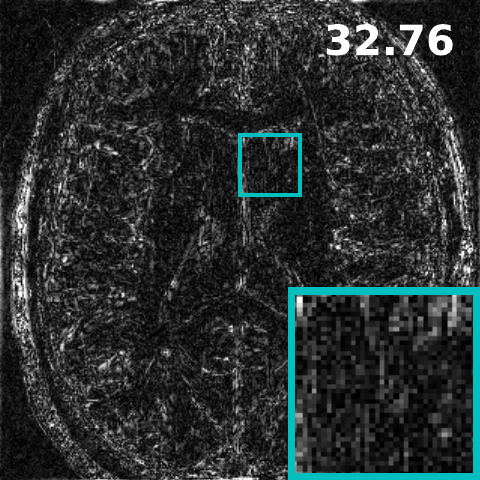} & \includegraphics[width=0.166\linewidth]{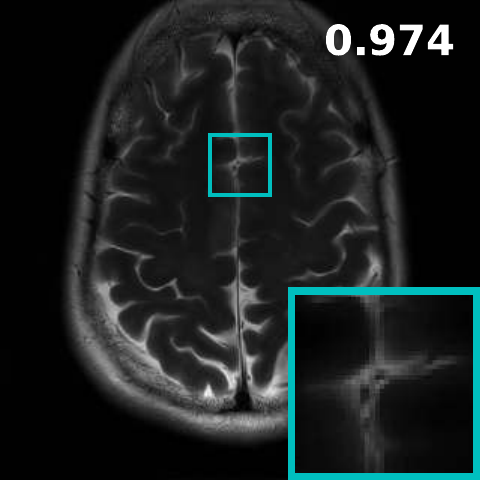} & \includegraphics[width=0.166\linewidth]{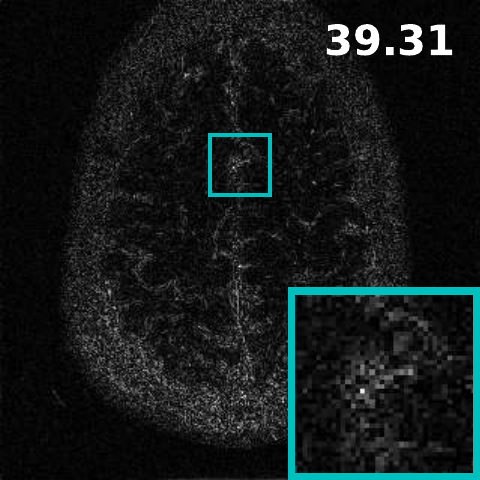}  \\[-\dp\strutbox]
    \includegraphics[width=0.166\linewidth]{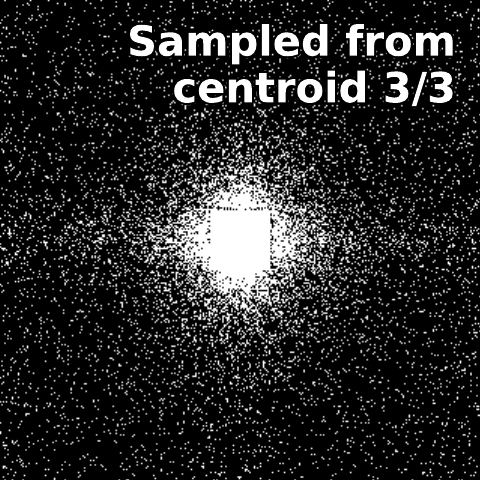} & \includegraphics[width=0.166\linewidth]{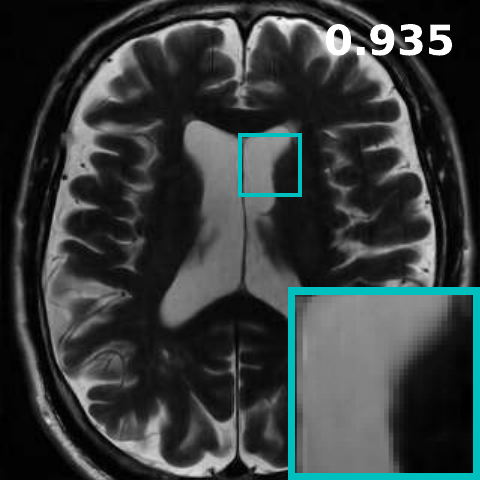} & \includegraphics[width=0.166\linewidth]{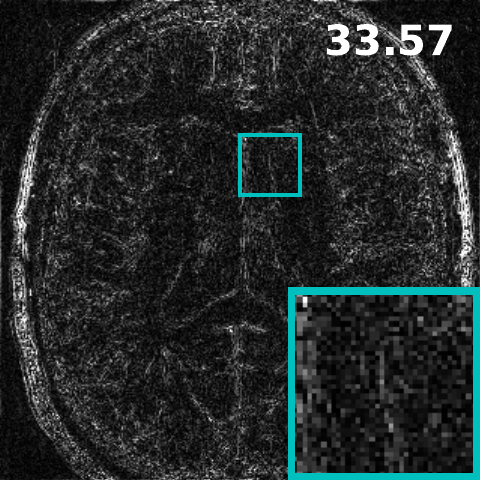} & \includegraphics[width=0.166\linewidth]{figs/mr_2d8x/rb3.pdf} & \includegraphics[width=0.166\linewidth]{figs/mr_2d8x/eb3.pdf}  \\[-\dp\strutbox]
\end{tabularx}
\caption{Comparison of the sampling-reconstruction pairs $(M_j,\theta_j)_{j=1}^3$ in our method $\mathcal{H}_{1.5}^J$.}
\end{subfigure}
  \caption{ Qualitative comparison of reconstruction and error map (a) of various methods, (b) obtained using the mask-reconstruction pairs ($(M_j, \theta_j)_{j=1}^3$) generated from Algorithm 1, at acceleration rate $8\times$ in the fastMRI dataset~\cite{zbontar2018fastmri}. 
  For comparison, we also show the results of the variable density (VD) \cite{wang2009variable}, LOUPE~\cite{bahadir2020deep}, and policy-based adaptive sampling~\cite{bakker2022learning} in (a).  SSIMs and PSNRs are included in the reconstructions and the error maps, respectively. 
  The images highlighted in \textcolor{red}{red} in (b) demonstrate that our Algorithm 2 estimated the HF Bayesian uncertainty of each image and selected an appropriate sampling-reconstruction pair $(M_j,\theta_j)$ using the uncertainty. The images thus selected become the final output of our method $\mathcal{H}_{1.5}^J$, as emphasized in \textcolor{red}{red} in (a).
}
  \label{sfig:5.3}

\end{figure*}

\begin{figure*}[h!]
\scriptsize
\centering
    \setlength{\tabcolsep}{0pt}
\begin{subfigure}{0.95\textwidth}
\begin{tabularx}{\linewidth}{cccccc}
Ground truth & $\mathcal{H}_1$: Random & $\mathcal{H}_1$: VD~\cite{wang2009variable} & $\mathcal{H}_1$:LOUPE\!\cite{bahadir2020deep} & $\mathcal{H}_2$: Policy~\cite{bakker2022learning} & $\mathcal{H}_{1.5}^J$ (ours) \\
\includegraphics[width=0.166\linewidth]{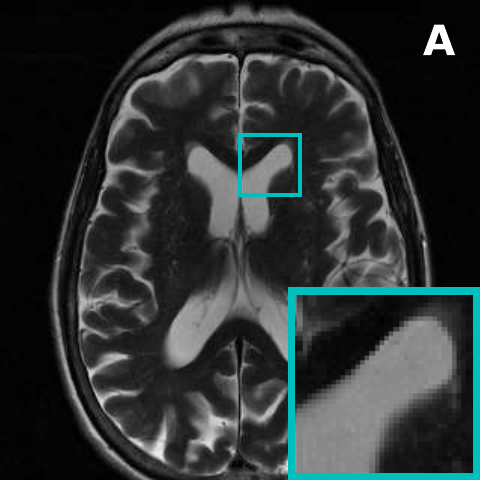} & \includegraphics[width=0.166\linewidth]{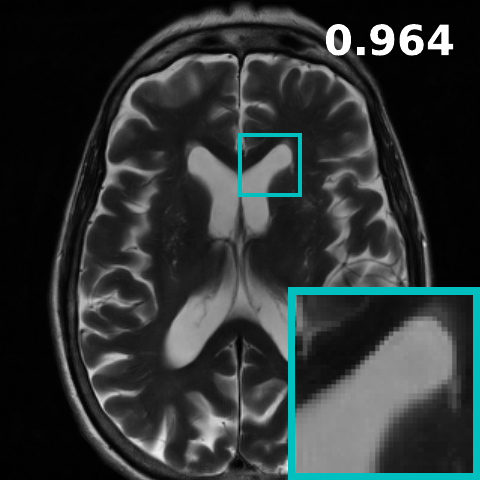} & \includegraphics[width=0.166\linewidth]{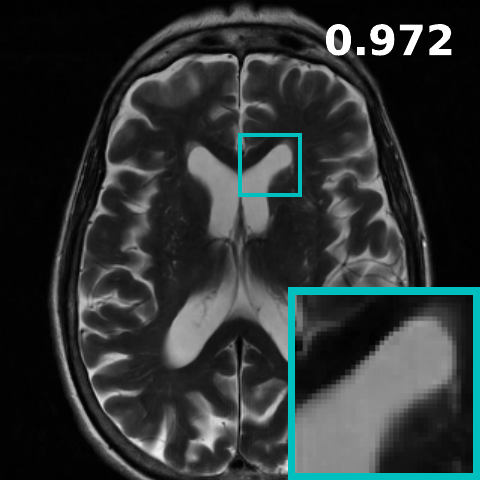} & \includegraphics[width=0.166\linewidth]{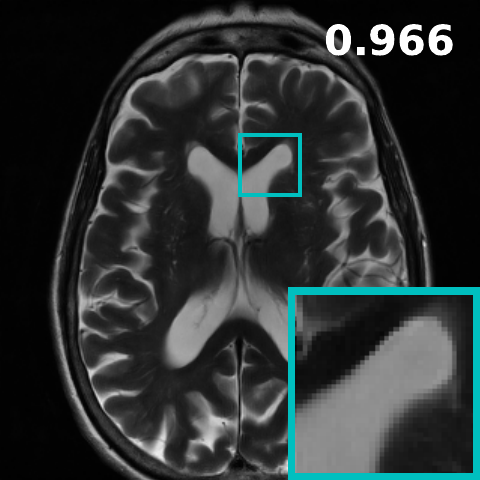} &  \includegraphics[width=0.166\linewidth]{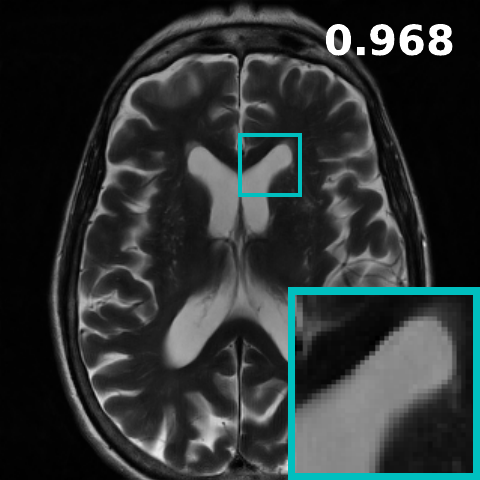} &  \includegraphics[width=0.166\linewidth]{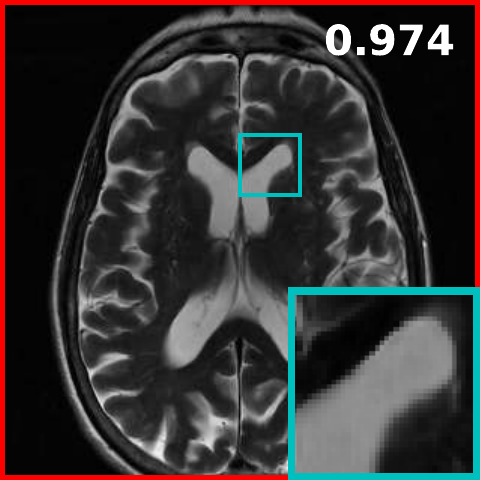}\\[-\dp\strutbox]
\includegraphics[width=0.166\linewidth]{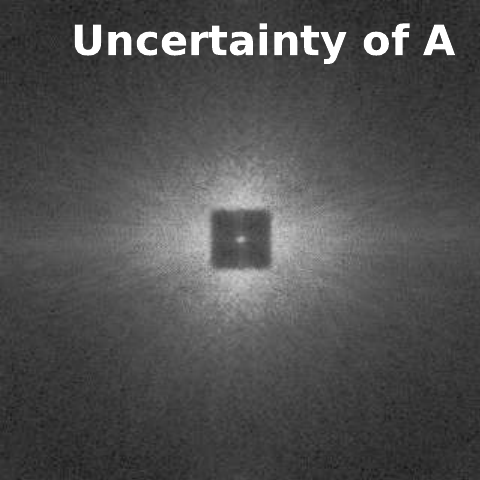} & \includegraphics[width=0.166\linewidth]{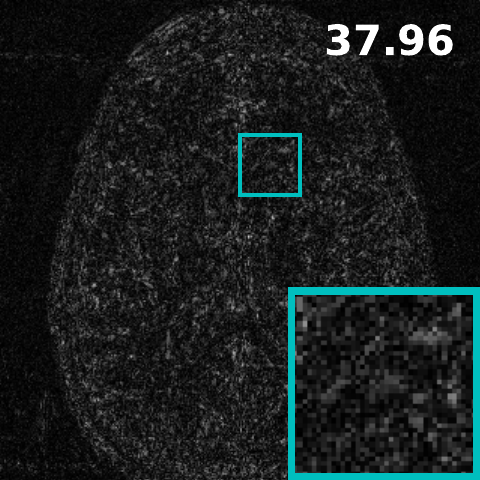} & \includegraphics[width=0.166\linewidth]{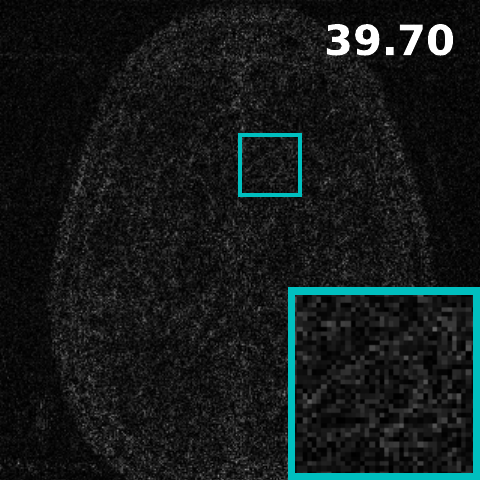} & \includegraphics[width=0.166\linewidth]{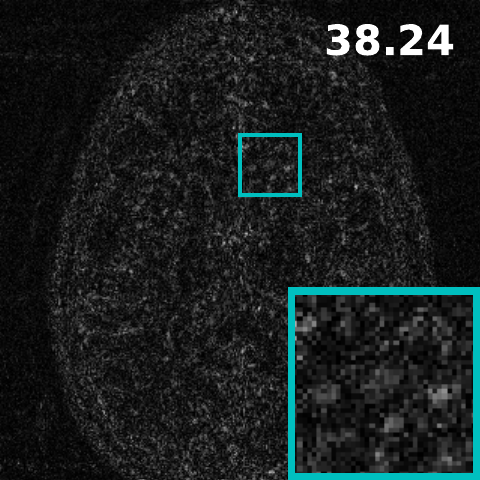} & \includegraphics[width=0.166\linewidth]{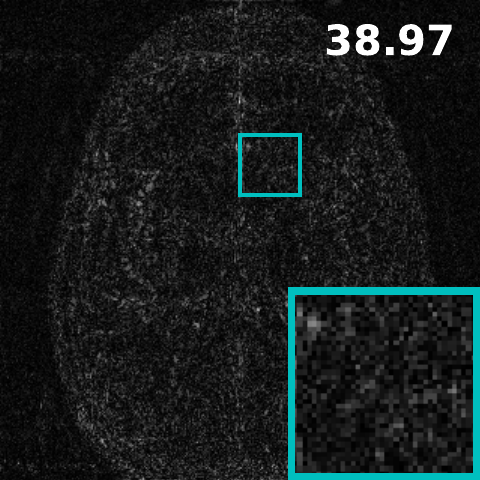} &  \includegraphics[width=0.166\linewidth]{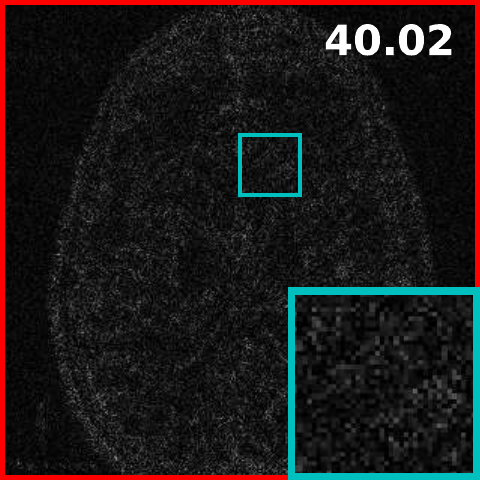}\\[-\dp\strutbox]
\includegraphics[width=0.166\linewidth]{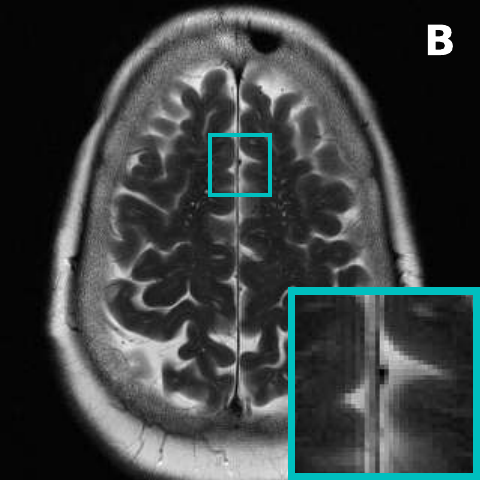} & \includegraphics[width=0.166\linewidth]{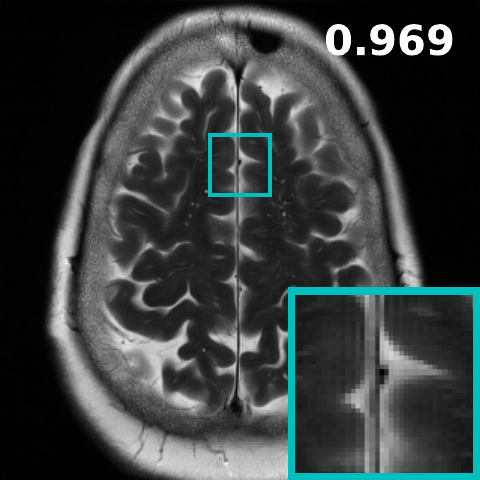} & \includegraphics[width=0.166\linewidth]{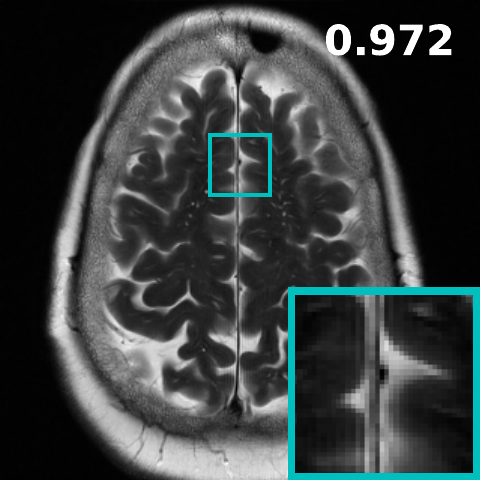} & \includegraphics[width=0.166\linewidth]{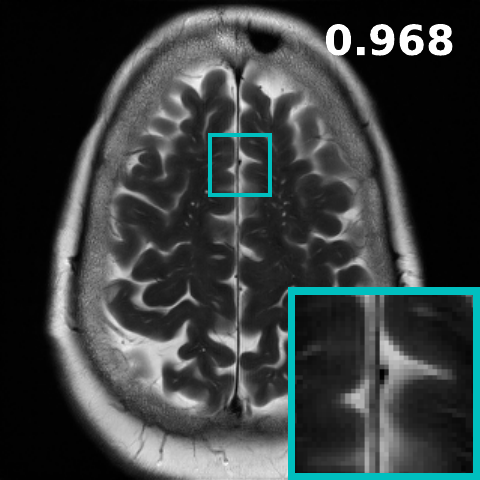} & \includegraphics[width=0.166\linewidth]{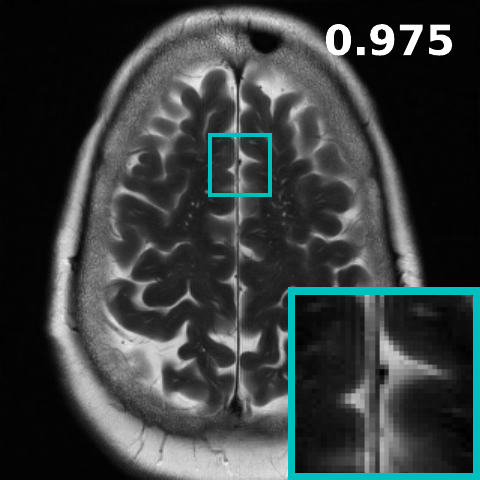} &  \includegraphics[width=0.166\linewidth]{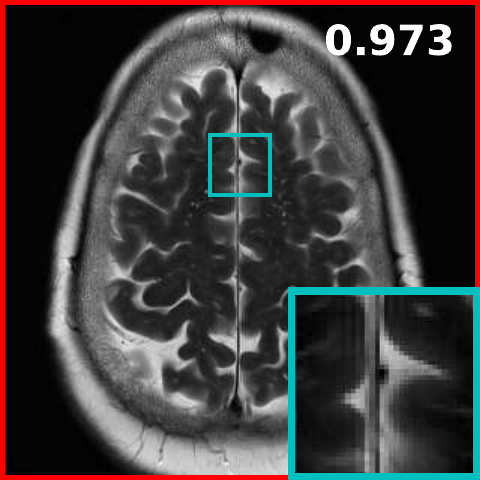} \\[-\dp\strutbox]
\includegraphics[width=0.166\linewidth]{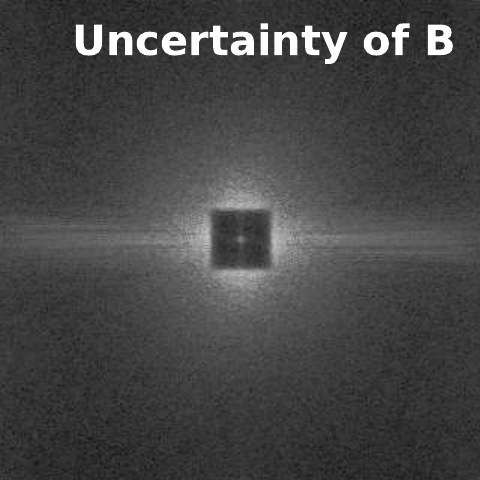} & \includegraphics[width=0.166\linewidth]{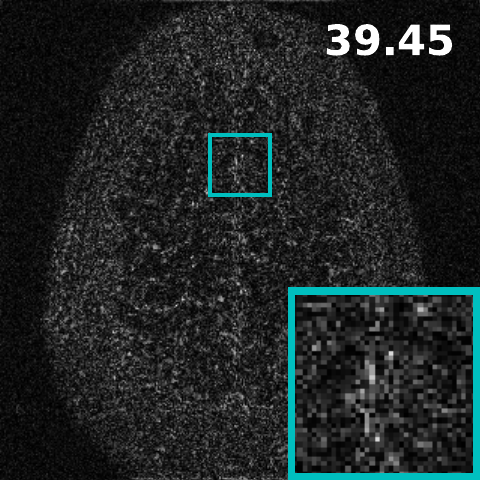} & \includegraphics[width=0.166\linewidth]{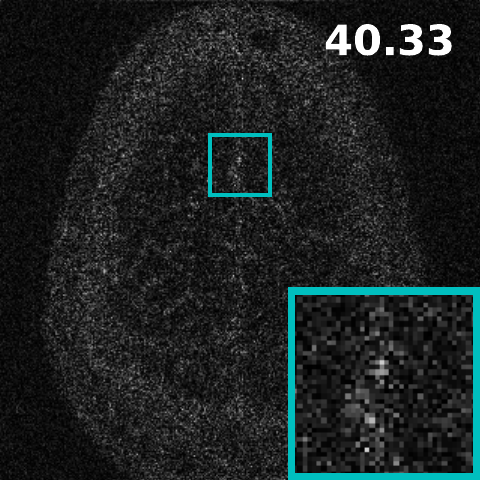} & \includegraphics[width=0.166\linewidth]{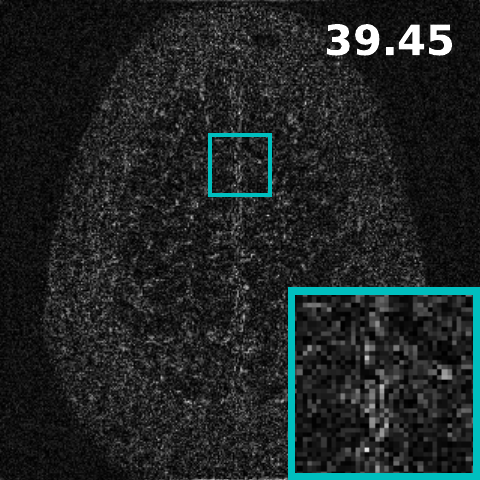} & \includegraphics[width=0.166\linewidth]{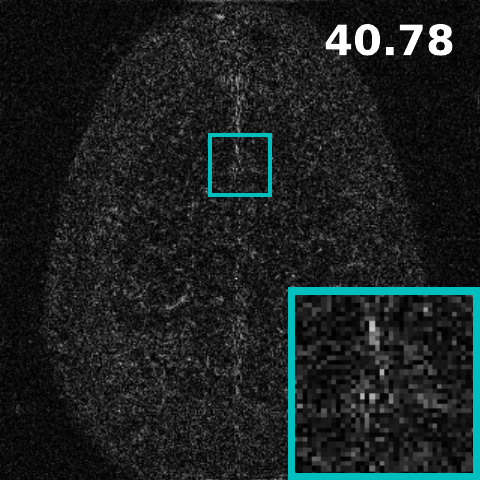} &  \includegraphics[width=0.166\linewidth]{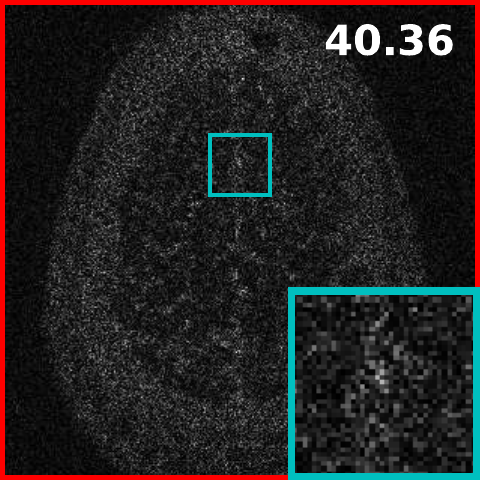} \\[-\dp\strutbox]
\end{tabularx}
\caption{Comparison of various methods.}
\end{subfigure}
\par\bigskip
\begin{subfigure}{0.95\textwidth}
\centering
\begin{tabularx}{0.833\linewidth}{ccccc}
    \includegraphics[width=0.166\linewidth]{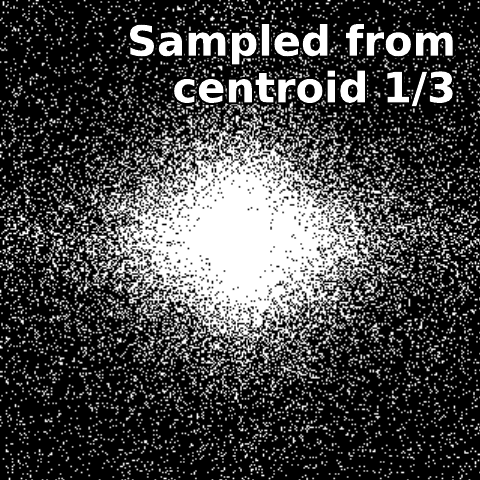} & \includegraphics[width=0.166\linewidth]{figs/mr_2d4x/ra1.pdf} & \includegraphics[width=0.166\linewidth]{figs/mr_2d4x/ea1.pdf} & \includegraphics[width=0.166\linewidth]{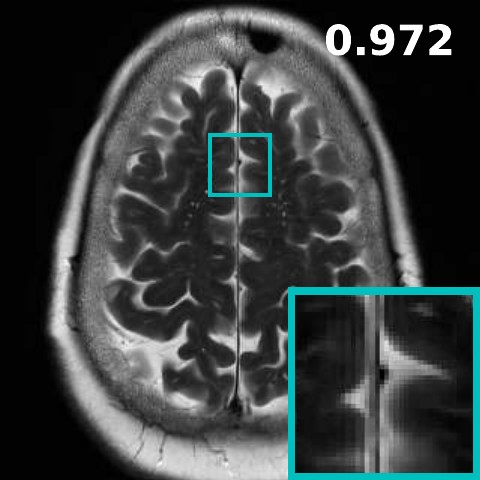} & \includegraphics[width=0.166\linewidth]{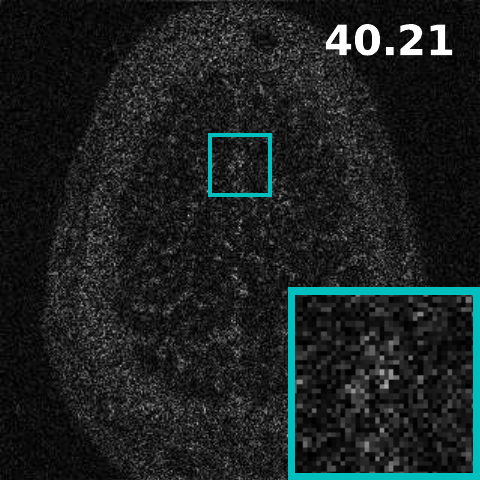} \\[-\dp\strutbox]
    \includegraphics[width=0.166\linewidth]{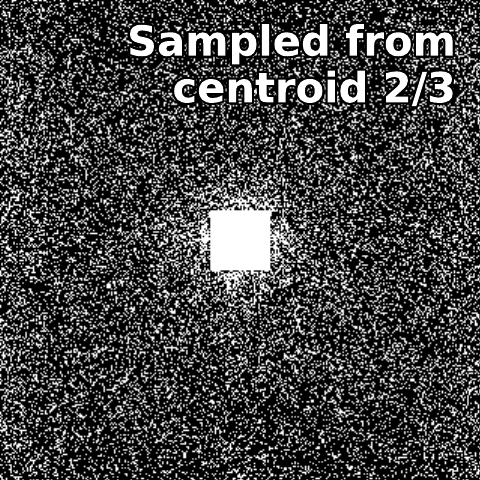} & \includegraphics[width=0.166\linewidth]{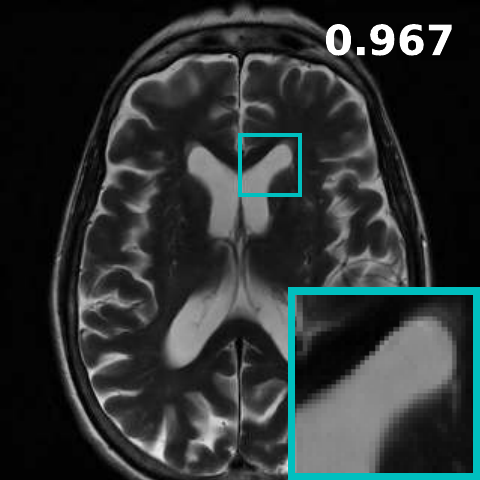} & \includegraphics[width=0.166\linewidth]{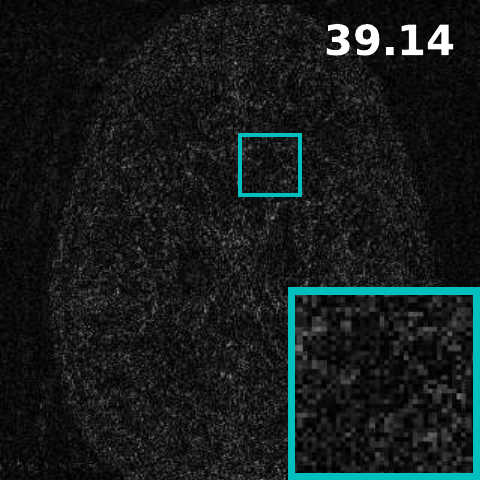} & \includegraphics[width=0.166\linewidth]{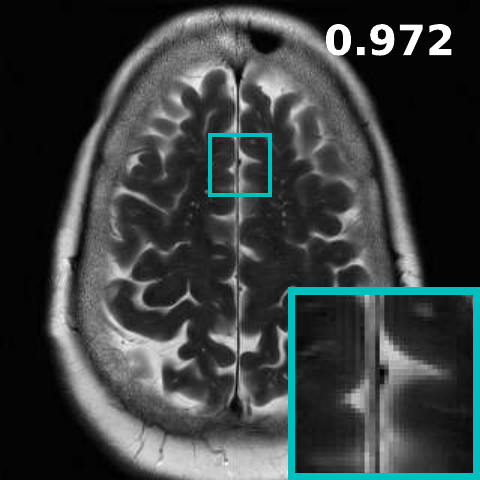} & \includegraphics[width=0.166\linewidth]{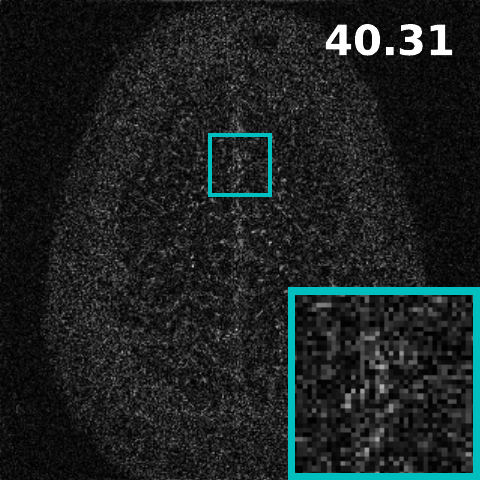}  \\[-\dp\strutbox]
    \includegraphics[width=0.166\linewidth]{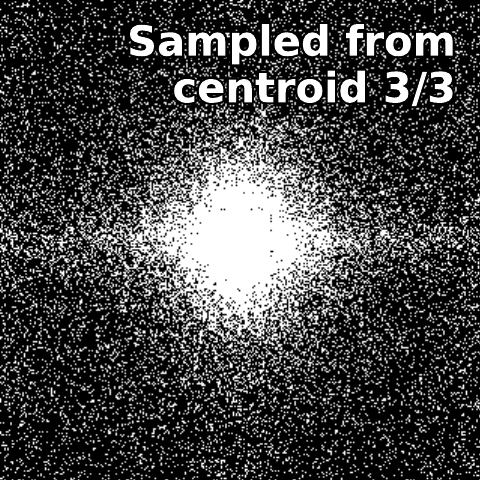} & \includegraphics[width=0.166\linewidth]{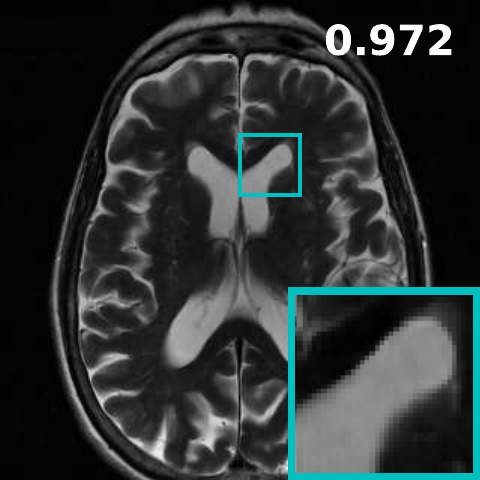} & \includegraphics[width=0.166\linewidth]{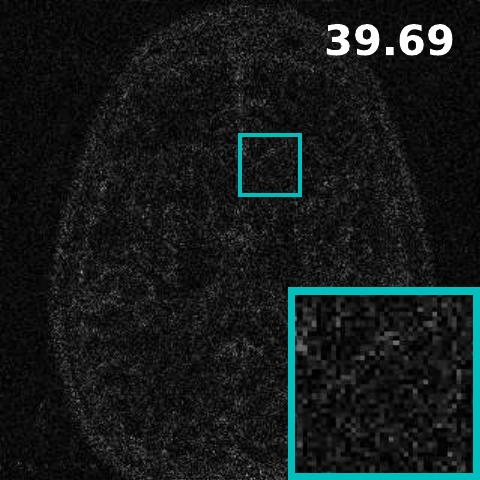} & \includegraphics[width=0.166\linewidth]{figs/mr_2d4x/rb3.pdf} & \includegraphics[width=0.166\linewidth]{figs/mr_2d4x/eb3.pdf}  \\[-\dp\strutbox]
\end{tabularx}
\caption{Comparison of the sampling-reconstruction pairs $(M_j,\theta_j)_{j=1}^3$ in our method $\mathcal{H}_{1.5}^J$.}
\end{subfigure}
  \caption{ Qualitative comparison of reconstruction and error map (a) of various methods, (b) obtained using the mask-reconstruction pairs ($(M_j, \theta_j)_{j=1}^3$) generated from Algorithm 1, at acceleration rate $4\times$ in the fastMRI dataset~\cite{zbontar2018fastmri}. 
  For comparison, we also show the results of the variable density (VD) \cite{wang2009variable}, LOUPE~\cite{bahadir2020deep}, and policy-based adaptive sampling~\cite{bakker2022learning} in (a).  SSIMs and PSNRs are included in the reconstructions and the error maps, respectively. 
  The images highlighted in \textcolor{red}{red} in (b) demonstrate that our Algorithm 2 estimated the HF Bayesian uncertainty of each image and selected an appropriate sampling-reconstruction pair $(M_j,\theta_j)$ using the uncertainty. The images thus selected become the final output of our method $\mathcal{H}_{1.5}^J$, as emphasized in \textcolor{red}{red} in (a).
}
  \label{sfig:5.4}

\end{figure*}

\paragraph{On other datasets}
In the main paper, we have shown qualitative results on 16$\times$ CelebA, and 4$\times$ fastMRI with 1D sampling. In Figs. \ref{sfig:5.1}, \ref{sfig:5.2}, \ref{sfig:5.3}, and \ref{sfig:5.4}, we provide qualitative results on other settings. The figures demonstrate that our method achieves optimal reconstruction performance by adaptively selecting an appropriate mask. While $\mathcal{H}_2$ occasionally shows a slight advantage in 2D CS-MRI, the overall superiority of our method $\mathcal{H}_{1.5}^J$ is evident, as confirmed by the results in Tab. 2 of the main text.

\begin{figure*}[h!]
  \centering
  \resizebox{0.6\linewidth}{!}{
    \setlength{\tabcolsep}{0pt}
    \begin{tabular}{cccc}
  \includegraphics[width=0.24\linewidth]{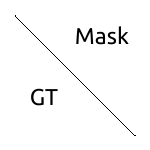}
  & \includegraphics[width=0.24\linewidth]{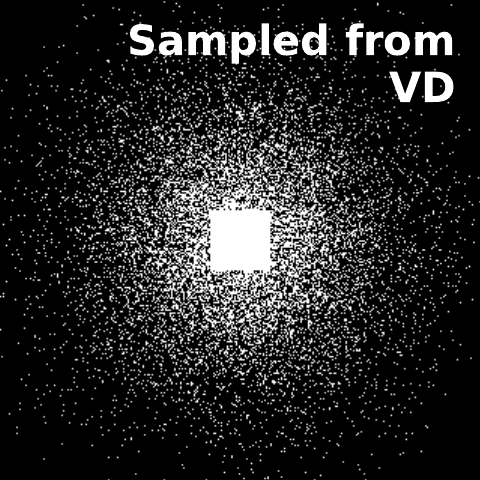} & \includegraphics[width=0.24\linewidth]{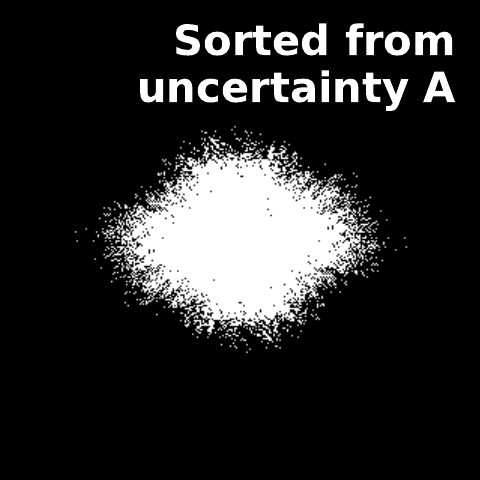} & \includegraphics[width=0.24\linewidth]{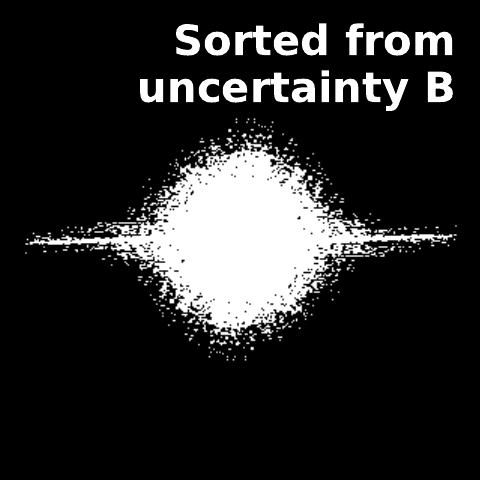} \\[-\dp\strutbox]
\includegraphics[width=0.24\linewidth]{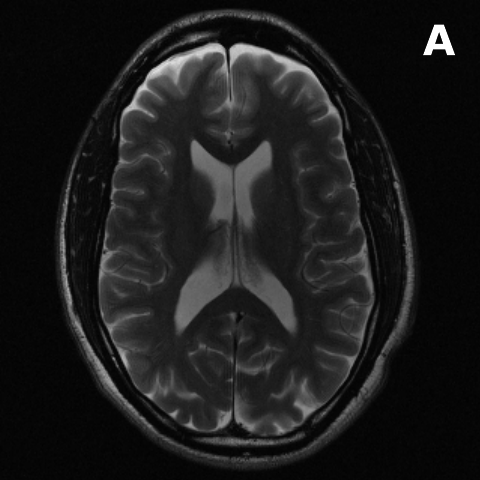} & \includegraphics[width=0.24\linewidth]{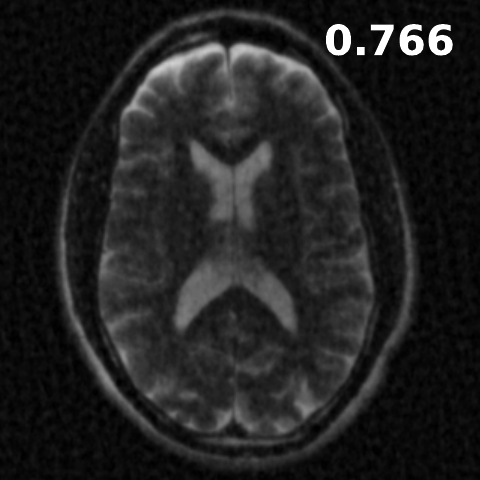} & \includegraphics[width=0.24\linewidth]{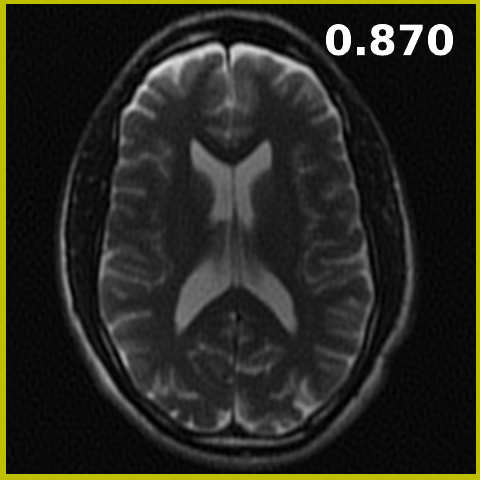} & \includegraphics[width=0.24\linewidth]{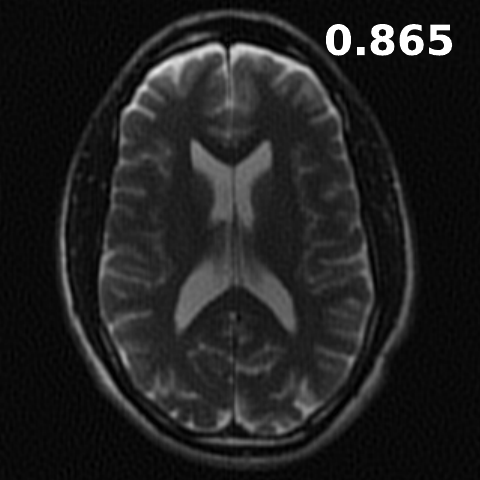} \\[-\dp\strutbox]
\includegraphics[width=0.24\linewidth]{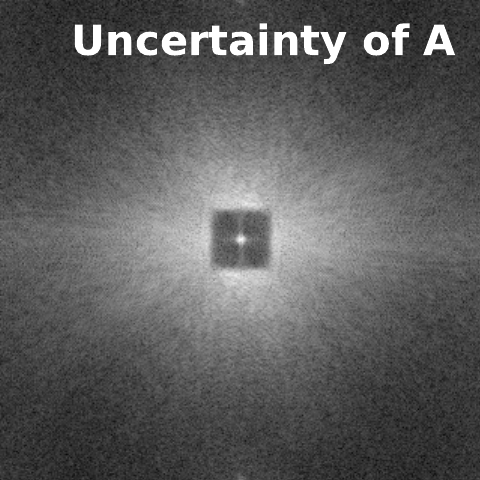} & \includegraphics[width=0.24\linewidth]{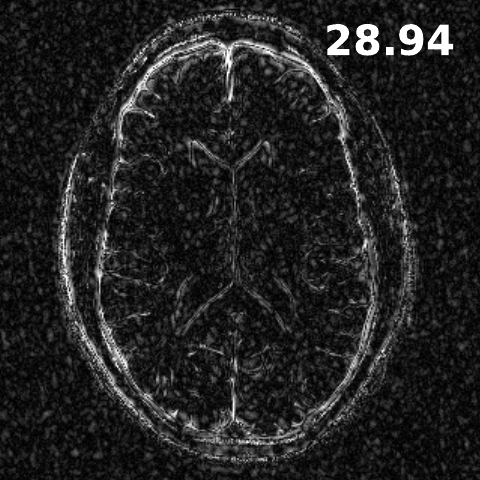} & \includegraphics[width=0.24\linewidth]{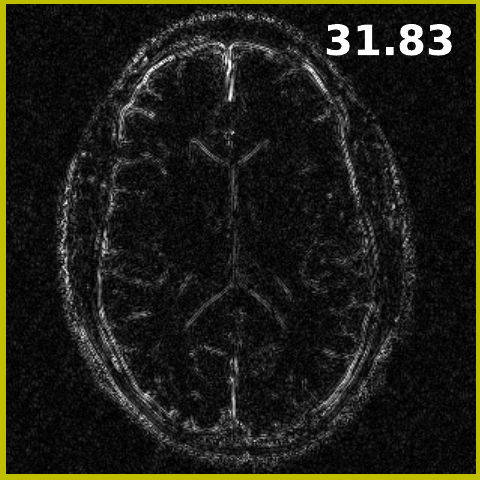} & \includegraphics[width=0.24\linewidth]{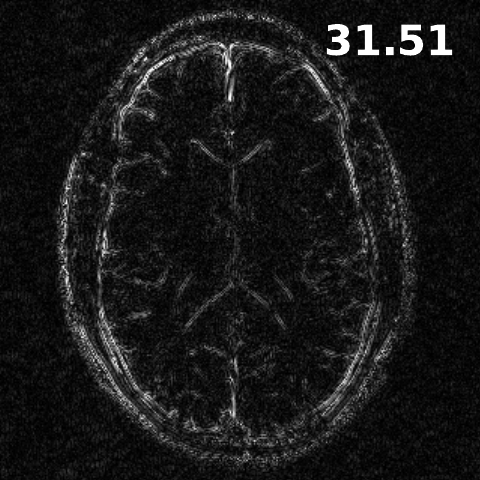} \\[-\dp\strutbox]
\includegraphics[width=0.24\linewidth]{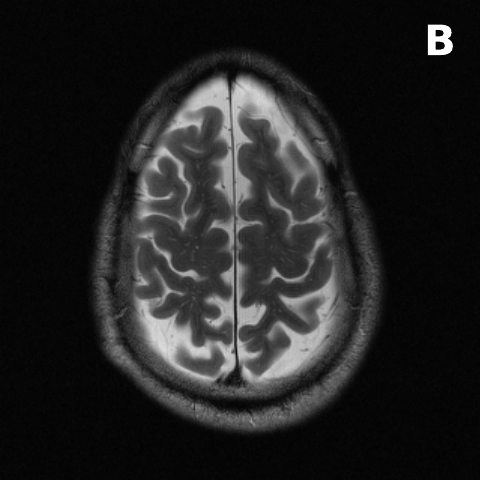} & \includegraphics[width=0.24\linewidth]{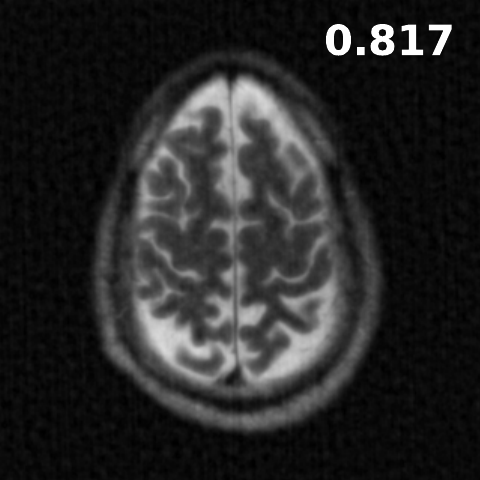} & \includegraphics[width=0.24\linewidth]{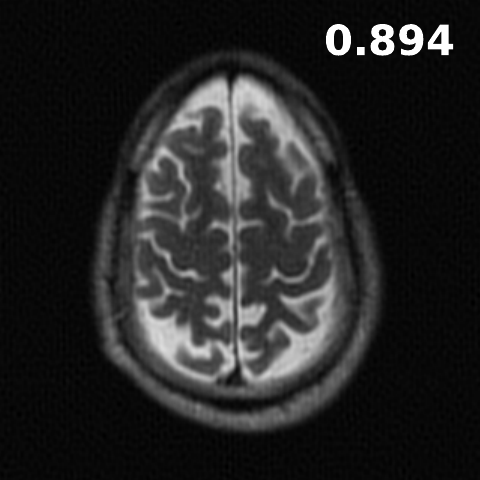} & \includegraphics[width=0.24\linewidth]{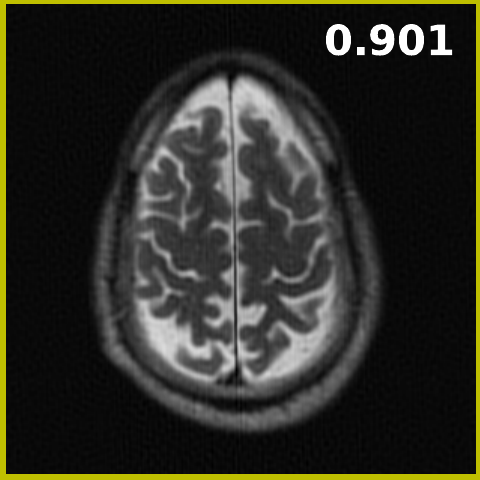} \\[-\dp\strutbox]
\includegraphics[width=0.24\linewidth]{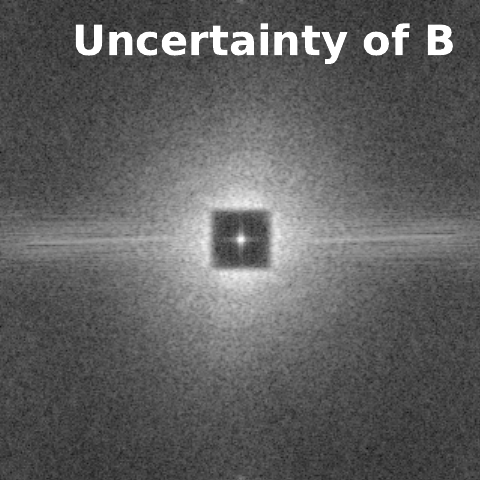} & \includegraphics[width=0.24\linewidth]{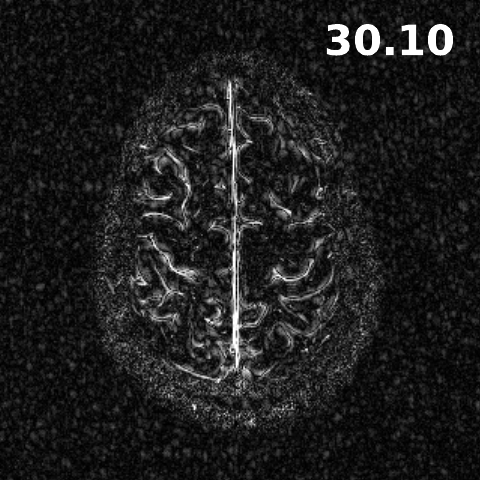} & \includegraphics[width=0.24\linewidth]{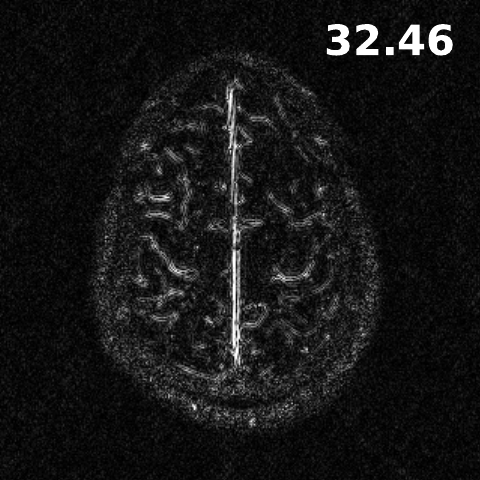} & \includegraphics[width=0.24\linewidth]{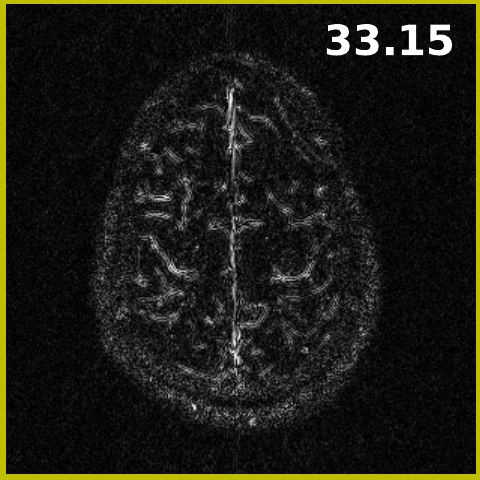} \\[-\dp\strutbox]
\end{tabular}
  }
  \caption{The SR space generation model~\cite{Song_2022_CVPR} quantifies the HF uncertainty well, and this is evidenced by the superior performance of the sampling mask obtained through the sorting of individual high-frequency (HF) uncertainties. We created a sampling mask with an acceleration rate $8\times$ by sorting the sample variance (i.e., HF Bayesian uncertainty) obtained from SR space generation on two different slices (A, B). Here, we show the results of reconstructing each of the two slices using their respective masks (and also VD~\cite{wang2009variable}) using the zero-filling reconstruction. SSIMs and PSNRs are inset in the reconstructions and the error maps, respectively. %
  }
  \label{sfig:5.5}
\end{figure*}

\paragraph{Does the SR space generation model quantify the HF uncertainty well?}
In \cref{sfig:5.5}, we analyze the HF Bayesian uncertainty quantification for two data points and sort them to generate masks. %
We additionally generate mask sampled from VD~\cite{wang2009variable} for comparison. We then compare the classical reconstructed results from the generated masks at acceleration rate $8\times$ with the metrics of SSIM and PSNR. The yellow boxes in \cref{sfig:5.5} highlight that sorting the self-variance results in the highest PSNR and SSIM scores. In particular, the error map reconstructed by sorting B's own uncertainty exhibits significantly smaller errors in distinguishing the left and right hemispheres (i.e., the longitudinal fissure). Note that this is a qualitative result example of the first paragraph of the disccusion section in the main paper.

\paragraph{Robustness is improved in $J>2$}
In \cref{tab:r1}, we provide more results that $\overline{\scriptstyle{\text{SSIM}}}^{\text{ low}}_{5\%}$ and $\overline{\scriptstyle{\text{SSIM}}}^{\text{ low}}_{10\%}$ were significantly improved in $J>2$, \ie, stronger robustness than the others.

\begin{table}[]
\small
\setlength{\tabcolsep}{8pt}
\caption{The average of the lowest 5\% and 10\% of SSIM values}\label{tab:r1}
\centering
\begin{tabular}{ll|cc|cc}
 & & \multicolumn{2}{c|}{CS-MRI 1D 8$\times$} & \multicolumn{2}{c}{CS-MRI 2D 8$\times$}\\
Method &  & $\overline{\scriptstyle{\text{SSIM}}}^{\text{ low}}_{5\%}$ & $\overline{\scriptstyle{\text{SSIM}}}^{\text{ low}}_{10\%}$ & $\overline{\scriptstyle{\text{SSIM}}}^{\text{ low}}_{5\%}$ & $\overline{\scriptstyle{\text{SSIM}}}^{\text{ low}}_{10\%}$ \\ \hline
\multirow{3}{*}{$\mathcal{H}_1$} & Random & 0.8636 & 0.8770 & 0.8941 & 0.9108 \\
& VD & 0.8791 & 0.8936 & \cellcolor{tabthird} 0.9111 & 0.9266\\
& LOUPE & 0.8574 & 0.8699 & 0.8971 & 0.9140\\ \hline
\multirow{1}{*}{$\mathcal{H}_2$} & Policy & 0.8636 & 0.8755 & 0.8999 & 0.9183\\
\hline
\multirow{4}{*}{$\mathcal{H}_{1.5}^J$(ours)}& 
$J=1$  & 0.8806 & 0.8968 & 0.9089 & 0.9249 \\
& $J=2$ & \cellcolor{tabthird} 0.8810 & \cellcolor{tabthird} 0.8990 & \cellcolor{tabsecond}0.9112 & \cellcolor{tabsecond}0.9269\\
& $J=3$ & \cellcolor{tabsecond}0.8839 & \cellcolor{tabsecond}0.8992 & 0.9109 & \cellcolor{tabthird}0.9268 \\
& $J=4$ & \cellcolor{tabfirst}0.8845 & \cellcolor{tabfirst}0.8999 & \cellcolor{tabfirst} 0.9116 & \cellcolor{tabfirst} 0.9273 \\
\end{tabular}
\end{table}

\bibliographystyle{splncs04}
\bibliography{refs}

\begin{thebibliography}{10}
\providecommand{\url}[1]{\texttt{#1}}
\providecommand{\urlprefix}{URL }
\providecommand{\doi}[1]{https://doi.org/#1}

\bibitem{abu2022image}
Abu-Hussein, S., Tirer, T., Chun, S.Y., Eldar, Y.C., Giryes, R.: Image restoration by deep projected {GSURE}. In: WACV. pp. 3602--3611 (2022)

\bibitem{arthur2007k}
Arthur, D., Vassilvitskii, S.: K-means++ the advantages of careful seeding. In: ACM-SIAM symposium on Discrete algorithms. pp. 1027--1035 (2007)

\bibitem{bahadir2020deep}
Bahadir, C.D., Wang, A.Q., Dalca, A.V., Sabuncu, M.R.: Deep-learning-based optimization of the under-sampling pattern in {MRI}. IEEE Transactions on Computational Imaging  \textbf{6},  1139--1152 (2020)

\bibitem{bakker2020experimental}
Bakker, T., van Hoof, H., Welling, M.: Experimental design for {MRI} by greedy policy search. NeurIPS  \textbf{33},  18954--18966 (2020)

\bibitem{bakker2022learning}
Bakker, T., Muckley, M., Romero-Soriano, A., Drozdzal, M., Pineda, L.: On learning adaptive acquisition policies for undersampled multi-coil {MRI} reconstruction. In: MIDL. pp. 63--85 (2022)

\bibitem{bengio2013estimating}
Bengio, Y., L{\'e}onard, N., Courville, A.: Estimating or propagating gradients through stochastic neurons for conditional computation. arXiv:1308.3432  (2013)

\bibitem{candes2006cs}
Candes, E., Romberg, J., Tao, T.: Robust uncertainty principles: exact signal reconstruction from highly incomplete frequency information. IEEE Transactions on Information Theory  \textbf{52}(2),  489--509 (2006)

\bibitem{chaithya2021learning}
Chaithya, G., Ramzi, Z., Ciuciu, P.: Learning the sampling density in {2D SPARKLING MRI} acquisition for optimized image reconstruction. In: EUSIPCO. pp. 960--964 (2021)

\bibitem{chaithya2022optimizing}
Chaithya, G., Weiss, P., Daval-Fr{\'e}rot, G., Massire, A., Vignaud, A., Ciuciu, P.: Optimizing full {3D} sparkling trajectories for high-resolution magnetic resonance imaging. IEEE Transactions on Medical Imaging  \textbf{41}(8),  2105--2117 (2022)

\bibitem{de2019compressed}
De~Maio, A., Eldar, Y.C., Haimovich, A.M.: Compressed sensing in radar signal processing. Cambridge University Press (2019)

\bibitem{deshpande2017learning}
Deshpande, A., Lu, J., Yeh, M.C., Jin~Chong, M., Forsyth, D.: Learning diverse image colorization. In: CVPR. pp. 6837--6845 (2017)

\bibitem{DBLP:conf/iclr/DinhSB17}
Dinh, L., Sohl{-}Dickstein, J., Bengio, S.: Density estimation using real {NVP}. In: {ICLR} (Poster) (2017)

\bibitem{donoho2006compressed}
Donoho, D.L.: Compressed sensing. IEEE Transactions on information theory  \textbf{52}(4),  1289--1306 (2006)

\bibitem{hong2023advanced}
Hong, S., Kim, S.C., Lee, S.: Advanced direction of arrival estimation using step-learnt iterative soft-thresholding for frequency-modulated continuous wave multiple-input multiple-output radar. IET Radar, Sonar \& Navigation  \textbf{17}(1),  2--14 (2023)

\bibitem{Hong_2023_ICCV}
Hong, S., Park, I., Chun, S.Y.: On the robustness of normalizing flows for inverse problems in imaging. In: ICCV. pp. 10745--10755 (October 2023)

\bibitem{kar2021fast}
Kar, A., Biswas, P.K.: Fast {Bayesian} uncertainty estimation and reduction of batch normalized single image super-resolution network. In: CVPR. pp. 4957--4966 (2021)

\bibitem{kingma2014adam}
Kingma, D.P., Ba, J.: Adam: A method for stochastic optimization. In: ICLR (2015)

\bibitem{DBLP:journals/corr/KingmaW13}
Kingma, D.P., Welling, M.: Auto-encoding variational bayes. In: {ICLR} (2014)

\bibitem{lazarus2019sparkling}
Lazarus, C., Weiss, P., Chauffert, N., Mauconduit, F., El~Gueddari, L., Destrieux, C., Zemmoura, I., Vignaud, A., Ciuciu, P.: {SPARKLING}: variable-density k-space filling curves for accelerated t2*-weighted {MRI}. Magnetic resonance in medicine  \textbf{81}(6),  3643--3661 (2019)

\bibitem{li2022srdiff}
Li, H., Yang, Y., Chang, M., Chen, S., Feng, H., Xu, Z., Li, Q., Chen, Y.: Srdiff: Single image super-resolution with diffusion probabilistic models. Neurocomputing  \textbf{479},  47--59 (2022)

\bibitem{liu2015deep}
Liu, Z., Luo, P., Wang, X., Tang, X.: Deep learning face attributes in the wild. In: ICCV. pp. 3730--3738 (2015)

\bibitem{lugmayr2020srflow}
Lugmayr, A., Danelljan, M., Gool, L.V., Timofte, R.: Srflow: Learning the super-resolution space with normalizing flow. In: ECCV. pp. 715--732 (2020)

\bibitem{lugmayr2021ntire}
Lugmayr, A., Danelljan, M., Timofte, R.: {NTIRE} 2021 learning the super-resolution space challenge. In: CVPRW. pp. 596--612 (2021)

\bibitem{lugmayr2022ntire}
Lugmayr, A., Danelljan, M., Timofte, R., Kim, K.w., Kim, Y., Lee, J.y., Li, Z., Pan, J., Shim, D., Song, K.U., et~al.: {NTIRE} 2022 challenge on learning the super-resolution space. In: CVPRW. pp. 786--797 (2022)

\bibitem{lustig2007sparse}
Lustig, M., Donoho, D., Pauly, J.M.: Sparse {MRI}: The application of compressed sensing for rapid {MR} imaging. Magnetic Resonance in Medicine  \textbf{58}(6),  1182--1195 (2007)

\bibitem{lustig2008compressed}
Lustig, M., Donoho, D.L., Santos, J.M., Pauly, J.M.: Compressed sensing {MRI}. IEEE signal processing magazine  \textbf{25}(2),  72--82 (2008)

\bibitem{pineda2020active}
Pineda, L., Basu, S., Romero, A., Calandra, R., Drozdzal, M.: Active {MR} k-space sampling with reinforcement learning. In: MICCAI. Springer (2020)

\bibitem{ronneberger2015u}
Ronneberger, O., Fischer, P., Brox, T.: U-net: Convolutional networks for biomedical image segmentation. In: MICCAI. pp. 234--241 (2015)

\bibitem{sanchez2022learning}
Sanchez, T.: Learning to sample in Cartesian {MRI}. Phd thesis, EPFL, Lausanne (June 2022)

\bibitem{Song_2022_CVPR}
Song, K.U., Shim, D., Kim, K.w., Lee, J.y., Kim, Y.: {FS-NCSR}: Increasing diversity of the super-resolution space via frequency separation and noise-conditioned normalizing flow. In: CVPRW. pp. 968--977 (June 2022)

\bibitem{sriram2020end}
Sriram, A., Zbontar, J., Murrell, T., Defazio, A., Zitnick, C.L., Yakubova, N., Knoll, F., Johnson, P.: End-to-end variational networks for accelerated {MRI} reconstruction. In: MICCAI (2020)

\bibitem{van2021active}
Van~Gorp, H., Huijben, I., Veeling, B.S., Pezzotti, N., Van~Sloun, R.J.: Active deep probabilistic subsampling. In: ICML. pp. 10509--10518 (2021)

\bibitem{wang2022b}
Wang, G., Luo, T., Nielsen, J.F., Noll, D.C., Fessler, J.A.: B-spline parameterized joint optimization of reconstruction and k-space trajectories ({Bjork}) for accelerated 2d {MRI}. IEEE Transactions on Medical Imaging  \textbf{41}(9),  2318--2330 (2022)

\bibitem{wang2009variable}
Wang, Z., Arce, G.R.: Variable density compressed image sampling. IEEE Transactions on image processing  \textbf{19}(1),  264--270 (2009)

\bibitem{yang2022l2sr}
Yang, P., Dong, B.: {L2SR}: Learning to sample and reconstruct for accelerated {MRI}. arXiv:2212.02190  (2022)

\bibitem{yin2021end}
Yin, T., Wu, Z., Sun, H., Dalca, A.V., Yue, Y., Bouman, K.L.: {End-to-End} sequential sampling and reconstruction for {MR} imaging. In: Machine Learning for Health Conference (2021)

\bibitem{zbontar2018fastmri}
Zbontar, J., Knoll, F., Sriram, A., Murrell, T., Huang, Z., Muckley, M.J., Defazio, A., Stern, R., Johnson, P., Bruno, M., et~al.: fast{MRI}: An open dataset and benchmarks for accelerated {MRI}. arXiv:1811.08839  (2018)

\bibitem{zhang2020extending}
Zhang, J., Zhang, H., Wang, A., Zhang, Q., Sabuncu, M., Spincemaille, P., Nguyen, T.D., Wang, Y.: Extending {LOUPE} for k-space under-sampling pattern optimization in multi-coil {MRI}. In: MLMIR (2020)

\bibitem{zhang2017beyond}
Zhang, K., Zuo, W., Chen, Y., Meng, D., Zhang, L.: Beyond a gaussian denoiser: Residual learning of deep cnn for image denoising. IEEE transactions on image processing  \textbf{26}(7),  3142--3155 (2017)

\bibitem{zhang2015denoising}
Zhang, X., Xu, Z., Jia, N., Yang, W., Feng, Q., Chen, W., Feng, Y.: Denoising of 3d magnetic resonance images by using higher-order singular value decomposition. Medical image analysis  \textbf{19}(1),  75--86 (2015)

\bibitem{zhang2019reducing}
Zhang, Z., Romero, A., Muckley, M.J., Vincent, P., Yang, L., Drozdzal, M.: Reducing uncertainty in undersampled {MRI} reconstruction with active acquisition. In: CVPR. pp. 2049--2058 (2019)

\end{thebibliography}
\end{document}